\documentclass[superscriptaddress,twocolumn,pre,nofootinbib]{revtex4-1}

\usepackage{amsmath}
\usepackage{graphicx,color}

\newcommand{\be}{\begin{equation}}
\newcommand{\ee}{\end{equation}}
\newcommand{\bea}{\begin{eqnarray}}
\newcommand{\eea}{\end{eqnarray}}

\begin{document}
\sloppy


\title{Models of universe with a polytropic equation of state \\
 I. The early universe}

\author{Pierre-Henri Chavanis}
\affiliation{Laboratoire de Physique Th\'eorique (IRSAMC), CNRS and UPS, Universit\'e de Toulouse, France}

\begin{abstract}

We construct models of universe with a generalized equation of state
$p=(\alpha \rho+k\rho^{1+1/n})c^2$ having a linear component and a
polytropic component. The linear equation of state $p=\alpha\rho c^2$
describes radiation ($\alpha=1/3$), pressureless matter ($\alpha=0$),
stiff matter ($\alpha=1$), and vacuum energy ($\alpha=-1$). The
polytropic equation of state $p=k\rho^{1+1/n} c^2$ may be due to
Bose-Einstein condensates with repulsive ($k>0$) or attractive ($k<0$)
self-interaction, or have another origin. In this paper, we consider
positive indices $n>0$. In that case, the polytropic component
dominates in the early universe where the density is high. For
$\alpha=1/3$, $n=1$ and $k=-4/(3\rho_P)$, we obtain a model of early
universe describing the transition from a pre-radiation era to the
radiation era.  The universe exists at any time in the past and there
is no singularity. However, for $t<0$, its size is less than the
Planck length $l_P=1.62\, 10^{-35}\, {\rm m}$. In this model, the
universe undergoes an inflationary expansion with the Planck density
$\rho_P=5.16 \, 10^{99}\, {\rm g}/{\rm m}^3$ that brings it to a size
$a_1=2.61\, 10^{-6}\, {\rm m}$ at $t_1=1.25\, 10^{-42}\, {\rm s}$
(about $20$ Planck times $t_P$). For $\alpha=1/3$, $n=1$ and
$k=4/(3\rho_P)$, we obtain a model of early universe with a new form
of primordial singularity: The universe starts at $t=0$ with an
infinite density and a finite radius $a=a_1$. Actually, this universe
becomes physical at a time $t_i=8.32\, 10^{-45}\, {\rm s}$ from which
the velocity of sound is less than the speed of light. When $a\gg
a_1$, the universe evolves like in the standard model. We describe the
transition from the pre-radiation era to the radiation era by analogy
with a second order phase transition where the Planck constant $\hbar$
plays the role of finite size effects (the standard Big Bang theory is
recovered for $\hbar=0$).

\end{abstract}

\maketitle

\section{Introduction}

The discovery of the expansion of the universe and the concept of a
singularity from which the universe emerged (Big Bang) has an interesting
history \cite{luminet,nb}. In this introduction, we provide a brief review of the early development of modern cosmology, describing  very elementary (yet fundamental) cosmological models that will play an important role in our study.

In 1917, Einstein \cite{einsteincosmo} applied the equations of general relativity to cosmology, assuming that the universe is homogeneous and isotropic. On the ground of cultural and philosophical beliefs, he considered a static universe. In order to accommodate for such a solution, he had to introduce a cosmological constant $\Lambda$ in the equations of general relativity\footnote{As a preamble to his paper, Einstein \cite{einsteincosmo} considered the Newtonian approximation and replaced the Poisson equation by an equation of the form $\Delta\Phi-(\Lambda/c^2)\Phi=4\pi G\rho$, so as to obtain a static solution $\Phi=-4\pi G\rho c^2/\Lambda$. In Einstein's belief, the Newtonian effect of the cosmological constant was to shield the gravitational interaction on a distance $c/\sqrt{\Lambda}$. This is actually  incorrect. Lema\^itre \cite{lemaitrecosmo} was the first to understand that the cosmological constant can be interpreted as a force of ``cosmic repulsion''. In the Newtonian context, the  modified Poisson equation including the cosmological constant is $\Delta\Phi=4\pi G\rho-\Lambda$, leading to the static solution $\rho=\Lambda/4\pi G$.}. He obtained a model of the universe with a positive curvature $k=+1$ that is finite (yet unbounded) with a ``radius'' $a_E=c/\sqrt{\Lambda}$, a density $\rho_E=\Lambda/4\pi G$, and a mass $M_E=\pi c^3/2G\sqrt{\Lambda}$.  The same year, de Sitter \cite{deSitter1,deSitter2}  discovered  another static solution of the Einstein equations (according to the review of Robertson \cite{robertson}, this solution was suggested to de Sitter by Ehrenfest). This universe is empty ($\rho=0$) and has a radius $a_{dS}=c\sqrt{3/\Lambda}$.  In this model, the stars and nebulae are just ``test particles'' that do not curve the universe.  De Sitter managed to predict that the light emitted  by these sources should be redshifted because (i) they move away from the observer due to the effect of the cosmological constant, and (ii) there is a slowing down of time with increasing distance from the observer (de Sitter effect).  He mentioned that some observations (presumably those made by Slipher since 1912) seem to confirm that claim. He also predicted a linear, or quadratic, relationship between the redshift and the distance. However, his interpretation of the redshifts remained enigmatic for a long time. Therefore, the early cosmological debate revolved around two deficient models: The model of Einstein which was unable to explain redshifts, and the model of de Sitter which contained no matter and was rather mysterious. Both were built on the assumption that the universe is static.

In 1922 and 1924, Friedmann \cite{friedmann1,friedmann2} discovered non-static cosmological solutions of the Einstein equations describing the temporal variation of space. He considered the case of spherical (positive  curvature $k=+1$) and hyperbolic (negative curvature $k=-1$) universes. He described expanding and contracting solutions, as well as cyclic solutions, and foresaw the possible beginning of the universe in a singularity. He obtained for the first time an estimate of the age of the universe of the order of $10$ billion years. At that time, Einstein did not believe in the significance of these non stationary solutions. He first argued that the Friedmann solutions were not compatible with the field equations of general relativity \cite{einsteincritique1}, then retracted his statement \cite{einsteincritique2} but remained convinced that the Friedmann solutions are not physically relevant \cite{luminet,nb}. Apart from Einstein, who criticized them, the papers of Friedmann remained unnoticed by the physical community until 1930.

In 1922, Lanczos \cite{lanczos1922} found a non-stationary form of de Sitter's solution in which the scale factor increases exponentially rapidly. He had the key to an expanding universe but did not recognize its physical significance.  In 1923, Weyl \cite{weyl} published a paper in which he tried to find his own way to de Sitter's redshift. He noted that the particles in de Sitter's universe move away from each other with a velocity proportional to their distance, but did not provide an explicit system of coordinates to make his claim more precise. In 1925, Lema\^itre \cite{lemaitre1925} found a system of coordinates in which de Sitter's model is non-static and leads to a linear relationship between the velocity (redshift) and the distance. This model explains the redshifts of distant objects in an unambiguous way. This model is Euclidean $k=0$ (which was problematic for Lema\^itre) and the scale factor increases exponentially rapidly with time due to the cosmological constant.

In 1927,  Lema\^itre \cite{lemaitre1927} rediscovered independently from Friedmann the non-stationary cosmological solutions of the Einstein equations. In addition, having spent two years in Harvard and at the MIT, he was aware of cosmological observations and related the expansion of space predicted by the theory of general relativity to the recession motion of galaxies observed by Slipher, Hubble and  Str\"omberg. He interpreted the redshift of the nebulae as a consequence of the expansion of the universe, predicted a linear relationship between the recession velocity and the distance (the nowdays called ``Hubble relationship'') and, using available data \cite{stromberg}, obtained an estimate of the ``Hubble constant'' $H=\dot a/a=v/r$ two years before Hubble \cite{hubble}\footnote{As pointed out by Luminet \cite{luminet} and Nussbaumer \& Bieri \cite{nb}, the Hubble law and the discovery of the expansion of the universe should be attributed in part to Lema\^itre, especially if we consider that Hubble did not relate his observations to the expansion of space in the theory of general relativity (and was later reluctant in accepting this idea), while this was Lema\^itre's main motivation. Indeed, Hubble interpreted the redshifts in terms of de Sitter's effect in a static universe, while Lema\^itre was the first to connect them to the expansion of the universe. Unfortunately, the paper of Lema\^itre was totally ignored at that time because it was published in French in a rather inaccessible Belgian journal. His paper was translated in English in 1931 \cite{lemaitre1931}, but for unknown reasons, the important paragraph of the paper where Lema\^itre gives the relation of proportionality between the recession velocity and the distance, and extracts the Hubble constant, is condensed into a single sentence. Because of this omission, Lema\^itre is not recognized on the same footing as Hubble for being the discoverer of the expansion of the universe \cite{luminet,nb}.}.  The work of Lema\^itre, like the one of his predecessor Friedmann, was still not appreciated by Einstein who found it ``abominable'' \cite{luminet,nb}.

In  1928 and 1929, Robertson \cite{robertson1928,robertson1929} reinterpreted de Sitter's solution dynamically. However, like Lanczos \cite{lanczos1922} and Weyl \cite{weyl}, he did not ``see'' the expanding universe.

In 1930, Eddington, who was working with McVittie on possible dynamical solutions of the cosmological Einstein equations (following a discussion that he had with de Sitter during a meeting of the Royal Astronomical Society \cite{deSitter1930a}), was alerted by Lema\^itre about his own results.  He publicized them in a paper \cite{eddington} in which he showed that the Einstein static universe is unstable (a result that was implicit in the paper of Lema\^itre). He also emphasized a  non-singular solution of Lema\^itre in which the universe starts from the static solution of Einstein for $t\rightarrow -\infty$, then expands (this model is often called the Eddington-Lema\^itre model). In parallel, de Sitter advertised Lema\^itre's discovery of an expanding universe in several publications \cite{sitter1930b,sitter1930c}. Together with the supporting observations of Hubble \cite{hubble}, and independent theoretical work by Robertson \cite{robertson1928,robertson1929}, this really meant the demise of the static universe.

In 1931, Einstein was forced to admit the theoretical and observational reality of the expansion of the universe. He took an extreme position and renounced to the cosmological constant \cite{einsteinzero} that he considered  being  the ``biggest blunder'' of his life (in contrast, Lema\^itre \cite{lemaitrecosmo} always defended the cosmological constant as one of the most fundamental ingredients of modern cosmology).  In 1932, Einstein published with  de Sitter a one-page paper \cite{eds}  in which they consider an  expanding universe in a Euclidean space (zero curvature $k=0$) without pressure ($p=0$) and without cosmological constant ($\Lambda=0$). This is the celebrated Einstein-de Sitter (EdS) solution that describes the matter era. The main motivation of these authors was to show that an expansion of the universe is possible even without the cosmological constant.

In the meantime, Lema\^itre had gone one step further. If the universe is expanding, it may
have started from a {\it singularity} in the past. In 1931, he proposed a model \cite{lemaitresingular} in which the universe begins its expansion from a singular initial state\footnote{Eddington \cite{eddington} also found singular solutions to the cosmological equations but rejected them as being unphysical.  In 1931, he wrote: ``Philosophically, the notion of a beginning of the present order of Nature is repugnant to me'' \cite{eddington1931}.}, or more precisely, from a near-singularity that  he called the ``primeval atom'' \cite{lemaitreNewton}\footnote{Lema\^itre \cite{lemaitre1933} also considered a model of universe in perpetual oscillation, undergoing cycles of expansion and contraction. He poetically  called it ``phoenix-universe''. However, he finally rejected this model at regret because he found that the duration of a cycle is too short. Cyclic universes were also discussed by Friedmann \cite{friedmann1,friedmann2} and Einstein \cite{einsteinzero}.}. This expansion is followed by a phase of stagnation during which the universe remains close to the static Einstein solution. In Lema\^itre's model, it is during this phase of stagnation that galaxies form. At later times, the universe enters in the de Sitter phase of exponential expansion. A new controversy started with Einstein who did not believe in the singular model of Lema\^itre that implies that the universe has a ``beginning''. This controversy remained until the end of their life. On the other hand, in 1960, Fred Hoyle, who promoted  a ``steady state theory'' \cite{hoyle} in which the universe has always been identical to itself and in which matter is created spontaneously, and continuously, made fun of Lema\^itre during a meeting in Pasadena by introducing him as the ``Big Bang man'' \cite{luminet}. This is how the term ``Big Bang'' was introduced in cosmology and became popular.

Up to the mid-1960's, it was not clear whether the early universe had been hot or cold.  The discovery  by Penzias and Wilson \cite{pw} of the $2.7\, {\rm K}$ cosmic microwave background radiation (CMB), whose existence had been predicted by the hot universe theory \cite{gamow,dn,dicke}, immediately led widespread acceptance to the Big Bang theory. According to that theory, the early universe was filled with an ultrarelativistic gas of photons, electrons, positrons, quarks, antiquarks, etc. This is described by an equation of state $p=\rho c^2/3$ from which it is easily deduced that the scale factor evolves as $a\propto t^{1/2}$, the density as $\rho\propto t^{-2}$, and the temperature as  $T\propto t^{-1/2}$. At $t=0$, the scale factor vanishes while the energy density and the temperature become infinite. This is why the point $t=0$ is known as the point of the initial cosmological singularity (Big Bang). As the universe expands, the density and the temperature decrease. When the universe is sufficiently cooled, particles and antiparticles annihilate each other, the photon-gas energy density falls off relatively rapidly and the main contribution of the matter density starts to come from the small excess of baryons over antibaryons, as well as from other fields and particles which now comprise the dark matter.

However, the hot universe theory suffers from a series of difficulties. The first difficulty is the singularity problem. The energy density goes to $+\infty$ at $t=0$ and the solution cannot be formally continued for $t<0$.  A natural question is: ``What was before $t=0$?''. Of course, one can argue that space and time started with the Big Bang, so that this question is meaningless. However, this answer is not quite satisfactory. On the other hand, close to the Big Bang, the density can be arbitrarily large. This is in conflict with the laws of physics. When the density becomes high enough, quantum effects should be taken into account in the problem\footnote{Remarkably, the question of the origin of space and time, and the importance of quantum mechanics in the early universe were discussed early by Lema\^itre \cite{lemaitrenature}: ``If the world has begun with a single quantum, the notion of space and time would altogether fail to have any meaning at the beginning''.}.  As a result, the Planck density $\rho_P=c^5/G^2\hbar=5.16\, 10^{99}\, {\rm g/m^3}$, the Planck length $l_P=(G\hbar/c^3)^{1/2}=1.62\, 10^{-35}\, {\rm m}$, the Planck mass $M_P=(\hbar c/G)^{1/2}=2.17\, 10^{-5}\, {\rm g}$, the Planck temperature $T_P=M_P c^2/k_B=1.42\, 10^{32}\, {\rm K}$, and the Planck time $t_P=(\hbar G/c^5)^{1/2}=5.39\, 10^{-44}\, {\rm s}$  should play a fundamental role in the very early universe.

The standard Big Bang theory suffers from other problems that are known as the flatness problem (or the fine-tuning problem), the horizon problem (or the causality problem), and the monopole problem. These problems can be solved at one stroke in the framework of inflationary cosmology \cite{guth,linde}. In that scenario, the universe was in an unstable vacuum-like state described by some hypothetical scalar field $\phi$ with its origin in quantum fluctuations. This  leads to an equation of state $p=-\rho c^2$, implying a constant energy density $\rho$, of the order of the Planck density $\rho_P$, called the vacuum energy. As a result of this vacuum energy, the universe expands exponentially rapidly in a tiniest fraction of a second. This ``inflation'' can solve the above-mentioned problems. At the same time, the inflation theory predicts the scale invariant spectrum of the primordial density fluctuations and explains why the universe is spatially flat (or behaves {\it as if} it were spatially flat due to the smallness of $k/a^2$). This is therefore a very attractive theory.

In this paper, we propose a model of the early universe based on a polytropic equation of state with index $n=1$. Before that, we first explain how we came to such a description.

In previous works \cite{c1,c2,c3}, we explored the possibility that the universe is filled with self-gravitating Bose-Einstein condensates (BEC) with short-range interactions (see a detailed list of references in \cite{c1} and recent additional references in \cite{recent}). In the strong coupling limit (the so-called Thomas-Fermi approximation), the equation of state of the BEC is $p=(2\pi a_s\hbar^2/m^3)\rho^2$, where $a_s$ is the scattering length \cite{dalfovo}. This is the equation of state of a polytrope with index $n=1$ \cite{chandra}. Initially, this model was introduced in order to describe dark matter halos. A nice feature of the BEC model is that it avoids the presence of cusps at the center of the halos  because of the effective repulsion due to the Heisenberg principle or because of the self-interaction of the bosons which plays the same role as the Pauli exclusion principle for fermions\footnote{If dark matter halos are made of self-gravitating fermions, the cusps are prevented by the Pauli exclusion principle; see \cite{c1} for discussions and references.}. Since cusps are not observed in real dark matter halos, the BEC model (or the fermion model) is favored as compared to the standard cold dark matter (CDM) model that predicts cuspy density profiles. Of course, at large scales, these models become equivalent.

The BEC model was then applied to cosmology  in order to describe the evolution of the universe as a whole \cite{harko,harko2,c4,kl,vw,harko3}. In \cite{c4}, we investigated the evolution of a ``BEC universe'' with an equation of state of the form $p=k\rho^2c^2$. We considered the case of repulsive ($k>0$) and attractive ($k<0$) self-interaction.  More generally, we studied a ``cosmic fluid'' described by an equation of state of the form $p=(\alpha \rho +k \rho^2)c^2$. When $k=0$, we recover the standard linear equation of state $p=\alpha \rho c^2$ describing radiation ($\alpha=1/3$), pressureless matter ($\alpha=0$), stiff matter ($\alpha=1$), and vacuum energy ($\alpha=-1$). Therefore, the generalized equation of state $p=(\alpha \rho +k\rho^2)c^2$ can be viewed as the sum of a linear term describing a ``classical'' universe and a quadratic term due to the BEC. At late times, when the density is low, the effect of the BEC is negligible and we recover the classical universe. On the contrary, the effect of the BEC becomes important in the early universe where the density is high. Therefore, it is interesting to study how the BEC  affects the early evolution of the universe. In our previous paper \cite{c4}, we assumed that BECs form after the radiation era, when the temperature has sufficiently decreased. This prevented us from extrapolating the solution before the radiation era (see the Remark in Sec. 8.2 of \cite{c4}). In the present paper, we explore the possibility that the equation of state   $p=k\rho^2c^2$ may hold {\it before} the radiation era. Therefore, we assume that the early universe is described by an equation of state of the form $p=(\alpha \rho +k \rho^2)c^2$ with $\alpha=1/3$. Actually, the study of this generalized equation of state is interesting even if its origin is not connected to BECs. Furthermore, for this class of models, the Friedmann equations can be solved analytically, which is an additional source of motivation. First results were given in \cite{harko,c4} where it was shown that the repulsive and attractive models behave very differently. For repulsive self-interactions ($k>0$), the universe starts at $t=0$ from a singularity at which the density is infinite but the radius finite. For attractive self-interactions ($k<0$), the universe  always existed in the past. For $t\rightarrow -\infty$, its density tends to a constant value while its radius tends to zero exponentially rapidly.

In the present paper, we complete the study of the generalized equation of state $p=(\alpha \rho +k \rho^2)c^2$. First, we analyze its thermodynamical properties and study the temporal evolution of the temperature. Secondly, for $k<0$, we show that the exponential growth of the scale factor can  account for a phase of inflation with constant density that we identify with the Planck density $\rho_P$. Finally, taking $\alpha=1/3$, we study the transition between a pre-radiation era and the radiation era. We mention the analogy with a second order phase transition where the control parameter is the time $t$, the order parameter is the scale factor $a(t)$, and the Planck constant $\hbar$ plays the role of finite size effects (the standard Big Bang theory is recovered for $\hbar=0$). For $k<0$, the radius of the universe passes from $a_i=l_P=1.62\, 10^{-35}\, {\rm m}$ at $t=t_i=0$ to $a_1=2.61\, 10^{-6}\, {\rm m}$ at $t=t_1=1.25\, 10^{-42}\, {\rm s}$ before entering in the radiation era and increasing algebraically as $t^{1/2}$. During the inflationary phase, the density remains approximately constant with the Planck value ($\rho\simeq \rho_P=5.16\, 10^{99}\, {\rm g}/{\rm m}^3$) before decreasing as $t^{-2}$. The temperature passes from $T_i=5.54\, 10^{-173}\, {\rm K}$ at $t=t_i=0$ to $T_1=3.93\, 10^{31}\, {\rm K}$ at $t=t_1$ before decreasing algebraically as $\sim t^{-1/2}$. It achieves its maximum value $T_e=7.40\, 10^{31}\, {\rm K}$ at $t=t_e=1.27\, 10^{-42}\, {\rm s}$.  For $k<0$, the initial radius of the universe $a(0)=a_1=2.61\, 10^{-6}\, {\rm m}$ is already ``large'' so there is no inflation. The velocity of sound is less than the speed of light only for $t>t_i=8.32\, 10^{-45}\, {\rm s}$, corresponding to $a_i=3.90\, 10^{-6}\, {\rm m}$, $\rho_i=1.29\, 10^{99}\, {\rm g}/{\rm m}^3$, and $T_i=1.64\, 10^{32}\, {\rm K}$. This marks the beginning of the physical universe in this model.

The paper is organized as follows. In Sec. \ref{sec_basic}, we recall the basic equations of cosmology and discuss famous cosmological models.  In Secs. \ref{sec_ges} and \ref{sec_dark}, we study the generalized equation of state  $p=(\alpha \rho +k\rho^{1+1/n})c^2$  for any value of the parameters $\alpha$, $k$ and $n>0$. We discuss two classes of solutions depending whether $k$ is positive or negative. In Secs. \ref{sec_inf} and \ref{sec_ni}, we consider a specific model of physical interest corresponding to $\alpha=1/3$, $n=1$, and $k=\pm 4/(3\rho_P)$.  For $k<0$ (negative polytropic pressure), we get a model of inflationary universe without singularity. It describes in a unified manner the transition from a pre-radiation era to the radiation era. For $k>0$ (positive polytropic pressure), we get a model of non-inflationary universe with a new type of initial singularity. In Sec. \ref{sec_analogy}, we develop an analogy with second order phase transitions where the Planck constant plays the role of finite size effects.

\section{Basic equations of cosmology}
\label{sec_basic}

In this section, we recapitulate the basic equations of cosmology \cite{weinberg} and discuss some famous models of universe that were mentioned in the Introduction. This will prepare the ground for the next sections in which we discuss generalized cosmological models. Some readers may directly go to Sec. \ref{sec_ges}.

\subsection{The Einstein equations}
\label{sec_einstein}

In a space with uniform curvature, the line element is given by the
Friedmann-Lema\^itre-Roberston-Walker (FLRW) metric
\begin{eqnarray}
\label{e1}
ds^2=c^2 dt^2-a(t)^2\left\lbrace \frac{dr^2}{1-k r^2}+r^2\,  (d\theta^2+ \sin^2\theta\,  d\phi^2)\right \rbrace,\nonumber\\
\end{eqnarray}
where $a(t)$ represents the radius of curvature of the $3$-dimensional space, or the scale factor. By an abuse of language, we shall call it the ``radius of the universe''. On the other hand, $k$ determines the curvature of space. The universe can be closed ($k>0$), flat ($k=0$), or open ($k<0$).

If the universe is isotropic and homogeneous at all points in conformity with the line element (\ref{e1}), and contains a uniform perfect fluid of energy density $\epsilon(t)=\rho(t) c^2$ and isotropic pressure $p(t)$, the energy-momentum tensor $T^{i}_{j}$ is
\begin{eqnarray}
\label{e2}
T^0_0=\rho c^2,\qquad T^1_1=T^2_2=T^3_3=-p.
\end{eqnarray}
The Einstein equations
\begin{eqnarray}
\label{e3}
R^i_j-\frac{1}{2}g^i_j R-\Lambda g^i_j=-\frac{8\pi G}{c^2}T^i_j,
\end{eqnarray}
relate the geometrical structure of the spacetime ($g_{ij}$) to the material content of the universe ($T_{ij}$). For the sake of generality, we have included the cosmological constant $\Lambda$. Given (\ref{e1}) and (\ref{e2}), these equations reduce to
\begin{eqnarray}
\label{e4}
8\pi G\rho+\Lambda=3 \frac{{\dot a}^2+kc^2}{a^2},
\end{eqnarray}
\begin{eqnarray}
\label{e5}
\frac{8\pi G}{c^2}p-\Lambda=-\frac{2 a \ddot a+{\dot a}^2+kc^2}{a^2},
\end{eqnarray}
where dots denote differentiation with respect to time. These are the well-known cosmological equations first derived by Friedmann \cite{friedmann1,friedmann2}.

\subsection{The Friedmann equations}
\label{sec_friedmann}

The Friedmann equations are usually written in the form
\begin{equation}
\label{f1}
\frac{d\rho}{dt}+3\frac{\dot a}{a}\left (\rho+\frac{p}{c^2}\right )=0,
\end{equation}
\begin{equation}
\label{f2}
\frac{\ddot a}{a}=-\frac{4\pi G}{3} \left (\rho+\frac{3p}{c^2}\right )+\frac{\Lambda}{3},
\end{equation}
\begin{equation}
\label{f3}
H^2=\left (\frac{\dot a}{a}\right )^2=\frac{8\pi G}{3}\rho-\frac{kc^2}{a^2}+\frac{\Lambda}{3},
\end{equation}
where we have introduced the Hubble parameter $H=\dot a/a$.  Among these three equations, only two are independent. The first equation, which can be viewed as an ``equation of continuity'', can be directly derived from the conservation of the energy momentum tensor $\partial_i T^{ij}=0$ which results from the Bianchi identities. For a given barotropic equation of state $p=p(\rho)$, it determines the relation between the density and the scale factor. Then, the temporal evolution of the scale factor is given by Eq. (\ref{f3}).

Equivalent expressions of the ``equation of continuity'' are
\begin{equation}
\label{f4}
a\frac{d\rho}{da}=-3\left (\rho+\frac{p}{c^2}\right ),
\end{equation}
\begin{equation}
\label{f5}
\frac{d}{da}(\rho a^3)=-3\frac{p}{c^2}a^2,
\end{equation}
\begin{equation}
\label{f6}
a^3\frac{dp}{dt}=\frac{d}{dt}\left \lbrack a^3 (p+\rho{c^2})\right \rbrack.
\end{equation}
Introducing the volume $V=(4/3)\pi a^3$ and the energy $E=\rho c^2 V$, Eq. (\ref{f5}) becomes $dE=-pdV$. This can be viewed as the first principle of thermodynamics for an adiabatic evolution of the universe $dS=0$. This thermodynamical interpretation of the Friedmann equation (\ref{f1}) was first given by Lema\^itre \cite{lemaitre1927,lemaitre1931}.

In most parts of this paper (except in Sec. \ref{sec_fm}), we shall take $\Lambda=0$. This is because we are interested in the very early universe in which the pressure of radiation dominates over the dark energy (cosmological constant). On the other hand, we consider a flat universe ($k=0$) in agreement with the observations of the cosmic microwave background (CMB) \cite{bt}. Then, the Friedmann equations reduce to
\begin{equation}
\label{f7}
\frac{d\rho}{dt}+3\frac{\dot a}{a}\left (\rho+\frac{p}{c^2}\right )=0,
\end{equation}
\begin{equation}
\label{f8}
\frac{\ddot a}{a}=-\frac{4\pi G}{3} \left (\rho+\frac{3p}{c^2}\right ),
\end{equation}
\begin{equation}
\label{f9}
H^2=\left (\frac{\dot a}{a}\right )^2=\frac{8\pi G}{3}\rho.
\end{equation}

The Friedmann equations can also be derived in the context of
Newtonian cosmology. In this framework, Milne \cite{milne} and McCrea
\& Milne \cite{mcCreamilne} first obtained the Friedmann equations
without pressure ({\it i.e.} for $c\rightarrow +\infty$). Later,
McCrea \cite{mcCrea} introduced relativistic effects in Newtonian
cosmology and obtained the complete set of Friedmann equations with
pressure. Another derivation was given by Harrisson \cite{harri} who
obtained a relativistic generalization of the Euler-Poisson system in
the context of Newtonian cosmology. Finally, the continuity equation
was corrected in a recent paper by Lima {\it et al.} \cite{lima}.
Actually, the derivation of the Friedmann equations with relativistic
effects (pressure) from Newtonian cosmology was first exposed by
Lema\^itre \cite{lemaitreNewton} as early as 1931 (see also
\cite{lemaitre1934}).  However, for Lema\^itre, the Newtonian
derivation of the cosmological equations is just a pedagogical means
and does not reflect the true (relativistic) nature of the problem. In
modern textbooks of astrophysics \cite{peeblesbook,bt}, Newtonian
cosmology with pressure is presented as a rigorously valid description
of the universe in a region small compared to the Hubble length $c/H$.

{\it Remark:} Considering Eq. (\ref{f2}), we see that the cosmological constant corresponds to a force $F=\Lambda a/3$ in the Newtonian interpretation. A positive cosmological constant (de Sitter) acts as a force of cosmic repulsion and a  negative cosmological constant (anti de Sitter) acts as a force of cosmic attraction whose effect is similar to that of a harmonic oscillator. On the other hand, a positive pressure has an attractive character while a negative pressure has a repulsive character. Gravity is attractive (the universe decelerates) when $\rho+3p/c^2>0$ and repulsive (the universe accelerates) when $\rho+3p/c^2<0$.

\subsection{The thermodynamical equation}
\label{sec_t}

From the first principle of thermodynamics
\begin{equation}
\label{t1}
dS(V,T)=\frac{1}{T}\left\lbrack d(\rho(T)c^2 V)+p(T)dV\right\rbrack,
\end{equation}
one can derive the thermodynamical equation \cite{weinberg}:
\begin{equation}
\label{t2}
\frac{dp}{dT}=\frac{1}{T}(\rho c^2+p).
\end{equation}
For a given barotropic equation of state $p=p(\rho)$, this equation can be used to obtain the relation $T=T(\rho)$ between the temperature and the density. Combining Eq. (\ref{t2}) with Eq. (\ref{f6}),  we get
\begin{equation}
\label{t3}
\frac{dS}{dt}=0,
\end{equation}
where
\begin{equation}
\label{t4}
S=\frac{a^3}{T}(p+\rho c^2).
\end{equation}
is the entropy of the universe in a volume $a^3$. This confirms that the Friedmann equations (\ref{f1})-(\ref{f3}) imply the conservation of the entropy \cite{weinberg}.

\subsection{Famous models of the universe}
\label{sec_fm}

In 1917, Einstein \cite{einsteincosmo} constructed a static model of the universe. He
assumed that the universe if filled with a pressureless gas ($p=0$)
and introduced a cosmological constant $\Lambda$ in the equations of
general relativity\footnote{Actually, there exist another static
solution of the Einstein equations without cosmological constant
corresponding to $p=-(1/3)\rho c^2$, $k=+1$ and $\rho=3c^2/(8\pi
Ga^2)$.}. Setting $\dot a =\ddot a=0$ in Eqs. (\ref{f1})-(\ref{f3}), it is
necessary that $\Lambda>0$ and $k=+1$. Then,
\begin{equation}
\label{fm1}
a_E=\frac{c}{\sqrt{\Lambda}}, \quad \rho_E=\frac{\Lambda}{4\pi G}=\frac{c^2}{4\pi G a_E^2}.
\end{equation}
The static Einstein universe is finite (though unbounded) with a positive curvature and a density fixed by $\Lambda$ and $G$. The volume of this universe is $V_E=2\pi^2 a_E^3$. Its mass $M_E=\rho_E V_E$ is therefore
\begin{equation}
\label{fm2}
M_E=\frac{\pi c^3}{2G\sqrt{\Lambda}}=\frac{\pi c^2 a_E}{2G}=2\pi^2\left (\frac{c^2}{4\pi G}\right )^{3/2}\frac{1}{\sqrt{\rho_E}}.
\end{equation}

The same year, de Sitter \cite{deSitter1,deSitter2} found another
cosmological solution of the Einstein field equations corresponding to
an empty universe ($p=\rho=0$) with a cosmological constant
$\Lambda>0$.  In his original article, he aimed at constructing a
static solution. He obtained a radius
$a_{dS}=c\sqrt{3/\Lambda}$. However, it was understood later by
Lema\^itre \cite{lemaitre1925} that, by using a proper change of
variables in the metric, the de Sitter solution actually describes a
dynamical universe whose radius increases exponentially rapidly with
time. Setting $p=\rho=0$ and $k=0$ in Eqs. (\ref{f1})-(\ref{f3}), we
get
\begin{equation}
\label{fm3}
a(t)=a(0)e^{\sqrt{\Lambda/3}t}.
\end{equation}
In this model, the Hubble parameter is constant: $H=\dot a/a=\sqrt{\Lambda/3}$. It can be written $H=c/a_{dS}$ where $a_{dS}=c\sqrt{3/\Lambda}$ is the radius of the original (static) de Sitter model. The solution (\ref{fm3}) describes an empty universe ($\rho=0$) that expands because of the repulsive nature of the cosmological constant. Any model with $\Lambda>0$ goes to a de Sitter model for $a\rightarrow +\infty$.

\begin{figure}[!h]
\begin{center}
\includegraphics[clip,scale=0.3]{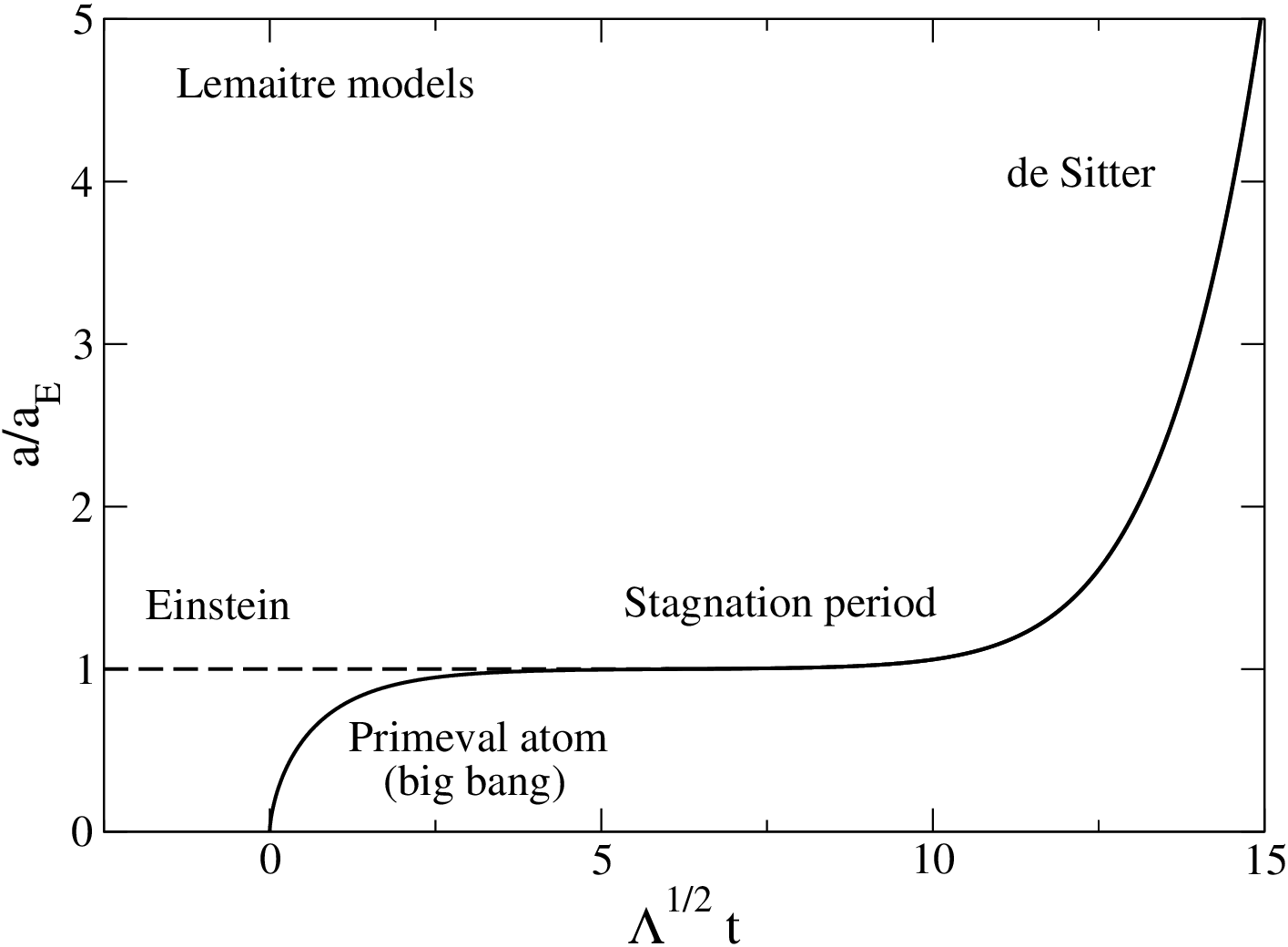}
\caption{Evolution of the scale factor in the Eddington-Lema\^itre universe (dashed-line) and in the Lema\^itre universe (full line). We have taken $\alpha=1.00001$.}
\label{lemaitre}
\end{center}
\end{figure}

The Lema\^itre model of 1927 corresponds to $k=+1$, $\Lambda>0$, $p=0$ and $\rho>0$ \cite{lemaitre1927,lemaitre1931}. The equation of continuity (\ref{f1}) implies $\rho a^3\propto 1$ which can be interpreted as the conservation of mass-energy for a pressureless universe. We can write this relation in the form $\rho=\alpha M_E/(2\pi^2 a^3)$, where $M_E$ is Einstein's mass and $\alpha$ is a dimensionless constant. Introducing the dimensionless scale factor $R=a/a_E$ and the dimensionless time $\tau=\sqrt{\Lambda}t$, we can rewrite the Friedmann equation  (\ref{f3}) in the form
\begin{equation}
\label{fm4}
\left (\frac{\dot R}{R}\right )^2=\frac{2\alpha}{3R^3}-\frac{1}{R^2}+\frac{1}{3}.
\end{equation}
For $\alpha=1$, this equation admits the static Einstein solution $R=1$ ({\it i.e.} $a=a_E$). However, as shown by Lema\^itre \cite{lemaitre1927,lemaitre1931}, and  emphasized by Eddington \cite{eddington}, this solution is unstable and, for sufficiently large times, the radius of the universe expands. This solution, according to which the radius of the universe is equal to the Einstein radius for $\tau\rightarrow -\infty$ and increases for large times (behaving ultimately like the de Sitter solution (\ref{fm3}) for $\tau\rightarrow +\infty$) is called the Eddington-Lema\^itre model (see Fig. \ref{lemaitre}). This model does not have an initial singularity.

For $\alpha=1$, the differential equation (\ref{fm4}) has the analytical solution
\begin{eqnarray}
\label{fm5}
\tau=(R-1)\sqrt{\frac{2+R}{R^3-3R+2}}\biggl \lbrack 2\sqrt{3}\, {\rm Argsinh}\left (\sqrt{\frac{R}{2}}\right )\nonumber\\
+\ln\left (\frac{\sqrt{R}-1}{\sqrt{R}+1}\right )+\ln\left (\frac{\sqrt{3(2+R)}+2-\sqrt{R}}{\sqrt{3(2+R)}+2+\sqrt{R}}\right )\biggr\rbrack,\qquad
\end{eqnarray}
where $\tau$ is defined up to an additional constant. For $\tau\rightarrow -\infty$, we obtain
\begin{eqnarray}
\label{fm6}
R-1\sim e^{\tau-C_1},\qquad ({\rm Einstein})
\end{eqnarray}
with $C_1=2\sqrt{3}\, {\rm Argsinh}(1/\sqrt{2})-\ln(6)\simeq 0.489279$. For $\tau\rightarrow +\infty$, we get
\begin{eqnarray}
\label{fm7}
R\sim \frac{1}{2}e^{(\tau+C_2)/\sqrt{3}} \qquad ({\rm de \, Sitter}),
\end{eqnarray}
with $C_2=\ln\lbrack (\sqrt{3}+1)/(\sqrt{3}-1)\rbrack\simeq 1.31696$. To the best of our knowledge, these expressions have not been given previously.

In 1931, Lema\^itre \cite{lemaitresingular} considered a model in which $\alpha$ is slightly larger than one ({\it i.e.} it contains more matter than the static Einstein model). This model starts from a singularity at $t=0$. The radius first increases as $R\sim (3\alpha/2)^{1/3}\tau^{2/3}$ (this corresponds to the solution (\ref{dust1}) found later by Einstein \& de Sitter \cite{eds}), then slows down and reaches a minimum rate at $R\simeq \alpha^{1/3}$, after which it re-expands rapidly, ultimately approaching the de Sitter solution (\ref{fm3}) for $\tau\rightarrow +\infty$. The main feature of this model is the existence of a ``stagnation period''\footnote{Lema\^itre thought that galaxies formed during this stagnation period, and that this ``instability'' triggered the expansion of the universe.} during which the radius remains close to $R\simeq \alpha^{1/3}$. During the stagnation period, the differential equation (\ref{fm4}) takes the approximate form
\begin{equation}
\label{fm8}
{\dot R}^2\simeq \alpha^{2/3}-1+(R-\alpha^{1/3})^2,
\end{equation}
whose  solution is \cite{weinberg}:
\begin{equation}
\label{fm9}
R=\alpha^{1/3}\left\lbrack 1+(1-\alpha^{-2/3})^{1/2}\sinh (\tau-\tau_m)\right\rbrack,
\end{equation}
where $\tau_m$ is the time at which $\dot R$ reaches its minimum. If
$\alpha$ is close to unity, then $a$ will remain close to the Einstein
static value $a_E$ for a long time $\Delta
\tau=|\ln(1-\alpha^{-2/3})|$.  For
$\alpha=1$, we recover the Eddington-Lema\^itre model. However, the
historical interest of the Lema\^itre model \cite{lemaitresingular} of
1931 is that it predicts for the first time that the universe may have
emerged from a {\it singularity} (or near-singularity) at $t=0$ in
which the radius is equal to zero and the density is infinite. This is
what Lema\^itre called the ``primeval atom'' \cite{lemaitreNewton}, and 
which became later the Big Bang theory.

\subsection{Linear equation of state}
\label{sec_leos}

In early works of cosmology, it was assumed that the universe is pressureless  ($p=0$). In later works, pressure effects have been taken into account. Many workers have considered a linear equation of state
\begin{equation}
\label{leos1}
p=\alpha \rho c^2,
\end{equation}
with $-1\le \alpha\le 1$. The pressure is positive for $\alpha>0$ and negative for $\alpha<0$. For this equation of state, the Friedmann equations  (\ref{f7}) leads to the relation
\begin{equation}
\label{leos5}
\rho a^{3(1+\alpha)}\propto 1.
\end{equation}
In the following, we assume $\alpha\neq -1$ (the case $\alpha=-1$ is treated in Sec. \ref{sec_vacuum}). Substituting Eq. (\ref{leos5}) in Eq. (\ref{f9}) and solving the resulting equation for $a(t)$, we find that the radius of the universe increases with time as
\begin{equation}
\label{leos6}
a\propto t^{2/\lbrack 3(1+\alpha)\rbrack}.
\end{equation}
We have considered expanding solutions and defined the origin of time $t=0$ such that $a(0)=0$. The Hubble parameter and the density decrease with time as
\begin{equation}
\label{leos8}
H=\frac{\dot a}{a}=\frac{2}{3(1+\alpha)t},\qquad \rho=\frac{1}{6\pi G(1+\alpha)^2 t^2}.
\end{equation}
In all these models, the universe is singular at $t=0$. The radius vanishes while the density and the Hubble parameter  are infinite (Big Bang singularity).

For the equation of state (\ref{leos1}), the thermodynamical equation (\ref{t2}) can be integrated into
\begin{equation}
\label{leos10}
T\propto \rho^{\alpha/(\alpha+1)}\propto \frac{1}{a^{3\alpha}}.
\end{equation}
The temperature decreases with time when $\alpha>0$ and increases with time when $\alpha<0$.

According to Eq. (\ref{f8}), the universe is decelerating for
$\alpha>-1/3$ and  accelerating for $\alpha<-1/3$. For $\alpha=-1/3$, the scale factor increases
linearly with time ($a\propto t$).

Finally, the square of the velocity of sound is $c_s^2=p'(\rho)=\alpha c^2$. The velocity of sound is real for $\alpha\ge 0$ (it is less than the speed of light for $\alpha\le 1$) and imaginary for $\alpha<0$. It vanishes for $\alpha=0$.

\subsubsection{Matter (Einstein-de Sitter universe)}
\label{sec_dust}

A pressureless universe ($p=0$) made of baryonic matter or dark matter (dust) corresponds to $\alpha=0$. This is the so-called Einstein-de Sitter (EdS) universe. In this model,
\begin{equation}
\label{dust1}
\rho \propto 1/a^3, \qquad a\propto t^{2/3},
\end{equation}
\begin{equation}
\label{dust2}
H=\frac{2}{3t},\qquad \rho=\frac{1}{6\pi G t^2}.
\end{equation}

\subsubsection{Radiation}
\label{sec_radiation}

The equation of state of a radiation (photon gas) is $p=\rho c^2/3$, corresponding to $\alpha=1/3$. This yields
\begin{equation}
\label{rad1}
\rho \propto 1/a^4, \qquad a\propto t^{1/2},
\end{equation}
\begin{equation}
\label{rad2}
H=\frac{1}{2t},\qquad \rho=\frac{3}{32\pi G t^2},
\end{equation}
\begin{equation}
\label{rad3}
T\propto \rho^{1/4}\propto \frac{1}{a}\propto \frac{1}{t^{1/2}}.
\end{equation}
The relation between the energy density and the temperature corresponds to the Stefan-Boltzmann law
\begin{equation}
\label{rad4}
\rho c^2=\sigma T^4, \qquad \sigma=\frac{\pi^2 k_B^4}{15 c^3 \hbar^3}.
\end{equation}

Introducing the Planck scales, we can write the foregoing equations in the form
\begin{equation}
\label{rad5}
\rho= \frac{3\rho_P}{32\pi}\left (\frac{t_P}{t}\right )^2,
\end{equation}
\begin{equation}
\label{rad6}
T= T_P \left (\frac{15}{\pi^2}\right )^{1/4} \left (\frac{\rho}{\rho_P}\right )^{1/4},
\end{equation}
\begin{equation}
\label{rad7}
T= T_P \left (\frac{45}{32\pi^3}\right )^{1/4} \left (\frac{t_P}{t}\right )^{1/2}.
\end{equation}

\subsubsection{Stiff equation of state}
\label{sec_stiff}

A stiff equation of state $p=\rho c^2$, for which the velocity of sound $c_s=(dp/d\rho)^{1/2}$ is equal to the speed of light $c$, corresponds to $\alpha=1$. In that case,
\begin{equation}
\label{stiff1}
\rho \propto 1/a^6, \qquad a\propto t^{1/3},
\end{equation}
\begin{equation}
\label{stiff2}
H=\frac{1}{3t},\qquad \rho=\frac{1}{24\pi G t^2},
\end{equation}
\begin{equation}
\label{stiff3}
T\propto \rho^{1/2}\propto \frac{1}{a^3}\propto \frac{1}{t}.
\end{equation}

\subsubsection{Vacuum energy and dark energy}
\label{sec_vacuum}

For the equation of state  $p=-\rho c^2$, corresponding to $\alpha=-1$, the density and the Hubble parameter are constant
\begin{equation}
\label{vacuum1}
\rho \propto 1, \qquad H=\sqrt{\frac{8\pi G\rho}{3}},
\end{equation}
and the scale factor increases exponentially rapidly as
\begin{equation}
\label{vacuum2}
a(t)=a(0)e^{H t}.
\end{equation}

Lema\^itre \cite{lemaitre1934} was the first to understand that the effect of the  cosmological constant $\Lambda$ is equivalent to that of a fluid with density  $\rho=\rho_\Lambda\equiv \Lambda/8\pi G$ [see Eq. (\ref{f3}) or (\ref{fm3})] described by an equation of state $p=-\rho c^2$. He interpreted $\rho_{\Lambda}$ as the vacuum energy.  The origin of the vacuum energy was later discussed by Sakharov \cite{sakharov} and Zeldovich \cite{zeldovich} in relation to quantum field theory. However, there is a fundamental difficulty in interpreting $\rho_{\Lambda}$ as the vacuum energy because quantum field theory predicts that the vacuum energy density should be of the order of $\rho_P$ which is $122$ orders of magnitude larger than $\rho_{\Lambda}$. This is known as the cosmological constant problem \cite{weinbergcosmo}. Therefore, $\rho_{\Lambda}$ may not be related to the vacuum energy (hence to quantum fluctuations) but could be a fundamental property of general relativity (see discussion in Paper II). In the modern terminology, the effect of the cosmological constant, or the effect of an exotic fluid, on the accelerated expansion of the universe is called dark energy \cite{cst}.

The equation of state $p=-\rho c^2$ also appeared in the paper of McCrea \cite{mcCrea} in his discussion of the steady state theory of Hoyle \cite{hoyle}. In this theory, in which there is a continuous creation of matter, the density and the Hubble parameter are constant, leading to Eq. (\ref{vacuum2}). According to Hoyle \cite{hoyle}, a continuous creation of matter can be achieved by a  proper  modification of the Einstein equations. McCrea \cite{mcCrea} noted that the same result could be obtained with an equation of state $p=-\rho c^2$  without invoking creation of matter and without having to modify the Einstein equations.

The equation of state $p=-\rho c^2$ also appeared in inflationary models of the early universe \cite{linde}. It leads to a constant density which is identified with the Planck density $\rho_P$. In that case, this equation of state may well be related to the vacuum energy density.

\subsection{Necessity of the inflation}
\label{sec_nec}

The previous models, based on a linear equation of state, lead to a finite time singularity at $t=0$ (except the model with $\alpha=-1$). This poses several problems. We briefly recall what the problems are and show in the next sections how a modification of the equation of state can resolve them.

We call $H_0$ the present value of the Hubble constant (the subscript  $0$ refers to present-day values). The Hubble time is $1/H_0$. It gives an estimate of the age of the universe.  The characteristic ``size'' of the universe (distance of the cosmological horizon) is the Hubble length $a_0={c}/{H_0}$. Using $H_0=2.27\, 10^{-18}\, {\rm s}^{-1}$, we find that $1/H_0=4.41\, 10^{17}\, {\rm s}$ and $a_0=1.32\, 10^{26}\, {\rm m}$. For a flat space ($k=0$), the density of the universe $\rho_0$ can be obtained from the Friedmann equation
\begin{equation}
\label{nec2}
\rho_0=\frac{3H_0^2}{8\pi G}=\frac{3}{8\pi}(H_0 t_P)^2\rho_P.
\end{equation}
Numerically, $\rho_0=9.20\, 10^{-24}\, {\rm g}/{\rm m}^3$. On the other hand, the temperature measured from the cosmic microwave
background (CMB) is $T_0=2.73\, {\rm K}$.

The early universe is dominated by radiation. In the standard model, the radiation phase leads to a finite time singularity at $t=0$ (Big Bang). It is expected that quantum mechanics will come into play at the Planck time $t_P$ when the universe is very ``small''. At that time, the density and the temperature are (see Sec. \ref{sec_radiation}):
\begin{equation}
\label{nec3}
\rho(t_P)= \frac{3}{32\pi}\rho_P=1.54\, 10^{98}\, {\rm g}/{\rm cm}^3,
\end{equation}
and
\begin{equation}
\label{nec4}
T(t_P)= \left (\frac{45}{32\pi^3}\right )^{1/4} T_P=6.54\, 10^{31}\, {\rm K}.
\end{equation}
The particle horizon at the Planck time is of the order of the Planck length $l_P=c t_P=1.62\, 10^{-35}\, {\rm m}$. Since $T(t)a(t)$ is invariant during the evolution of the universe, we have
\begin{equation}
\label{nec6}
\left (\frac{45}{32\pi^3}\right )^{1/4}T_P \, a(t_P)=T_0 a_0.
\end{equation}
If we take $a(t_P)=l_P$, we find that the present size of the universe is of order $3.87\, 10^{-4}\, {\rm m}$ which is obviously too small.  The same argument can be turned the other way round. Using the present size of the universe $a_0=1.32\, 10^{26}\, {\rm m}$, we find that its size at the Planck time is $a(t_P)=5.52\, 10^{-6}\, {\rm m}$ which is much larger than the Planck length $l_P$ by a factor $\sim 10^{29}$.

It is instructive to recover this result in a slightly different, but equivalent, manner. Since $\rho_{rad}(t)a(t)^4$ is conserved during the evolution of the universe, we have $\rho(t_P) a(t_P)^4=\rho_{rad,0}\, a_0^4$. Writing $\rho_{rad,0}=\Omega_{rad,0}\rho_0$, and using Eq. (\ref{nec2}), we obtain
\begin{equation}
\label{nec7}
\frac{a(t_P)}{a_0}=(4\Omega_{rad,0})^{1/4}\sqrt{H_0 t_P}=4.75\, 10^{-32}.
\end{equation}
Since
\begin{equation}
\label{nec8}
\frac{l_P}{a_0}={H_0 t_P}=1.22\, 10^{-61},
\end{equation}
the foregoing equation can be rewritten
\begin{equation}
\label{nec9}
\frac{a(t_P)}{l_P}=(4\Omega_{rad,0})^{1/4}\frac{1}{\sqrt{H_0 t_P}}=3.88\, 10^{29}.
\end{equation}
Since $H_0 t_P=1.22\, 10^{-61}$ (meaning that the Hubble time $1/H_0$ is tremendously larger than the Planck time $t_P$), we see that
\begin{equation}
\label{nec10}
l_P\ll a(t_P)\ll a_0.
\end{equation}
The fact that  $a(t_P)\gg l_P$ makes problem \cite{linde}. One must find a mechanism to bring the ``size'' of the universe from $l_P=1.62\, 10^{-35}\, {\rm m}$ to $a(t_P)=6.28\, 10^{-6}\, {\rm cm}$ in a very short lapse of time. This can be achieved with the inflationary scenario.  The usual inflationary scenario is based on the equation of state $p=-\rho c^2$ (vacuum energy). As we have seen in Sec. \ref{sec_vacuum}, this implies that $\rho$ and $H$ are constant so that the scale factor increases exponentially rapidly with time, as $e^{Ht}$ with $H=(8\pi G\rho/3)^{1/2}$. If we assume that $\rho\sim \rho_P$, we obtain $H=(8\pi/3)^{1/2}t_P^{-1}$ so that the exponential time scale is of the order of the Planck time. This solves the above-mentioned problem \cite{linde}.  In the following sections, we show that a polytropic equation of state with $n>0$ produces similar results and leads to simple cosmological models unifying the phases of inflation and radiation.

\section{Generalized equation of state with positive index}
\label{sec_ges}

We consider a generalized equation of state of the form
\begin{equation}
\label{ges0}
p=(\alpha \rho+k\rho^{1+1/n}) c^2.
\end{equation}
This equation of state was introduced in our previous paper \cite{c4} for the specific index $n=1$. This is the sum of a standard linear equation of state $p=\alpha\rho c^2$ and a polytropic equation of state $p=k\rho^{\gamma} c^2$, where $k$ is the polytropic constant and $\gamma=1+1/n$ is the polytropic index. Concerning the linear equation of state, we assume $-1\le \alpha\le 1$  (the case $\alpha=-1$ is treated specifically in Appendix \ref{sec_eosgm}). This equation of state describes radiation ($\alpha=1/3$), pressureless matter ($\alpha=0$), and  vacuum energy ($\alpha=-1$). Concerning the polytropic equation of state, we remain very general. This polytropic equation of state may correspond to self-gravitating BECs with repulsive ($k>0$) or attractive ($k<0$) self-interaction (the standard BEC model corresponds to $n=1$) \cite{c4}, but it may have another origin.

For an equation of state of the form (\ref{ges0}) with positive index $n>0$, the polytropic component dominates the linear component when the density is high. These models, studied in the present paper,  describe the early universe. Conversely, when $n<0$, the polytropic component dominates  the linear component when the density is low. These models, studied in Paper II, describe the late universe. As will be discussed in detail in Paper II, these two situations are strikingly similar. They reveal a form of ``symmetry'' between the early and the late universe. In Papers I and II, we assume that $\alpha+1+k\rho^{1/n}\ge 0$. This corresponds to the ``normal'' case where the density decreases with the radius (see below). The opposite case $\alpha+1+k\rho^{1/n}\le 0$, leading to a ``phantom universe'' where the density increases with the radius, is considered in Paper III.

The purpose of our series of papers  is to make an exhaustive study of all the possible cases $(\alpha,n,k)$, even if some of them seem to be unphysical or in conflict with the known  properties of our universe. Actually, by rejecting the cases leading to past or future singularities, we will be led naturally to the ``good'' model (see Paper II).

\subsection{The density}
\label{sec_gesdensity}

For the equation of state (\ref{ges0}), the Friedmann equation (\ref{f7}) becomes
\begin{equation}
\label{ges1}
\frac{d\rho}{dt}+3\frac{\dot a}{a}\rho (1+\alpha+k\rho^{1/n})=0.
\end{equation}
Assuming $\alpha+1+k\rho^{1/n}\ge 0$, this equation can be integrated into
\begin{equation}
\label{ges3}
\rho=\frac{\rho_*}{\left\lbrack (a/a_*)^{3(1+\alpha)/n}\mp 1\right\rbrack^n},
\end{equation}
where  $\rho_*=\lbrack (\alpha+1)/|k|\rbrack^n$ and $a_*$ is a constant of integration. The upper sign corresponds to $k>0$ (repulsive self-interaction) and the lower sign corresponds to $k<0$ (attractive self-interaction).  We may draw a parallel between these two types of distributions and the statistics of fermions ($+$) and bosons ($-$).

\begin{figure}[!h]
\begin{center}
\includegraphics[clip,scale=0.3]{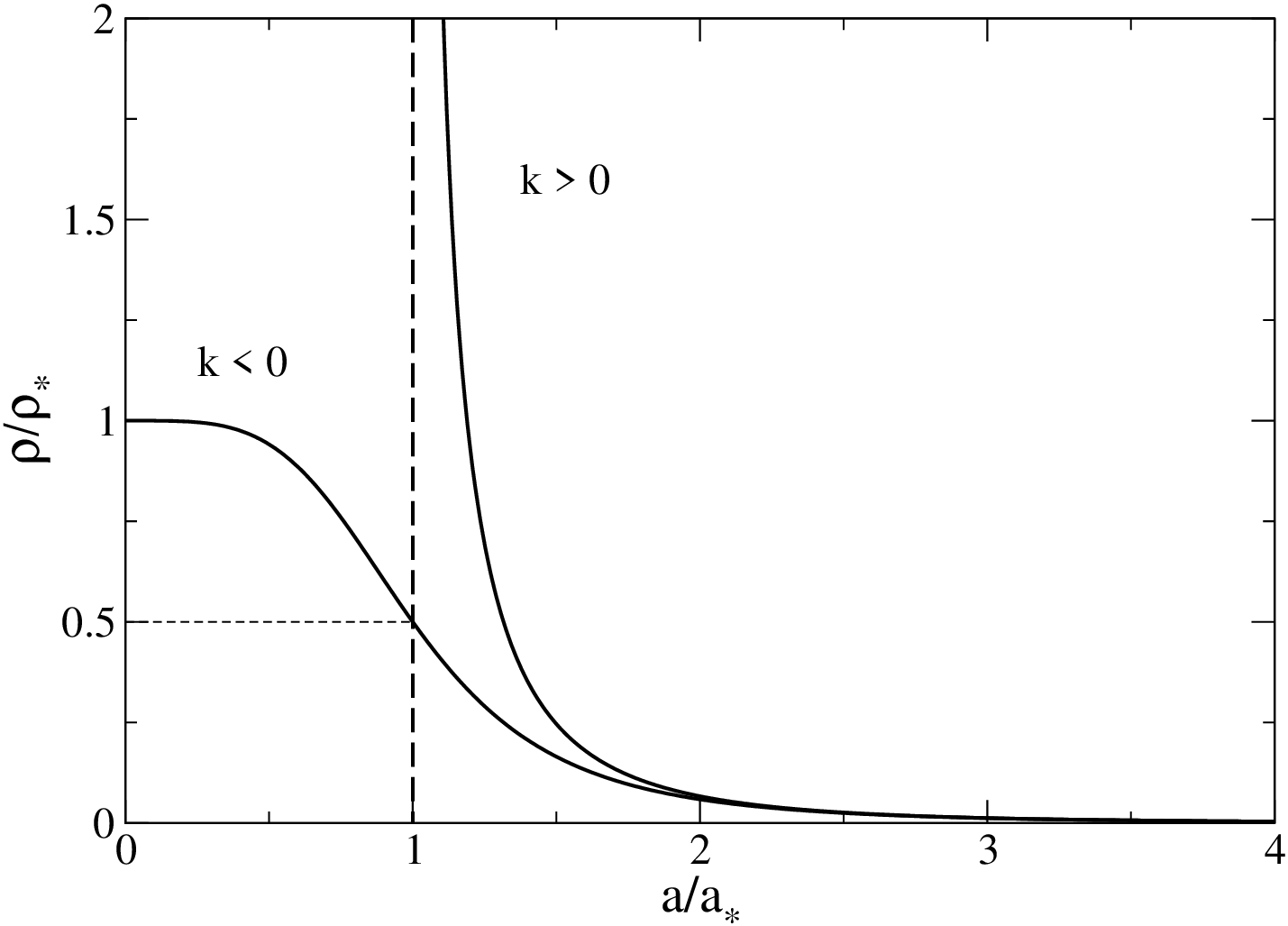}
\caption{Density as a function of the scale factor for $k>0$ and $k<0$. We have taken $n=1$ and $\alpha=1/3$.}
\label{rhoa}
\end{center}
\end{figure}

For $k>0$, the density is defined only for $a>a_*$.  When $a\rightarrow a_*$,
\begin{equation}
\label{ges3b}
\frac{\rho}{\rho_*}\sim \left\lbrack \frac{n}{3(1+\alpha)}\right \rbrack^n \frac{1}{(a/a_*-1)^n}\rightarrow +\infty.
\end{equation}
When $a\rightarrow +\infty$, $\rho/\rho_*\sim (a_*/a)^{3(1+\alpha)}\rightarrow 0$ corresponding to the linear equation of state. In the same limits, $p\rightarrow +\infty$ and $p\rightarrow 0$, respectively. 

For $k<0$, the density is defined for all $a$.  When $a\rightarrow 0$, the density tends to a finite value  $\rho_*$  When $a\rightarrow +\infty$, $\rho\sim \rho_* (a_*/a)^{3(1+\alpha)}\rightarrow 0$ corresponding to the linear equation of state. In the same limits,  $p\rightarrow -\rho_* c^2$ and $p\rightarrow 0$, respectively. 

Some curves giving the evolution of the density $\rho$  as a function of the scale factor $a$ are plotted in Fig. \ref{rhoa} for $k>0$ and $k<0$.

\subsection{The temperature}
\label{sec_gestemperature}

For the equation of state (\ref{ges0}), the thermodynamical equation (\ref{t2}) becomes
\begin{equation}
\label{ges4}
(\alpha+k\gamma\rho^{1/n}) \frac{d\rho}{dT}=\frac{1}{T}(\alpha+1+k\rho^{1/n})\rho.
\end{equation}
This equation can be integrated into
\begin{equation}
\label{ges5}
T=T_* \left \lbrack 1\pm ({\rho}/{\rho_*})^{1/n}\right \rbrack^{(\alpha+n+1)/(\alpha+1)}\left ({\rho}/{\rho_*}\right )^{\alpha/(\alpha+1)},
\end{equation}
where $T_*$ is a constant of integration. This relation between the temperature and the density can be viewed as a generalized Stefan-Boltzmann law. Combined with Eq. (\ref{ges3}), we obtain
\begin{equation}
\label{ges6}
T=T_* \frac{(a/a_*)^{3(\alpha+n+1)/n}}{\left \lbrack (a/a_*)^{3(1+\alpha)/n}\mp 1\right \rbrack^{n+1}}.
\end{equation}
Let us consider some asymptotic limits.

When $a\gg a_*$, Eq. (\ref{ges6}) reduces to
\begin{equation}
\label{ges7}
T/T_*\sim \frac{1}{(a/a_*)^{3\alpha}},
\end{equation}
returning Eq. (\ref{leos10}) for a linear equation of state. As we have already indicated, in this regime, the temperature decreases to zero when  $\alpha>0$ and increases to $+\infty$ when $\alpha<0$. When $\alpha=0$, it tends to a constant value $T_*$.

When $k>0$ and $a\rightarrow a_*$,
\begin{equation}
\label{ges8}
T/T_*\sim \left\lbrack \frac{n}{3(1+\alpha)}\right \rbrack^{n+1} \frac{1}{(a/a_*-1)^{n+1}}\rightarrow +\infty.
\end{equation}
When $k<0$ and $a\rightarrow 0$,
\begin{equation}
\label{ges9}
T/T_*\sim  (a/a_*)^{3(\alpha+n+1)/n} \rightarrow 0.
\end{equation}

\begin{figure}[!h]
\begin{center}
\includegraphics[clip,scale=0.3]{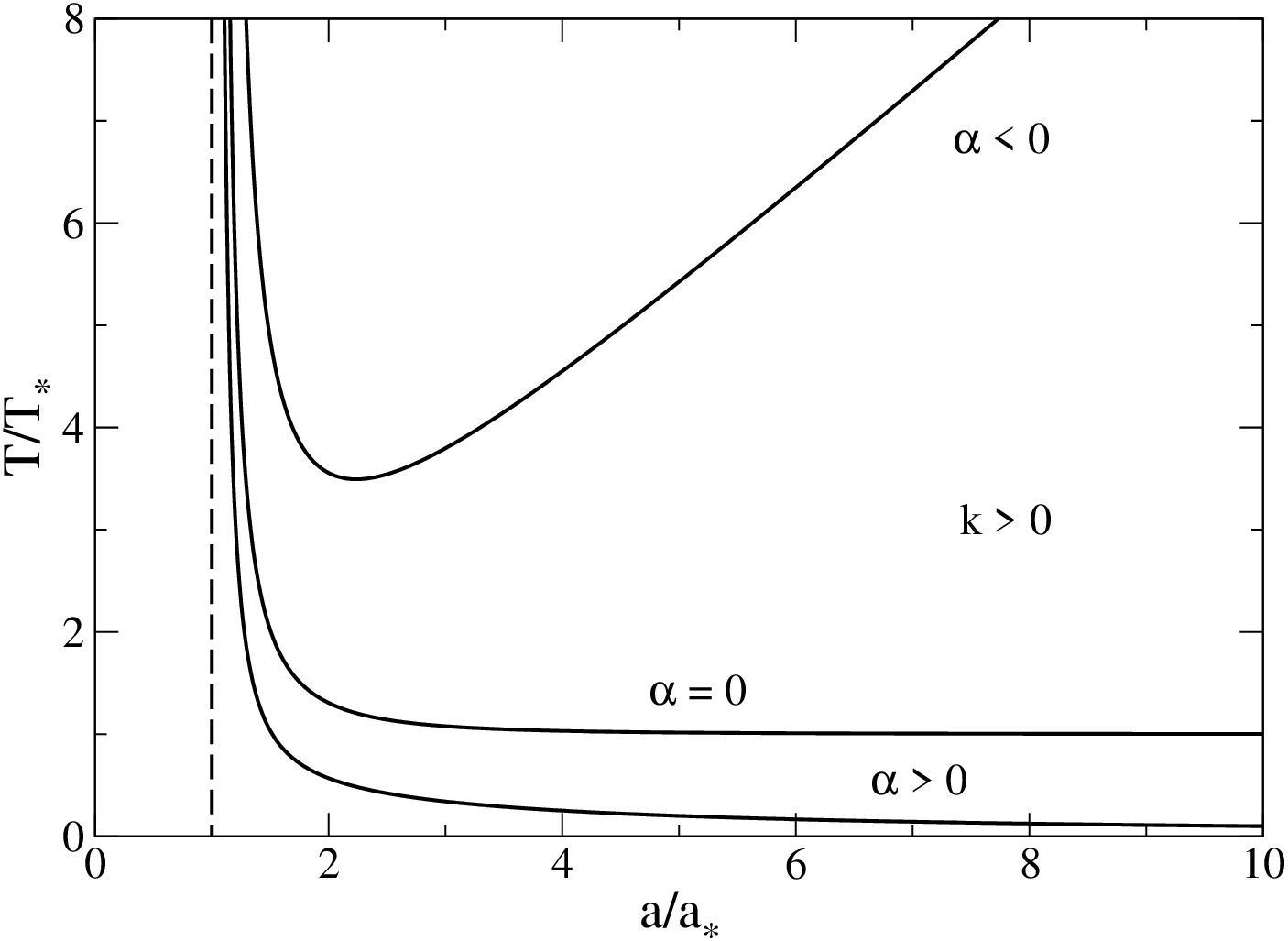}
\caption{Temperature as a function of the scale factor for $k>0$. We have taken $n=1$, $\alpha=-1/3$, $\alpha=0$ and $\alpha=1/3$. }
\label{atempKpos}
\end{center}
\end{figure}

\begin{figure}[!h]
\begin{center}
\includegraphics[clip,scale=0.3]{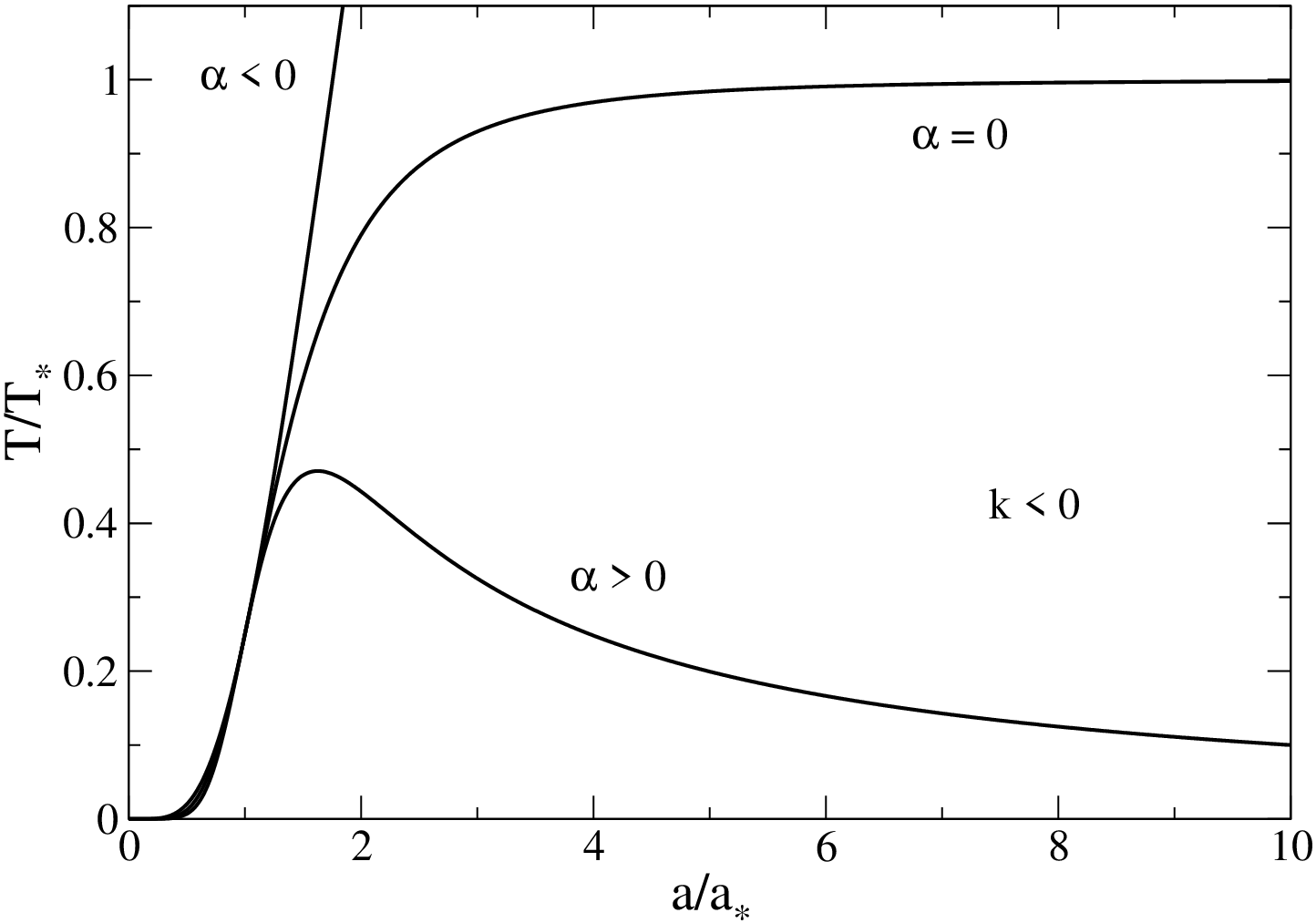}
\caption{Temperature as a function of the scale factor for $k<0$. We have taken $n=1$, $\alpha=-1/3$, $\alpha=0$ and $\alpha=1/3$.}
\label{atempKneg}
\end{center}
\end{figure}

The extremum of temperature (when it exists) is located at
\begin{equation}
\label{ges10a}
\frac{\rho_{e}}{\rho_*}=\left\lbrack \mp \frac{\alpha n}{(\alpha+1)(n+1)}\right\rbrack^n,
\end{equation}
\begin{equation}
\label{ges10b}
\frac{a_{e}}{a_*}=\left (\mp \frac{\alpha+n+1}{\alpha n}\right )^{n/\lbrack 3(\alpha+1)\rbrack},
\end{equation}
\begin{eqnarray}
\label{ges11}
\frac{T_{e}}{T_*}=\left\lbrack \mp \frac{n\alpha}{(\alpha+1)(n+1)}\right\rbrack^{\frac{\alpha n}{\alpha+1}}\left \lbrack \frac{n+\alpha+1}{(1+\alpha)(n+1)}\right \rbrack^{\frac{n+\alpha+1}{1+\alpha}}.\nonumber\\
\end{eqnarray}
Some curves giving the evolution of the temperature $T$  as a function of the scale factor $a$ are plotted in Figs. \ref{atempKpos} and \ref{atempKneg} for $k>0$ and $k<0$.

Finally, the entropy (\ref{t4}) is given by
\begin{equation}
\label{ges12}
S=(\alpha+1)\frac{a_*^3}{T_*}\rho_* c^2,
\end{equation}
and we explicitly check that it is a constant.

\subsection{The equation of state parameter $w(t)$}
\label{sec_gesw}

We can rewrite the equation of state (\ref{ges0}) as $p=w(t)\rho c^2$ with
\begin{equation}
\label{gesw2}
w(t)=\alpha\pm (\alpha+1)\left (\frac{\rho}{\rho_*}\right )^{1/n}.
\end{equation}
We define
\begin{equation}
\label{gesw3}
\frac{\rho_w}{\rho_*}=\left (\mp \frac{\alpha}{\alpha+1}\right )^n,\qquad \frac{a_w}{a_*}=\left (\mp \frac{1}{\alpha}\right )^{n/\lbrack 3(\alpha+1)\rbrack},
\end{equation}
\begin{equation}
\label{gesw4}
\frac{T_w}{T_*}=\frac{(\mp\alpha)^{\alpha n/(\alpha+1)}}{(\alpha+1)^{n+1}},
\end{equation}
corresponding to a possible point where the pressure vanishes ($w=0$).

\begin{figure}[!h]
\begin{center}
\includegraphics[clip,scale=0.3]{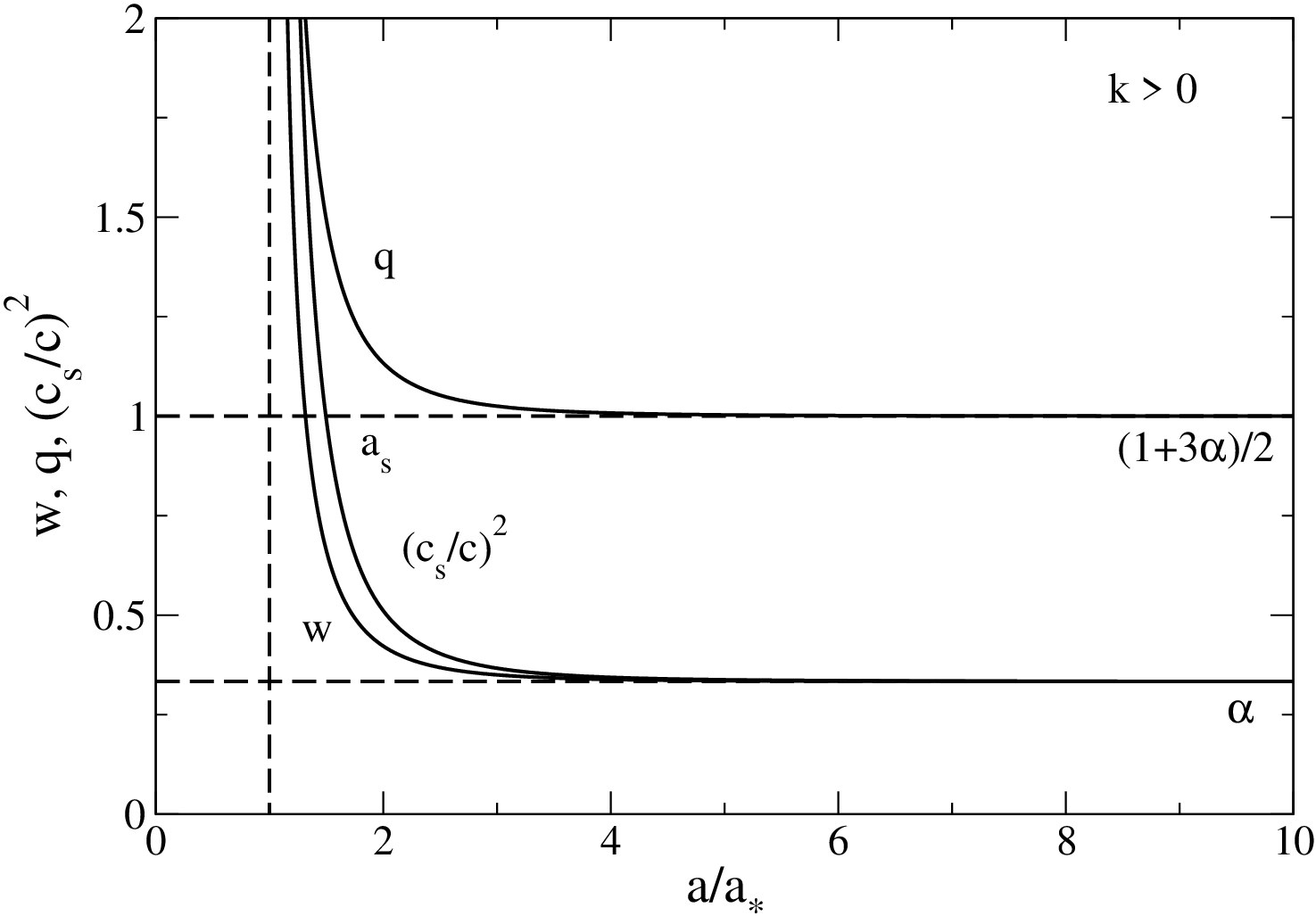}
\caption{Evolution of $w$, $q$ and $(c_s/c)^2$ as a function of the scale factor for $k>0$. We have taken $n=1$ and $\alpha=1/3$.}
\label{wqcPOS}
\end{center}
\end{figure}

\begin{figure}[!h]
\begin{center}
\includegraphics[clip,scale=0.3]{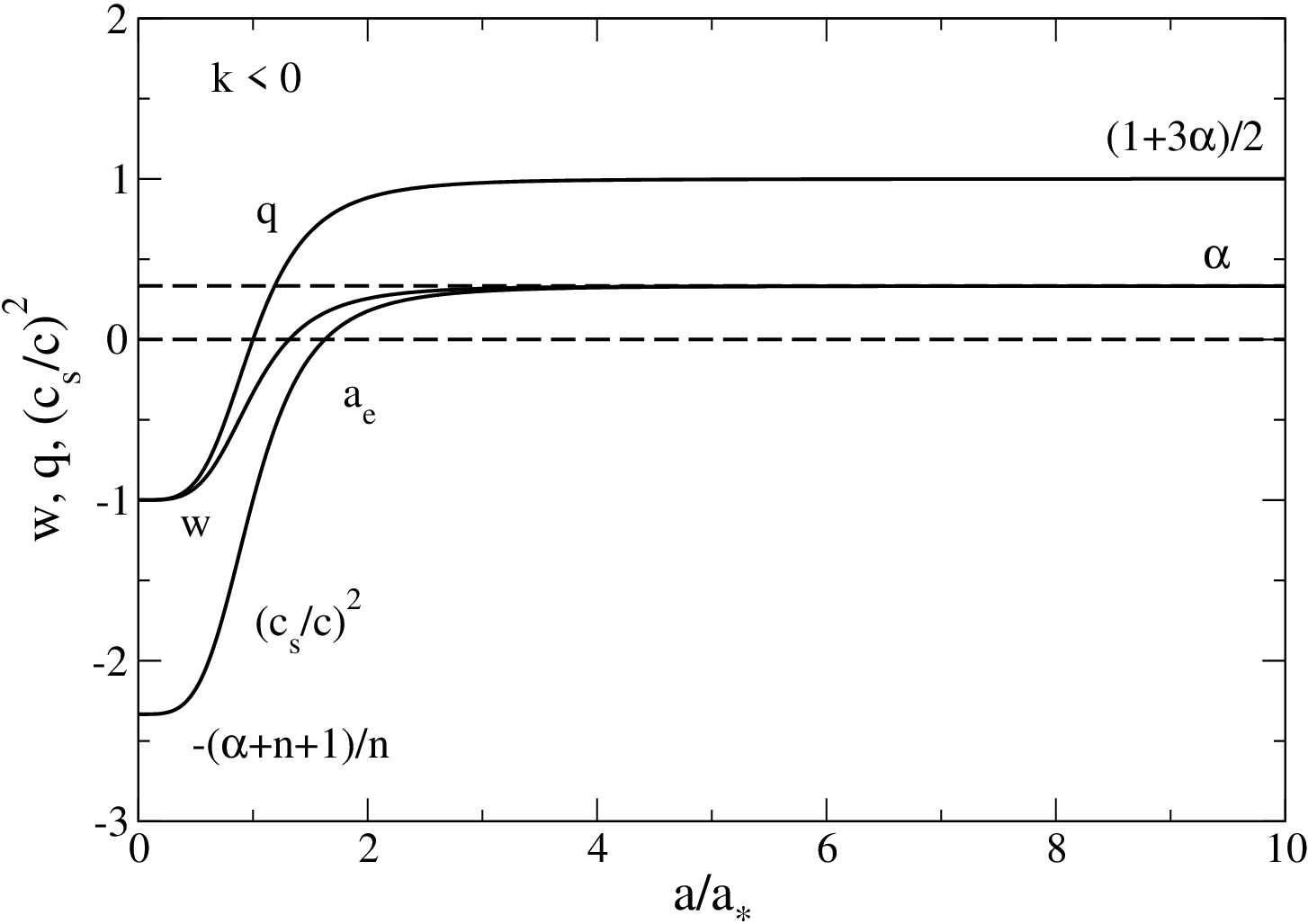}
\caption{Evolution of $w$, $q$ and $(c_s/c)^2$ as a function of the scale factor for $k<0$. We have taken $n=1$ and $\alpha=1/3$.}
\label{wqcNEG}
\end{center}
\end{figure}

For $k>0$,  $w\rightarrow +\infty$ when $a\rightarrow a_*$ and  $w\rightarrow \alpha$ when $a\rightarrow +\infty$. For $\alpha>0$, the pressure is always positive ($w>0$).  For $\alpha<0$, the pressure is positive when $a_*<a<a_w$ and negative
when  $a>a_w$.

For $k<0$, $w\rightarrow -1$ when $a\rightarrow 0$ and  $w\rightarrow \alpha$ when $a\rightarrow +\infty$. For $\alpha<0$,
the pressure  is always negative ($w<0$). For
$\alpha>0$, the pressure is negative when
$a<a_w$ and positive when  $a>a_w$.

Some curves giving the evolution of $w$  as a function of the scale factor $a$ are plotted in Figs. \ref{wqcPOS} and \ref{wqcNEG} for $k>0$ and $k<0$.

\subsection{The velocity of sound}
\label{sec_gessound}

For the equation of state (\ref{ges0}), the velocity of sound is given by
\begin{equation}
\label{sound1}
c_s^2=p'(\rho)=\left \lbrack \alpha\pm (\alpha+1)\frac{n+1}{n}\left (\frac{\rho}{\rho_*}\right )^{1/n}\right \rbrack c^2.
\end{equation}
The velocity of sound vanishes at the point defined by Eqs. (\ref{ges10a})-(\ref{ges11}) where the temperature is extremum. At that point the pressure is extremum with value
\begin{equation}
\label{sound1b}
\frac{p_e}{\rho_* c^2}=\frac{\alpha}{n+1}\left \lbrack \mp\frac{\alpha n}{(\alpha+1)(n+1)}\right \rbrack^n.
\end{equation}
The case $c_s^2<0$ corresponds to an imaginary velocity of sound.
We  define
\begin{equation}
\label{sound2}
\frac{\rho_s}{\rho_*}=\left \lbrack \pm\frac{(1-\alpha)n}{(1+\alpha)(n+1)}\right \rbrack^n,
\end{equation}
\begin{equation}
\label{sound3}
\frac{a_s}{a_*}= \left \lbrack \pm\frac{\alpha+2n+1}{n(1-\alpha)}\right \rbrack^{n/\lbrack 3(1+\alpha)\rbrack},
\end{equation}
\begin{eqnarray}
\label{sound4}
\frac{T_{s}}{T_*}=\left\lbrack \pm \frac{n(1-\alpha)}{(\alpha+1)(n+1)}\right\rbrack^{\frac{\alpha n}{\alpha+1}}\left \lbrack \frac{1+2n+\alpha}{(1+\alpha)(n+1)}\right \rbrack^{\frac{n+\alpha+1}{1+\alpha}},\nonumber\\
\end{eqnarray}
corresponding to a possible point where the velocity of sound is equal to the speed of light. Different cases have to be considered.

We first assume $k>0$.  When $a\rightarrow a_*$, $(c_s/c)^2\rightarrow +\infty$; when $a\rightarrow +\infty$, $(c_s/c)^2\rightarrow \alpha$.  For $\alpha>0$,  $c_s^2$  is always positive. For $\alpha<0$, $c_s^2$ is positive when $a_*<a<a_e$ and negative when $a>a_e$. The velocity of sound is larger than the speed of light when $a_*<a<a_s$ and smaller when $a>a_s$.

We now assume $k<0$. When $a\rightarrow 0$, $(c_s/c)^2\rightarrow -(\alpha+n+1)/n$; when $a\rightarrow +\infty$, $(c_s/c)^2\rightarrow \alpha$. For $\alpha<0$, $c_s^2$  is always negative. For $\alpha>0$, $c_s^2$ is negative when $a<a_e$ and positive when $a>a_e$. For $\alpha\le 1$, the velocity of sound is always smaller than the speed of light.

Some curves giving the evolution of $(c_s/c)^2$  as a function of the scale factor $a$ are plotted in Figs. \ref{wqcPOS} and \ref{wqcNEG} for $k>0$ and $k<0$.

\section{Evolution of the scale factor}
\label{sec_dark}

\subsection{The deceleration parameter}
\label{sec_dec}

The deceleration parameter is defined by
\begin{equation}
\label{dec1}
q(t)=-\frac{{\ddot a}a}{{\dot a}^2}.
\end{equation}
The universe is decelerating when $q>0$ and accelerating when $q<0$.  Using the Friedmann equations (\ref{f8}) and (\ref{f9}), we obtain for a flat universe
\begin{equation}
\label{dec2}
q(t)=\frac{1+3w(t)}{2}.
\end{equation}
For the equation of state (\ref{ges0}), using Eq. (\ref{gesw2}), we get
\begin{equation}
\label{dec3}
q(t)=\frac{1+3\alpha}{2}\pm \frac{3}{2}(\alpha+1)\left (\frac{\rho}{\rho_*}\right )^{1/n}.
\end{equation}
We define a critical density, scale factor, and temperature
\begin{equation}
\label{dec4}
\frac{\rho_c}{\rho_*}=\left \lbrack \mp\frac{1+3\alpha}{3(1+\alpha)}\right \rbrack^n,\qquad \frac{a_c}{a_*}=\left (\mp\frac{2}{1+3\alpha}\right )^{{n}/{\lbrack 3(1+\alpha)\rbrack}},
\end{equation}
\begin{equation}
\label{dec5}
\frac{T_c}{T_*}=2^{(\alpha+n+1)/(\alpha+1)}\frac{\left\lbrack \mp(1+3\alpha)\right \rbrack^{\alpha n/(\alpha+1)}}{\left\lbrack 3(1+\alpha)\right\rbrack^{n+1}}.
\end{equation}
corresponding to a possible inflexion point ($q=\ddot a=0$) in the curve
$a(t)$. Different cases have to be considered.

We first assume $k>0$. When $a\rightarrow a_*$, $q\rightarrow +\infty$; when $a\rightarrow +\infty$, $q\rightarrow  (1+3\alpha)/2$. For  $\alpha>-1/3$, the universe is always decelerating
($q>0$).  For $\alpha<-1/3$, the universe is decelerating when $a_*<a<a_c$ and accelerating when $a>a_c$.

We now assume $k<0$. When $a\rightarrow 0$, $q\rightarrow -1$; when $a\rightarrow +\infty$, $q\rightarrow (1+3\alpha)/2$. For $\alpha<-1/3$, the universe is always accelerating ($q<0$). For $\alpha>-1/3$, the universe is accelerating when $a<a_c$ and decelerating  when $a>a_c$.

Some curves giving the evolution of $q$  as a function of the scale factor $a$ are plotted in Figs. \ref{wqcPOS} and \ref{wqcNEG} for $k>0$ and $k<0$.

\subsection{The differential equation}
\label{sec_darkdiff}

The temporal  evolution of the scale factor $a(t)$ is determined by the Friedmann equation (\ref{f9}).
Introducing the normalized radius $R=a/a_*$, the density (\ref{ges3}) can be written
\begin{equation}
\label{dark1}
\rho=\frac{\rho_*}{\lbrack R^{3(1+\alpha)/n}\mp 1\rbrack^n}.
\end{equation}
Substituting this expression in Eq. (\ref{f9}), we obtain the differential equation
\begin{equation}
\label{dark2}
\dot R=\frac{\epsilon K R}{\lbrack R^{3(1+\alpha)/n}\mp 1\rbrack^{n/2}},
\end{equation}
where $K=({8\pi G\rho_*}/{3})^{1/2}$ and $\epsilon=\pm 1$. In general, we shall select the sign $\epsilon=+1$ corresponding to an expanding universe ($\dot R>0$). The solution can be written as
\begin{equation}
\label{dark4}
\epsilon Kt=\int \left\lbrack R^{3(1+\alpha)/n}\mp 1\right\rbrack^{n/2}\, \frac{dR}{R},
\end{equation}
or, after a change of variables $x=R^{3(1+\alpha)/n}$, as
\begin{equation}
\label{dark5}
\frac{3(\alpha+1)}{n}\epsilon Kt=\int^{R^{3(\alpha+1)/n}} (x\mp 1)^{n/2}\, \frac{dx}{x}.
\end{equation}
The integral can be expressed in terms of hypergeometric functions. Some simple analytical expressions can be obtained for specific values of $n$. Actually, we can have a good idea of the behavior of the solution of Eq. (\ref{dark2}) by considering asymptotic limits (see below). The complete solution is represented in the figures by solving Eq. (\ref{dark2}) numerically.

\subsection{The case $k>0$}
\label{sec_darkrep}

The universe starts from a primordial singularity at $t=0$ (see below for a revision of this statement) with a finite radius $R=1$ and an infinite density, pressure, and temperature. When $t\rightarrow 0$,
\begin{eqnarray}
\label{dark6}
R\simeq 1+\left\lbrack \frac{n}{3(1+\alpha)}\right\rbrack^{n/(n+2)}\left (\frac{n+2}{2}Kt\right )^{2/(n+2)},
\end{eqnarray}
\begin{eqnarray}
\label{dark7}
\frac{\rho}{\rho_*}\sim \left\lbrack \frac{3}{2}(1+\alpha)\frac{n+2}{n}Kt\right \rbrack^{-2n/(n+2)},
\end{eqnarray}
\begin{eqnarray}
\label{dark8}
\frac{T}{T_*}\sim \left\lbrack \frac{3}{2}(1+\alpha)\frac{n+2}{n}Kt\right \rbrack^{-2(n+1)/(n+2)}.
\end{eqnarray}
Then, the universe expands indefinitely. When $t\rightarrow +\infty$, the density decreases to zero and we recover the solution corresponding to a  linear equation of state
\begin{eqnarray}
\label{dark9}
R\sim \left \lbrack\frac{3(\alpha+1)}{2}Kt\right \rbrack^{2/\lbrack 3(1+\alpha)\rbrack},
\end{eqnarray}
\begin{eqnarray}
\label{toto1}
\frac{\rho}{\rho_*}\sim \left\lbrack \frac{3}{2}(1+\alpha)Kt\right \rbrack^{-2},
\end{eqnarray}
\begin{eqnarray}
\label{toto2}
\frac{T}{T_*}\sim \left\lbrack \frac{3}{2}(1+\alpha)Kt\right \rbrack^{-2\alpha/(1+\alpha)}.
\end{eqnarray}
For $\alpha>-1/3$, the universe is always decelerating ($\ddot R<0$). For $\alpha<-1/3$, the universe is decelerating for $R<R_c= \lbrack {-2}/(1+3\alpha)\rbrack^{{n}/\lbrack {3(1+\alpha)}\rbrack}$ and accelerating for $R>R_c$. The velocity of sound is less than the speed of light when $R>R_s= \lbrace (\alpha+2n+1)/\lbrack n(1-\alpha)\rbrack \rbrace^{{n}/\lbrack {3(1+\alpha)}\rbrack}$.

For $n=1$,  using the identity
\begin{eqnarray}
\label{dark10}
\int \sqrt{x-1}\, \frac{dx}{x}=2\sqrt{x-1}-2\arctan\left (\sqrt{x-1}\right ),
\end{eqnarray}
the general solution of  Eq. (\ref{dark2}) is
\begin{eqnarray}
\label{dark11}
\sqrt{R^{3(1+\alpha)}-1}-\arctan \sqrt{R^{3(1+\alpha)}-1}=\frac{3(1+\alpha)}{2} K t.\nonumber\\
\end{eqnarray}
For $\alpha<-1/3$, the time $t_c$ at which the universe starts accelerating is given by
\begin{equation}
\label{dark12}
\frac{3(1+\alpha)}{2}Kt_c=\sqrt{\frac{-3(1+\alpha)}{1+3\alpha}}-\arctan \sqrt{\frac{-3(1+\alpha)}{1+3\alpha}}.
\end{equation}
The time $t_i$ at which the velocity of sound is equal to the speed of light is given by
\begin{equation}
\label{dark13}
\frac{3(1+\alpha)}{2}K t_i=\sqrt{\frac{2(1+\alpha)}{1-\alpha}}-\arctan \sqrt{\frac{2(1+\alpha)}{1-\alpha}}.
\end{equation}
This gives the time at which the physical universe begins. It may be meaningless to consider smaller times because the velocity of sound is greater than the speed of light. 

For $n=2$, the general solution of Eq. (\ref{dark2}) is
\begin{eqnarray}
\label{dark11bb}
R^{\frac{3}{2}(1+\alpha)}-\frac{3}{2}(\alpha+1)\ln R
=\frac{3(1+\alpha)}{2}K t+C.
\end{eqnarray}

\begin{figure}[!h]
\begin{center}
\includegraphics[clip,scale=0.3]{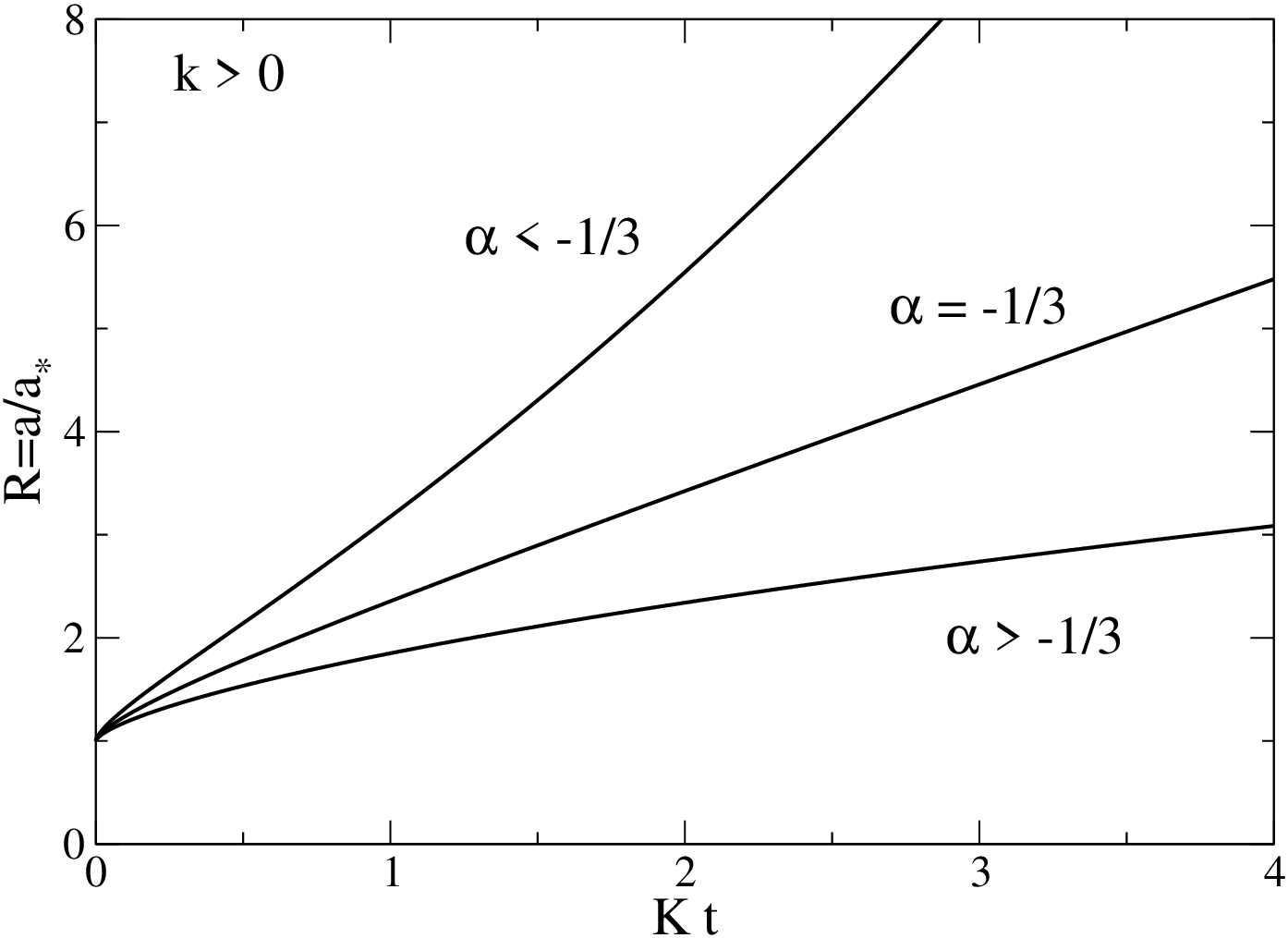}
\caption{Evolution of the scale factor in the case $k>0$ for different values of $\alpha$ (specifically $\alpha=-2/3$, $\alpha=-1/3$, and $\alpha=1/3$). We have taken $n=1$.}
\label{generalisationPOS}
\end{center}
\end{figure}

Typical evolutions of the scale factor $R(t)$ corresponding to $\alpha<-1/3$,
$\alpha=1/3$, and $\alpha>1/3$ are represented in Fig.  \ref{generalisationPOS}.

\subsection{The case $k<0$}
\label{sec_darkatt}

The universe starts from  $t\rightarrow -\infty$ with a vanishing radius $R=0$, a vanishing temperature $T=0$,  and a finite density $\rho=\rho_*$.  When $t\rightarrow -\infty$,
\begin{eqnarray}
\label{dark14}
R\propto e^{Kt},\qquad \frac{T}{T_*}\propto e^{\frac{3}{n}(\alpha+n+1)Kt}.
\end{eqnarray}
This corresponds to an exponential expansion (early inflation) due to the fact that the density is approximately constant. The universe expands indefinitely. When $t\rightarrow +\infty$, the density decreases to zero and we recover the solution (\ref{dark9})-(\ref{toto2}) corresponding to a  linear equation of state. For $\alpha< -1/3$, the universe is always accelerating ($\ddot R>0$). For $\alpha>-1/3$, the universe is accelerating when $R<R_c= \lbrack{2}/(1+3\alpha)\rbrack^{{n}/\lbrack {3(1+\alpha)}\rbrack}$ and decelerating when $R>R_c$. In this model, the velocity of sound is always less than the speed of light.

For $n=1$, using the identity
\begin{eqnarray}
\label{dark16}
\int \sqrt{x+1}\, \frac{dx}{x}=2\sqrt{x+1}+\ln\left (\frac{\sqrt{x+1}-1}{\sqrt{x+1}+1}\right ),
\end{eqnarray}
the general solution of Eq. (\ref{dark2}) is
\begin{eqnarray}
\label{dark17}
\sqrt{R^{3(1+\alpha)}+1}-\ln \left (\frac{1+\sqrt{R^{3(1+\alpha)}+1}}{R^{3(1+\alpha)/2}}\right )\nonumber\\
=\frac{3(1+\alpha)}{2}K t+C,
\end{eqnarray}
where $C$ is a constant of integration. For $\alpha>-1/3$, the time $t_c$ at which the universe starts decelerating is given by
\begin{eqnarray}
\label{dark18}
\frac{3(1+\alpha)}{2}Kt_c&+C&=\sqrt{\frac{3(1+\alpha)}{1+3\alpha}}\nonumber\\
&-&\ln \left (\frac{\sqrt{1+3\alpha}+\sqrt{3(1+\alpha)}}{\sqrt{2}}\right ).\qquad
\end{eqnarray}
When $t\rightarrow -\infty$, we obtain from Eq. (\ref{dark17}) the asymptotic behavior
\begin{equation}
\label{dark19}
R(t)\sim e^{Kt+D},
\end{equation}
with $D=2(C+\ln 2-1) /\lbrack 3(1+\alpha)\rbrack$.

For $n=2$, the general solution of Eq. (\ref{dark2}) is
\begin{eqnarray}
\label{dark17bb}
R^{\frac{3}{2}(1+\alpha)}+\frac{3}{2}(\alpha+1)\ln R
=\frac{3(1+\alpha)}{2}K t+C.
\end{eqnarray}

\begin{figure}[!h]
\begin{center}
\includegraphics[clip,scale=0.3]{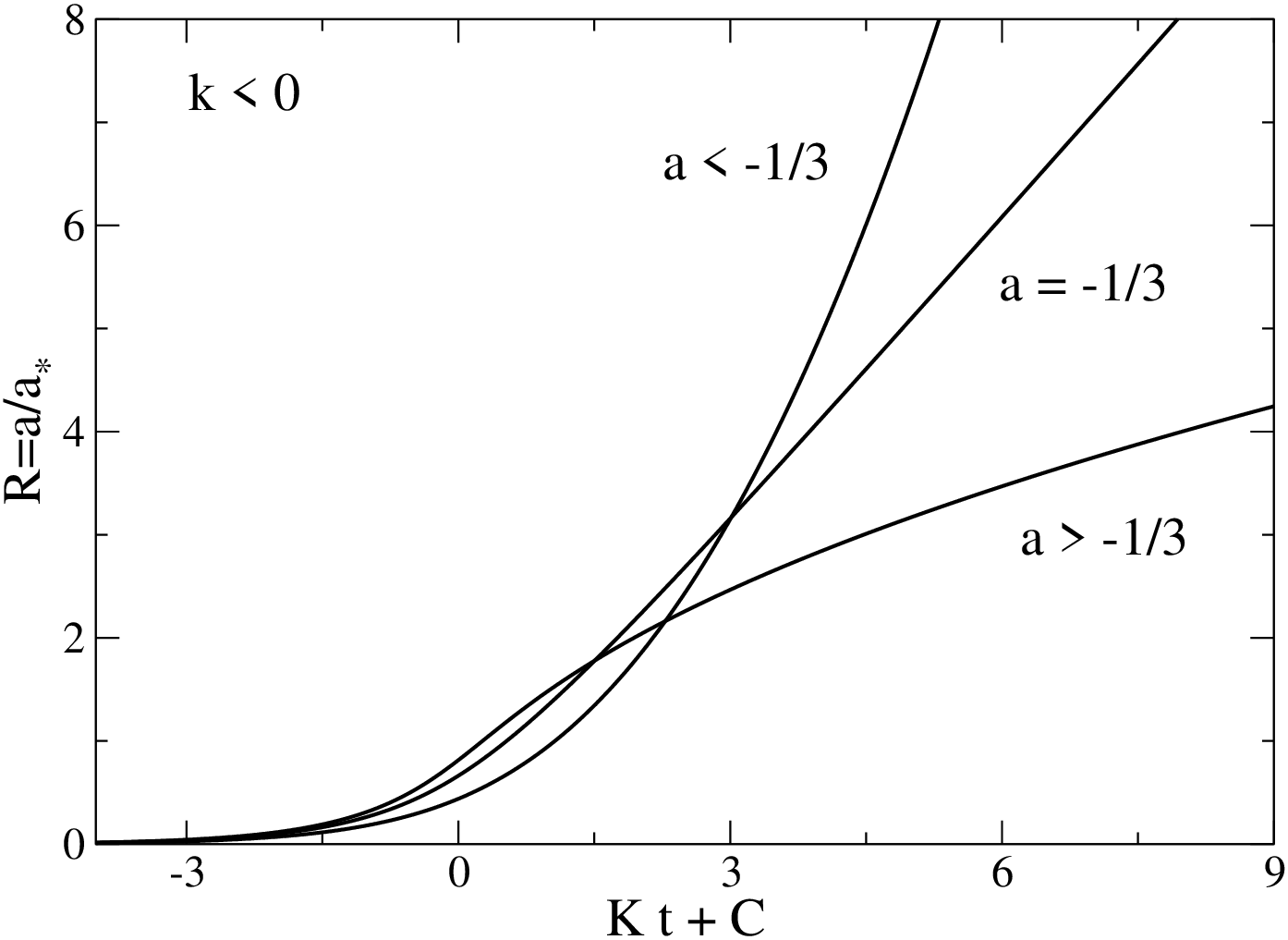}
\caption{Evolution of the scale factor in the case $k<0$ for different values of $\alpha$ (specifically $\alpha=-2/3$, $\alpha=-1/3$, and $\alpha=1/3$). We have taken $n=1$.}
\label{generalisationNEG}
\end{center}
\end{figure}

Typical evolutions of $R(t)$ corresponding to $\alpha<-1/3$, $\alpha=1/3$, and $\alpha>1/3$ are represented in Fig. \ref{generalisationNEG}.

\section{A model of non-singular inflationary universe}
\label{sec_inf}

The generalized equation of state (\ref{ges0}) with $n>0$ and $k<0$  can be used to describe the transition between a phase of early inflation where the density of the universe is constant ($\rho=\rho_*$) and a phase where the universe has a linear equation of state ($p=\alpha\rho c^2$). Therefore, if we take $\alpha=1/3$, it describes the transition between the pre-radiation era  and the radiation era. This provides a model of non-singular inflationary universe. For simplicity, we shall take $n=1$ (in the BEC model, this corresponds to the standard index \cite{c4}).

\subsection{The basic equations}
\label{sec_infb}

For $\alpha=1/3$, $n=1$, and $k<0$, Eqs. (\ref{ges0}), (\ref{ges3}), (\ref{ges5}), and (\ref{ges6}) become
\begin{equation}
\label{infb0}
p=(\rho/3+k\rho^2)c^2,
\end{equation}
\begin{equation}
\label{infb1}
\rho=\frac{\rho_*}{(a/a_1)^{4}+1},
\end{equation}
\begin{equation}
\label{infb2}
T=T_* \left (1- {\rho}/{\rho_*}\right )^{7/4}\left ({\rho}/{\rho_*}\right )^{1/4},
\end{equation}
\begin{equation}
\label{infb3}
T=T_*\frac{(a/a_1)^7}{\left\lbrack (a/a_1)^4+ 1\right\rbrack^2},
\end{equation}
where $\rho_*=4/(3|k|)$, and we have noted $a_1$ for $a_*$.

When $a\ll a_1$, the density has a constant value $\rho\simeq \rho_*$. This gives rise to a phase of early inflation. The value $\rho_*$ defines a fundamental upper bound $\rho_{max}$ for the density. It is natural to identify it with the Planck density $\rho_P=5.16\, 10^{99}\, {\rm g}/{\rm m}^3$ (see Sec. \ref{sec_nec}). This is how quantum mechanics is taken into account in our model. Since the Planck density is obtained from a pure dimensional analysis, we should write $\rho_*=\kappa\rho_P$, where $\kappa$ is a constant of order unity (actually, the Planck constant should be defined so as to correspond exactly to $\rho_{max}$). For simplicity, we shall take
\begin{equation}
\label{infb4}
\rho_*=\rho_{max}=\rho_P.
\end{equation}
This fixes the constant $k=-4/(3\rho_P)$.  If the pre-radiation era is made of BECs, this relation determines the ratio between the coupling constant $\lambda<0$ of the bosons and their mass $m$ (see Appendix \ref{sec_annbec}).

When $a\gg a_1$, we can make the approximation $\rho/\rho_P\sim (a_1/a)^4$. This correspond to the radiation era described by an equation of state $p=\rho c^2/3$. The conservation of $\rho_{rad}\,  a^4$ implies that $\rho_P a_1^4=\rho_{rad,0}\, a_0^4$. Writing $\rho_{rad,0}=\Omega_{rad,0}\rho_0$, where $\rho_0$ is the present density of the universe, and using Eq. (\ref{nec2}), we obtain
\begin{equation}
\label{infb5}
\frac{a_1}{a_0}=\left (\frac{3\Omega_{rad,0}}{8\pi}\right )^{1/4} \sqrt{H_0 t_P}=1.97\, 10^{-32}.
\end{equation}
Using Eq. (\ref{nec8}), we get
\begin{equation}
\label{infb6}
\frac{a_1}{l_P}=\left (\frac{3\Omega_{rad,0}}{8\pi}\right )^{1/4} \frac{1}{\sqrt{H_0 t_P}}=1.61\, 10^{29}.
\end{equation}
The value of the  characteristic length $a_1$ is
\begin{equation}
\label{infb7}
a_1=2.61\, 10^{-6}\, {\rm m}.
\end{equation}
It provides a lengthscale that is  intermediate between the Planck length $l_P=1.62\, 10^{-35}\, {\rm m}$ and the present size of the universe $a_0=1.32\, 10^{26}\, {\rm m}$. It corresponds to the typical size of the universe at the end of the inflationary phase (or at the beginning of the radiation era).
On the other hand, in the radiation era, Eq. (\ref{infb2}) reduces to  $\rho c^2\sim \rho_P c^2(T/T_*)^4$,
so we can identify ${\rho_P c^2}/{T_*^4}$ with the Stefan-Boltzmann constant $\sigma$. Using Eq. (\ref{rad6}), this yields
\begin{equation}
\label{infb8}
T_*=\left (\frac{15}{\pi^2}\right )^{1/4}T_P.
\end{equation}
Numerically, $T_*=1.11 T_P=1.57\, 10^{32}\, {\rm K}$.

Regrouping these results, we obtain the basic equations of the non-singular model
\begin{equation}
\label{infb8a}
p=\frac{1}{3}\rho (1-4\rho/\rho_P)c^2,
\end{equation}
\begin{equation}
\label{infb8b}
w=\frac{1}{3} (1-4\rho/\rho_P), \qquad q=1-2\rho/\rho_P,
\end{equation}
\begin{equation}
\label{infb9}
\rho=\frac{\rho_P}{(a/a_1)^{4}+ 1},
\end{equation}
\begin{equation}
\label{infb10}
T=T_P \left (\frac{15}{\pi^2}\right )^{1/4} \left (1- \frac{\rho}{\rho_P}\right )^{7/4}\left (\frac{\rho}{\rho_P}\right )^{1/4},
\end{equation}
\begin{equation}
\label{infb11}
T=T_P \left (\frac{15}{\pi^2}\right )^{1/4} \frac{(a/a_1)^7}{\left\lbrack (a/a_1)^4+ 1\right\rbrack^2}.
\end{equation}
The entropy (\ref{ges12}) is given by
\begin{equation}
\label{infb12}
S/k_B=\frac{4}{3}\left (\frac{3\Omega_{rad,0}}{8\pi}\right )^{3/4} \left (\frac{\pi^2}{15}\right )^{1/4}\frac{1}{(H_0 t_P)^{3/2}}.
\end{equation}
Numerically, $S/k_B=5.04\, 10^{87}$. Finally, we note that $K=\left ({8\pi}/{3}\right )^{1/2}{t_P^{-1}}$.

As the universe expands from $a=0$ to $a=+\infty$, the density decreases from $\rho_P$ to $0$, the equation of state parameter $w$ increases from $-1$ to $1/3$, the deceleration parameter $q$ increases from $-1$ to $1$, and the ratio $(c_s/c)^2$ goes from $-7/3$ to $1/3$. The temperature starts at $T=0$, reaches a maximum and decreases to zero.

\subsection{The inflationary phase}
\label{sec_ip}

During the inflationary phase ($a\ll a_1$),  the density has an approximately constant value $\rho\simeq \rho_P$. The corresponding Hubble parameter is
\begin{equation}
\label{ip1}
H=\frac{\dot a}{a}\simeq \left (\frac{8\pi G}{3}\rho_P\right )^{1/2}\simeq \left (\frac{8\pi}{3}\right )^{1/2}\frac{1}{t_P}.
\end{equation}
Numerically, $\rho=5.16\, 10^{99}\, {\rm g}/{\rm m}^3$ and $H=5.37\, 10^{43}\, {\rm s}^{-1}$. Integrating Eq. (\ref{ip1}), we find that the scale factor increases exponentially rapidly with time as
\begin{equation}
\label{ip2}
a(t)\sim l_P e^{({8\pi}/{3})^{1/2}t/t_P}.
\end{equation}
We have determined the constant of integration such that $a(t=0)=l_P$. This universe exists at any time in the past, so there is no singularity.

When $t<t_i=0$, the radius of the universe is smaller than the Planck length $l_P=1.62\, 10^{-35}\, {\rm m}$ and it tends to zero exponentially rapidly when $t\rightarrow -\infty$.  When $t\ge t_i=0$, the radius of the universe increases  exponentially rapidly on a timescale of the order of the Planck time $t_P=5.39\, 10^{-44}\, {\rm s}$. This phase of inflation connects the pre-radiation era to the radiation era on a very short interval of time (of the order of a few Planck times). We call $t_1$ the time at which the scale factor $a(t)$ reaches the value $a_1=2.61\, 10^{-6}\, {\rm m}$. Using the approximate expression (\ref{ip2}) of the scale factor during the inflationary phase, we obtain
\begin{equation}
\label{ip3}
t_1=\left (\frac{3}{8\pi}\right )^{1/2}\ln\left (\frac{a_1}{l_P}\right ) t_P.
\end{equation}
Numerically, $t_1=23.2\, t_P=1.25\, 10^{-42}\, {\rm s}$. This time corresponds typically to the end of the inflation. Therefore, during the inflation, the radius increases exponentially rapidly from $a_i=l_P=1.62\, 10^{-35}\, {\rm m}$ at $t=t_i=0$ to $a=a_1=2.61\, 10^{-6}\, {\rm m}$ at $t=t_1=1.25\, 10^{-42}{\rm s}$.

On the other hand, in the inflationary phase, the temperature is related to the density and to the scale factor by
\begin{eqnarray}
\label{ip5}
\frac{T}{T_P}\sim  \left (\frac{15}{\pi^2}\right )^{1/4} \left (1- \frac{\rho}{\rho_P}\right )^{7/4}\sim \left (\frac{15}{\pi^2}\right )^{1/4} \left (\frac{a}{a_1}\right )^7.\nonumber\\
\end{eqnarray}
Using the approximate expression (\ref{ip2}), we find that the temperature increases exponentially rapidly with time as
\begin{eqnarray}
\label{ip6}
T(t)\sim T_P \left (\frac{15}{\pi^2}\right )^{1/4} \left (\frac{l_P}{a_1}\right )^7 e^{7({8\pi}/{3})^{1/2}t/t_P}.
\end{eqnarray}
When $t<t_i=0$, the temperature is less than $T_i=3.91\, 10^{-205} \, T_P=5.54\, 10^{-173}\, {\rm K}$, and it tends to zero exponentially rapidly when $t\rightarrow -\infty$. Therefore, the pre-radiation era is extremely cold (this is consistent with the BEC model which assumes $T=0$ \cite{c4}). When $t\ge t_i=0$, the temperature increases exponentially rapidly. It passes  from $T=T_i=5.54\, 10^{-173}\, {\rm K}$ at $t=t_i=0$ to $T=T_1=1.11\, T_P=1.57\, 10^{32}\, {\rm K}$ at $t=t_1=1.25\, 10^{-42}{\rm s}$.

These values are based on the asymptotic results (\ref{ip2}) and (\ref{ip6}). They will be slightly revised in Sec. \ref{sec_gp} using the exact solution (\ref{gp1}) of the Friedmann equations.

\subsection{The radiation era}
\label{sec_rp}

After the inflation ($a\gg a_1$), the universe enters in the radiation era. In that case, the quadratic pressure ($\propto -\rho^2$) is negligible as compared to the pressure of radiation ($p\propto \rho$) and we recover the standard radiation model of Sec. \ref{sec_radiation}. The density is related to the scale factor by $\rho\sim \rho_P a_1^4/a^4$.  The Friedmann equation (\ref{f9}) becomes
\begin{equation}
\label{rp1}
H=\frac{\dot a}{a}\sim \left (\frac{8\pi G}{3}\frac{\rho_P a_1^4}{a^4}\right )^{1/2}\sim \left (\frac{8\pi}{3}\right )^{1/2}\frac{1}{t_P}\left (\frac{a_1}{a}\right )^2,
\end{equation}
yielding
\begin{equation}
\label{rp2}
a(t)\sim a_1\left (\frac{32\pi}{3}\right )^{1/4} \left (\frac{t}{t_P}\right )^{1/2}.
\end{equation}
Using the approximate expression (\ref{rp2}) of the scale factor during the radiation era, we find that the time at which $a=a_1$ is
\begin{equation}
\label{rp3}
t_1=\left (\frac{3}{32\pi}\right )^{1/2}t_P.
\end{equation}
Numerically, $t_1=0.173\, t_P=9.31\, 10^{-45}\, {\rm s}$. This marks the beginning of the radiation era. Again, this numerical value will be revised in Sec. \ref{sec_gp}.  We also have
\begin{equation}
\label{rp4}
\rho\sim \frac{\rho_P}{(a/a_1)^4}\sim \frac{3\rho_P}{32\pi}\left (\frac{t_P}{t}\right )^2,
\end{equation}
and
\begin{equation}
\label{rp5}
T\sim T_P \left (\frac{15}{\pi^2}\right )^{1/4} \frac{a_1}{a}\sim T_P \left (\frac{45}{32\pi^3}\right )^{1/4} \left (\frac{t_P}{t}\right )^{1/2}.
\end{equation}
During the radiation era, the density and the temperature decrease algebraically as the universe expands.

\subsection{The general solution}
\label{sec_gp}

The equation of state (\ref{infb8a}) interpolates smoothly between the pre-radiation era described by a constant density $\rho=\rho_P$ (vacuum energy) and the radiation era described by a density $\rho\propto a^{-4}$. It provides therefore a unified description of the early universe. Using the results of Sec. \ref{sec_darkatt}, the general solution of the Friedmann equation (\ref{f9}) is \cite{c4}:
\begin{eqnarray}
\label{gp1}
\sqrt{(a/a_1)^{4}+1}-\ln \left (\frac{1+\sqrt{(a/a_1)^{4}+1}}{(a/a_1)^{2}}\right )\nonumber\\
=2\left (\frac{8\pi}{3}\right )^{1/2} \frac{t}{t_P}+C,
\end{eqnarray}
where $C$ is a constant of integration. It is determined such that $a=l_P$ at $t=0$. Setting
\begin{eqnarray}
\label{gp2}
\epsilon=l_P/a_1,
\end{eqnarray}
we get
\begin{eqnarray}
\label{gp3}
C(\epsilon)=\sqrt{\epsilon^{4}+1}-\ln \left (\frac{1+\sqrt{\epsilon^{4}+1}}{\epsilon^{2}}\right ).
\end{eqnarray}
Numerically,  $\epsilon=6.20\, 10^{-30}$ and $C=-134$. For $t\rightarrow -\infty$, we have the exact asymptotic result
\begin{equation}
\label{gp3b}
a(t)\sim a_1 e^{({8\pi}/{3})^{1/2}t/t_P+D},
\end{equation}
with $D=(C+\ln 2-1)/2$.  Due to the smallness of $\epsilon$, a very good approximation of $C$ and $D$ is given by $C\simeq 1-\ln 2 +2\ln\epsilon$ and $D\simeq \ln\epsilon$. With this approximation, Eq. (\ref{gp3b}) returns Eq. (\ref{ip2}).

\begin{figure}[!h]
\begin{center}
\includegraphics[clip,scale=0.3]{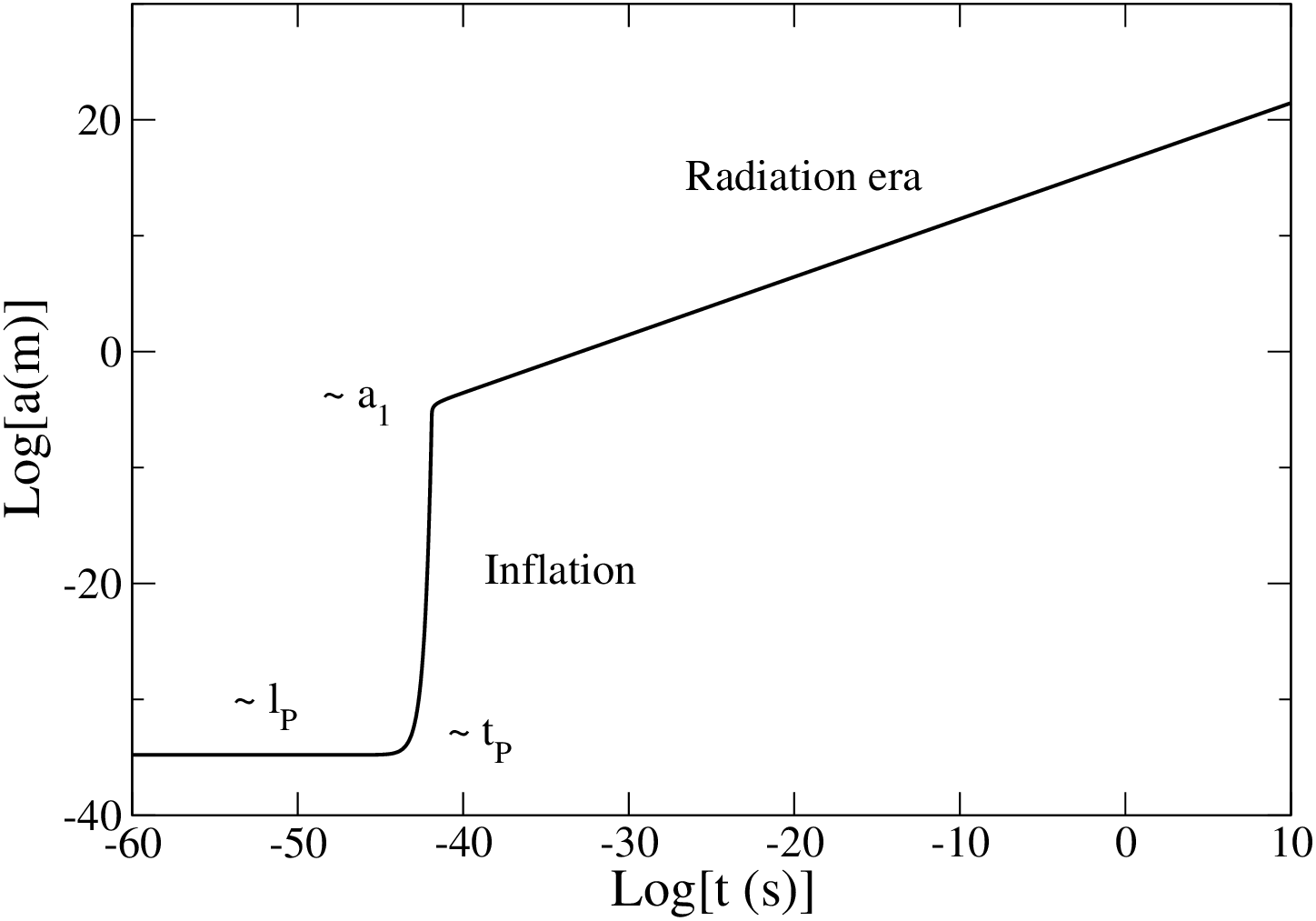}
\caption{Evolution of the scale factor $a$ with the time $t$ in logarithmic scales. This figure clearly shows the phase of inflation connecting the pre-radiation era to the radiation era. During the inflation, the scale factor increases by $29$ orders of magnitude in less than $10^{-42}\, {\rm s}$.}
\label{taLOGLOG}
\end{center}
\end{figure}

\begin{figure}[!h]
\begin{center}
\includegraphics[clip,scale=0.3]{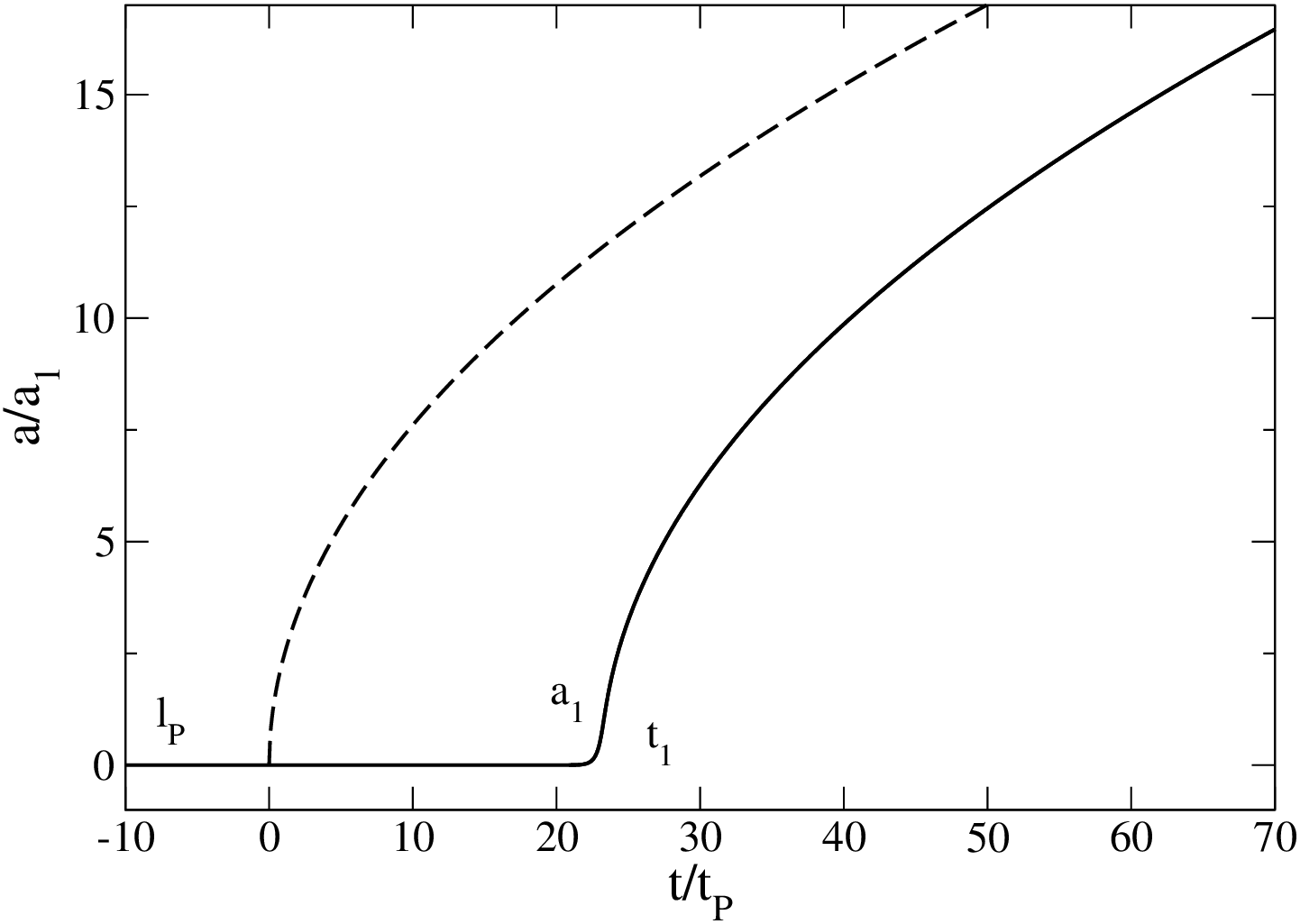}
\caption{Evolution of the scale factor $a$ with the time $t$ in linear scales. The dashed line corresponds to the standard radiation model of Sec. \ref{sec_radiation} exhibiting a finite time singularity at $t=0$ (Big Bang). When quantum mechanics is taken into account (as in our simple model), the initial singularity is smoothed-out and the scale factor at $t=0$ is equal to the Planck length $l_P=1.62\, 10^{-35}\, {\rm m}$. This is similar to a second order phase transition where the Planck constant plays the role of finite size effects (see Sec. \ref{sec_analogy} for a development of this analogy).}
\label{taLINLIN}
\end{center}
\end{figure}

\begin{figure}[!h]
\begin{center}
\includegraphics[clip,scale=0.3]{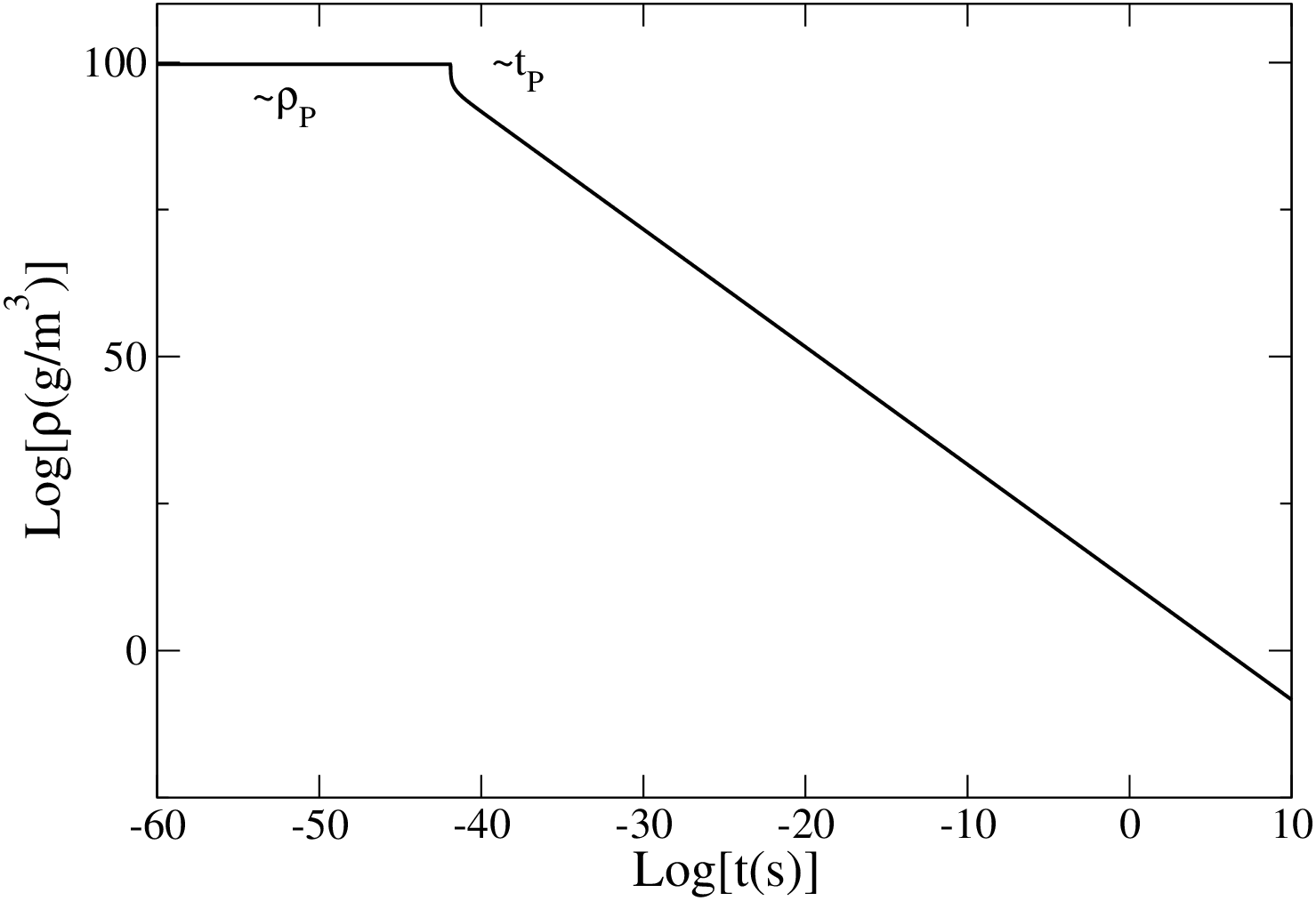}
\caption{Evolution of the density $\rho$ with the time $t$ in logarithmic scales. During the inflation, the density remains approximately constant with the Planck value $\rho_{\max}=\rho_P$ which represents an upper bound. }
\label{trhoLOGLOG}
\end{center}
\end{figure}

\begin{figure}[!h]
\begin{center}
\includegraphics[clip,scale=0.3]{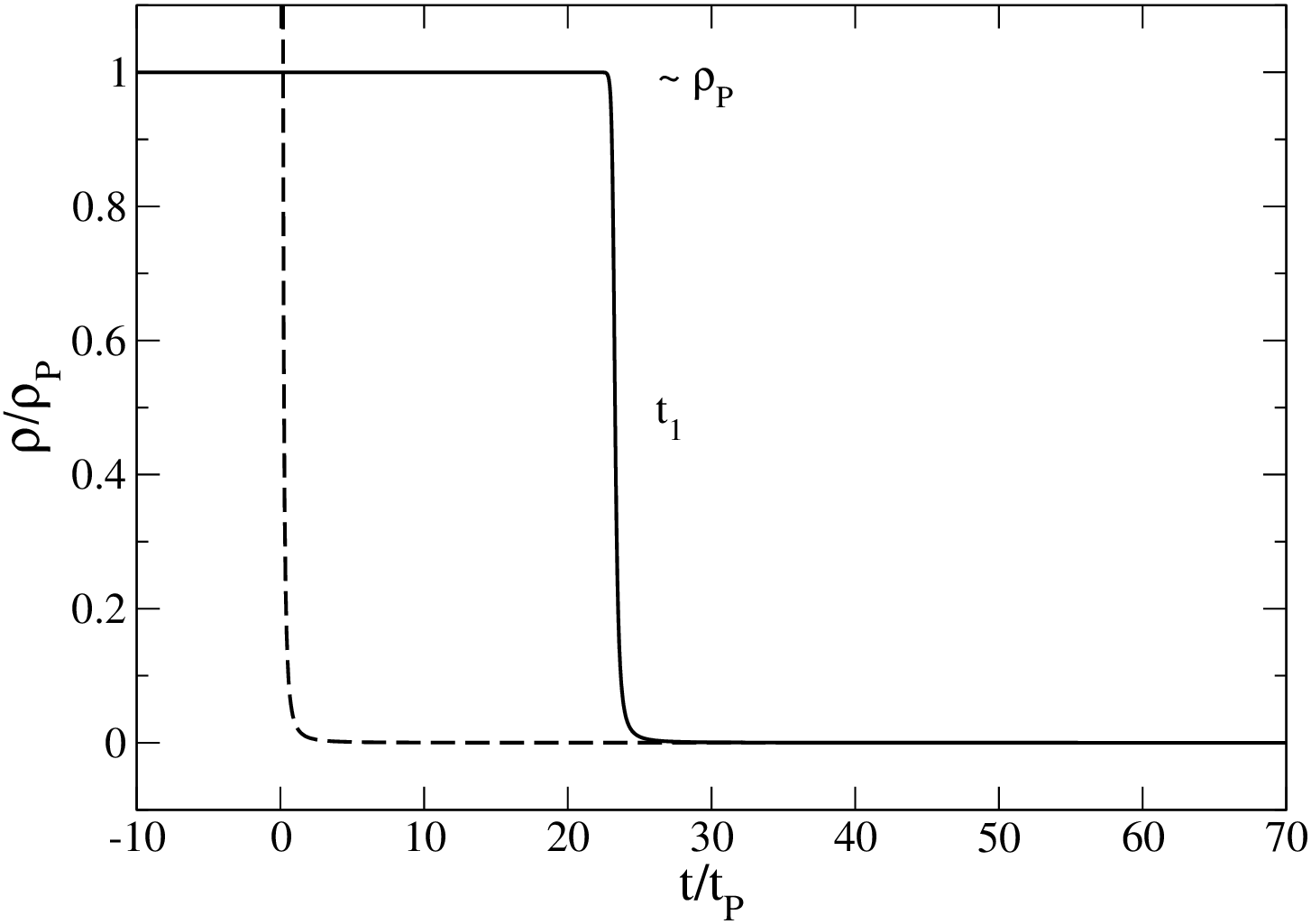}
\caption{Evolution of the density $\rho$ with the time $t$ in linear scales. The dashed line corresponds to the standard radiation model of Sec. \ref{sec_radiation} exhibiting a finite time singularity at $t=0$. Quantum mechanics limits the rise of the density to the Planck value $\rho_P=5.16\, 10^{99}\, {\rm g}/{\rm m}^3$.}
\label{trhoLINLIN}
\end{center}
\end{figure}

\begin{figure}[!h]
\begin{center}
\includegraphics[clip,scale=0.3]{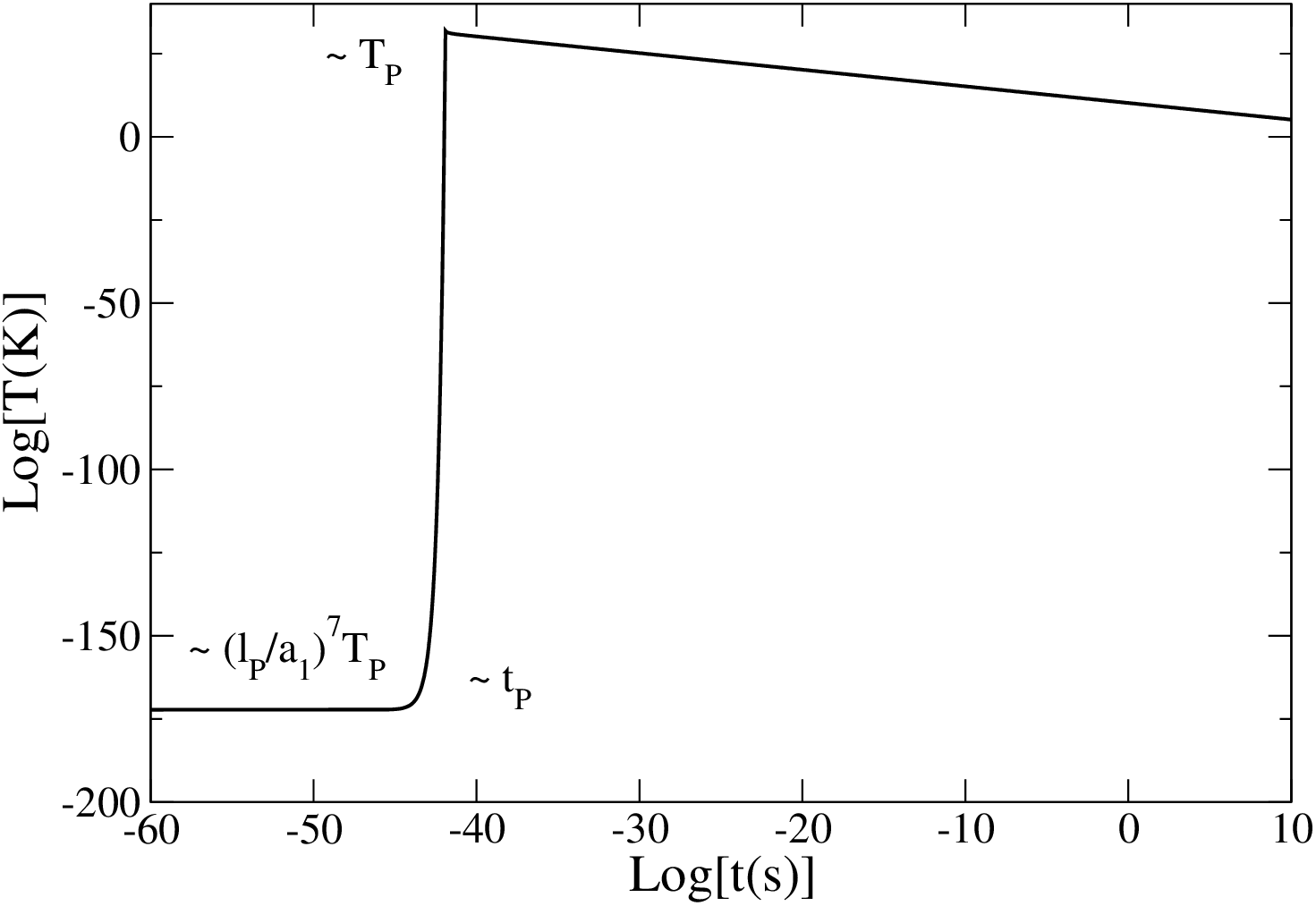}
\caption{Evolution of the temperature $T$ with the time $t$ in logarithmic scales. For $t\le 0$, the universe is extremely cold ($T<10^{-173}\, {\rm K}$). During the inflation, the temperature increases by $204$ orders of magnitude in less than $10^{-42}\, {\rm s}$. During the radiation era, the temperature decreases algebraically.}
\label{ttempLOGLOG}
\end{center}
\end{figure}

\begin{figure}[!h]
\begin{center}
\includegraphics[clip,scale=0.3]{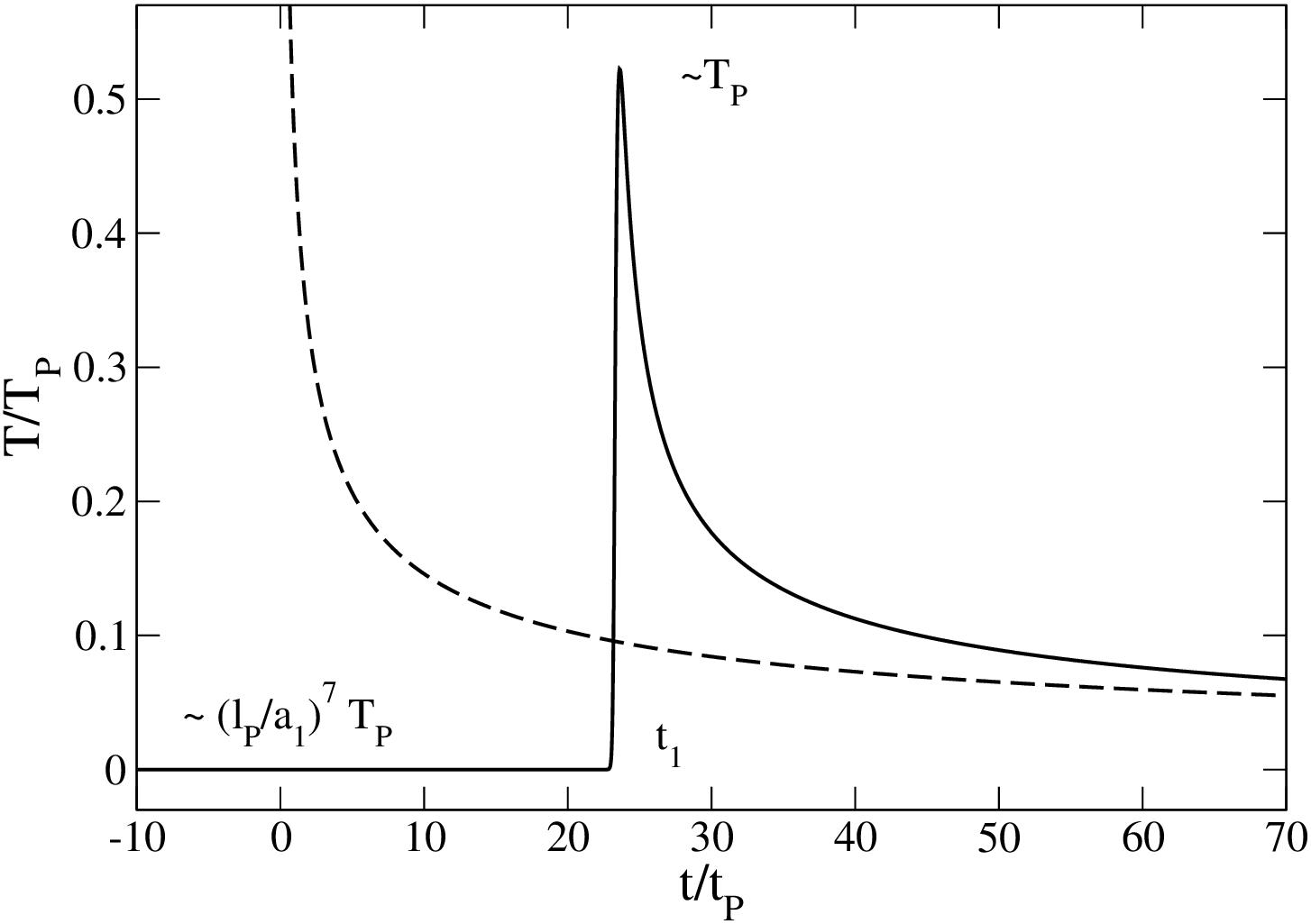}
\caption{Evolution of the temperature $T$ with the time $t$ in linear scales. The dashed line corresponds to the standard radiation model of Sec. \ref{sec_radiation} exhibiting a finite time singularity at $t=0$. Quantum mechanics limits the rise of the temperature to about half the Planck value $T_P=1.42\, 10^{32}\, {\rm K}$. In the pre-radiation era, the temperature drops to zero exponentially rapidly.}
\label{ttempLINLIN}
\end{center}
\end{figure}

At $t=t_i=0$,
\begin{equation}
\label{gp8}
a_i=l_P,\qquad \rho_i=\frac{\rho_P}{\epsilon^4+1},
\end{equation}
\begin{equation}
\label{gp10}
T_i=T_P\left (\frac{15}{\pi^2}\right )^{1/4}\frac{\epsilon^7}{(\epsilon^4+1)^2}.
\end{equation}
Numerically, $a_i=l_P=1.62\, 10^{-35}\, {\rm m}$, $\rho_i\simeq \rho_P=5.16\, 10^{99}\, {\rm g}/{\rm m}^3$, and $T_i=3.91\, 10^{-205} \, T_P=5.54\, 10^{-173}\, {\rm K}$.

Using the results of Sec. \ref{sec_dec}, we find that the universe is accelerating for $a<a_c$ ({\it i.e.} $\rho>\rho_c$) and decelerating for $a>a_c$ ({\it i.e.}  $\rho<\rho_c$) where
\begin{equation}
\label{gp4}
a_c=a_1, \qquad \rho_c=\frac{1}{2}\rho_P.
\end{equation}
The corresponding temperature is
\begin{equation}
\label{gp6}
T_c=\frac{1}{4}\left (\frac{15}{\pi^2}\right )^{1/4}T_P.
\end{equation}
The time $t_c$ at which the universe starts decelerating is given by
\begin{equation}
\label{gp7}
{t_c}=\left (\frac{3}{8\pi}\right )^{1/2}\frac{1}{2}\left\lbrack \sqrt{2}-\ln(1+\sqrt{2})-C\right\rbrack t_P.
\end{equation}
This is the time at which the curve $a(t)$ presents an inflexion point. It turns out that this inflexion point coincides with $a_1$ (this property is true only for $\alpha=1/3$). Therefore, it coincides  with our definition of the end of the inflation: $t_c=t_1$. Numerically, $\rho_c=\rho_1=0.5\,\rho_P=2.58\, 10^{99}\, {\rm g}/{\rm m}^3$, $a_c=a_1=2.61\, 10^{-6}\, {\rm m}$, $T_c=T_1=0.278\, T_P=3.93\, 10^{31}\, {\rm K}$, and $t_c=t_1=23.3\, t_P=1.25\, 10^{-42}\, {\rm s}$.

The temperature starts from $T=0$ at $t\rightarrow -\infty$, increases exponentially rapidly during the inflation, reaches a maximum value, and decreases algebraically during the radiation era. Using the results of Sec. \ref{sec_gestemperature}, we find that the point corresponding to the maximum temperature is
\begin{equation}
\label{gp11}
a_{e}=7^{1/4}a_1,\qquad \rho_{e}=\frac{1}{8}\rho_P,
\end{equation}
\begin{equation}
\label{gp13}
T_{e}=\left (\frac{7}{8}\right )^{7/4}\left (\frac{15}{8\pi^2}\right )^{1/4}T_P.
\end{equation}
It is achieved at a time
\begin{equation}
\label{gp14}
t_{e}=\left (\frac{3}{8\pi}\right )^{1/2}\frac{1}{2}\left\lbrack \sqrt{8}-\ln\left (\frac{1+\sqrt{8}}{\sqrt{7}}\right )-C\right\rbrack t_P.
\end{equation}
Numerically, $\rho_e=0.125\, \rho_P=6.44\, 10^{98}\, {\rm g}/{\rm m}^3$, $a_e=1.63 a_1=4.24\, 10^{-6}\, {\rm m}$, $T_e=0.523\, T_P=7.40\, 10^{31}\, {\rm K}$, and $t_e=23.6\, t_P=1.27\, 10^{-42}\, {\rm s}$.

The evolution of the scale factor, density and temperature as a function of time are represented in Figs. \ref{taLOGLOG}-\ref{ttempLINLIN} in logarithmic and linear scales.

\subsection{Summary}

In this model, the universe starts at  $t\rightarrow -\infty$ with a vanishing radius $a=0$, a finite density $\rho=\rho_P=5.16\, 10^{99}\, {\rm g/m^3}$, and a vanishing temperature $T=0$. The universe exists at any time in the past and does not present any singularity. It has been called ``aioniotic'' universe in \cite{c4} since it has no origin nor end. To make contact with the Big Bang theory, we define the ``original'' time $t=t_i=0$ as the time at which the radius of the universe is equal to the Planck length. Thus $a_i=l_P=1.62\, 10^{-35}\, {\rm m}$. The corresponding density and temperature are $\rho_i\simeq \rho_P=5.16\, 10^{99}\, {\rm g}/{\rm m}^3$ and $T_i=3.91\, 10^{-205} \, T_P=5.54\, 10^{-173}\, {\rm K}$. We note that quantum mechanics regularizes the finite time singularity that arises in the standard Big Bang theory. This is similar to finite size effects in second order phase transitions (see Sec. \ref{sec_analogy}). We also note that the universe is very cold at $t=t_i=0$, unlike what is predicted by the Big Bang theory (a naive extrapolation of the law $T\propto t^{-1/2}$ leads to $T(0)=+\infty$). The universe first undergoes a phase of inflation during which its radius and temperature increase exponentially rapidly [see Eqs. (\ref{ip2}) and (\ref{ip6})] while its density remains approximately constant [see Eq. (\ref{infb4})]. The inflation ``starts'' at $t_i=0$ and ends at
$t_1=23.3\, t_P=1.25\, 10^{-42}\, {\rm s}$. During this very short lapse of time,  the radius of the universe grows from $a_i=l_P=1.62\, 10^{-35}\, {\rm m}$ to $a_1=2.61\, 10^{-6}\, {\rm m}$, and the temperature grows from $T_i=3.91\, 10^{-205} \, T_P=5.54\, 10^{-173}\, {\rm K}$ to $T_1=0.278\, T_P=3.93\, 10^{31}\, {\rm K}$. By contrast, the density does not change significatively: It goes from  $\rho_i\simeq \rho_P=5.16\, 10^{99}\, {\rm g}/{\rm m}^3$ to $\rho_1=0.5\, \rho_P=2.58\, 10^{99}\, {\rm g/m^3}$. After the inflation, the universe enters in the radiation era and, from that point, we recover the standard model. The radius increases algebraically [see Eq. (\ref{rp2})] while the density and the temperature decrease algebraically [see Eqs. (\ref{rp4}) and (\ref{rp5})]. The temperature achieves its maximum value $T_{e}=0.523\, T_P=7.40\, 10^{31}\, {\rm K}$ at $t=t_{e}=23.6\, t_P=1.27\, 10^{-42}\, {\rm s}$.  At that moment, the density of the universe is $\rho_{e}=0.125\, \rho_P=6.44\, 10^{98}\, {\rm g}/{\rm m}^3$ and its radius $a_{e}=1.63\, a_1=4.24\, 10^{-6}\, {\rm m}$. During the inflation, the universe is accelerating and during the radiation era it is decelerating. The transition takes place at a time $t_c=t_1=23.3\, t_P=1.25\, 10^{-42}\, {\rm s}$ coinciding with the end of the inflation ({\it i.e.} $a_c=a_1$).

We note that the inflationary process described above is relatively different from the usual inflationary scenario \cite{guth,linde}.  In standard inflation, the universe is radiation dominated up to $t_i=10^{-35}{\rm s}$, but expands exponentially by a factor $10^{30}$ in the interval $t_i<t<t_f$ with $t_f=10^{-33}{\rm s}$. For $t>t_f$ the evolution is again radiation dominated.    At $t=t_i$ the temperature is about $10^{27} {\rm K}$ (this corresponds to the epoch at which most ``grand unified theories'' have a significant influence on the evolution of the universe). During the exponential inflation, the temperature drops drastically; however, the matter is expected to be reheated to the initial temperature of $10^{27} {\rm K}$ by various high energy processes \cite{paddybook}. In the inflationary process described previously, the evolution of the temperature is different.

{\it Remark:} Of course, our definition of the ``original'' time
$t=0$, corresponding to $a(0)=l_P$, is relatively arbitrary. What
essentially matters is the evolution of the different variables on a
tiny interval of time of the order of $20 t_P$.

\section{A model of non-inflationary universe with a new type of initial singularity}
\label{sec_ni}

The equation of state (\ref{ges0}) with $n>0$ and $k>0$ leads to a model of non-inflationary universe with a new type of initial singularity. We take the same values of $\alpha$, $n$, $\rho_*$, $a_1$ and $T_*$ as in the previous section. We also take $k=4/(3\rho_P)$. The basic  equations of the singular model are
\begin{equation}
\label{ni1a}
p=\frac{1}{3}\rho (1+4\rho/\rho_P)c^2,
\end{equation}
\begin{equation}
\label{ni1b}
w=\frac{1}{3} (1+4\rho/\rho_P), \qquad q=1+2\rho/\rho_P,
\end{equation}
\begin{equation}
\label{ni1}
\rho=\frac{\rho_P}{(a/a_1)^{4}- 1},
\end{equation}
\begin{equation}
\label{ni2}
T=T_P \left (\frac{15}{\pi^2}\right )^{1/4} \left (1+ \frac{\rho}{\rho_P}\right )^{7/4}\left (\frac{\rho}{\rho_P}\right )^{1/4},
\end{equation}
\begin{equation}
\label{ni3}
T=T_P \left (\frac{15}{\pi^2}\right )^{1/4} \frac{(a/a_1)^7}{\left\lbrack (a/a_1)^4- 1\right\rbrack^2}.
\end{equation}
The entropy is still given by Eq. (\ref{infb12}). As the universe expands from $a=a_1$ to $+\infty$, the density and the temperature decreases from $+\infty$ to $0$, the parameter $w$ decreases from $+\infty$ to $1/3$, the deceleration parameter $q$ decreases from $+\infty$ to $1$, and the ratio $(c_s/c)^2$ decreases from $+\infty$ to $1/3$.

\begin{figure}[!h]
\begin{center}
\includegraphics[clip,scale=0.3]{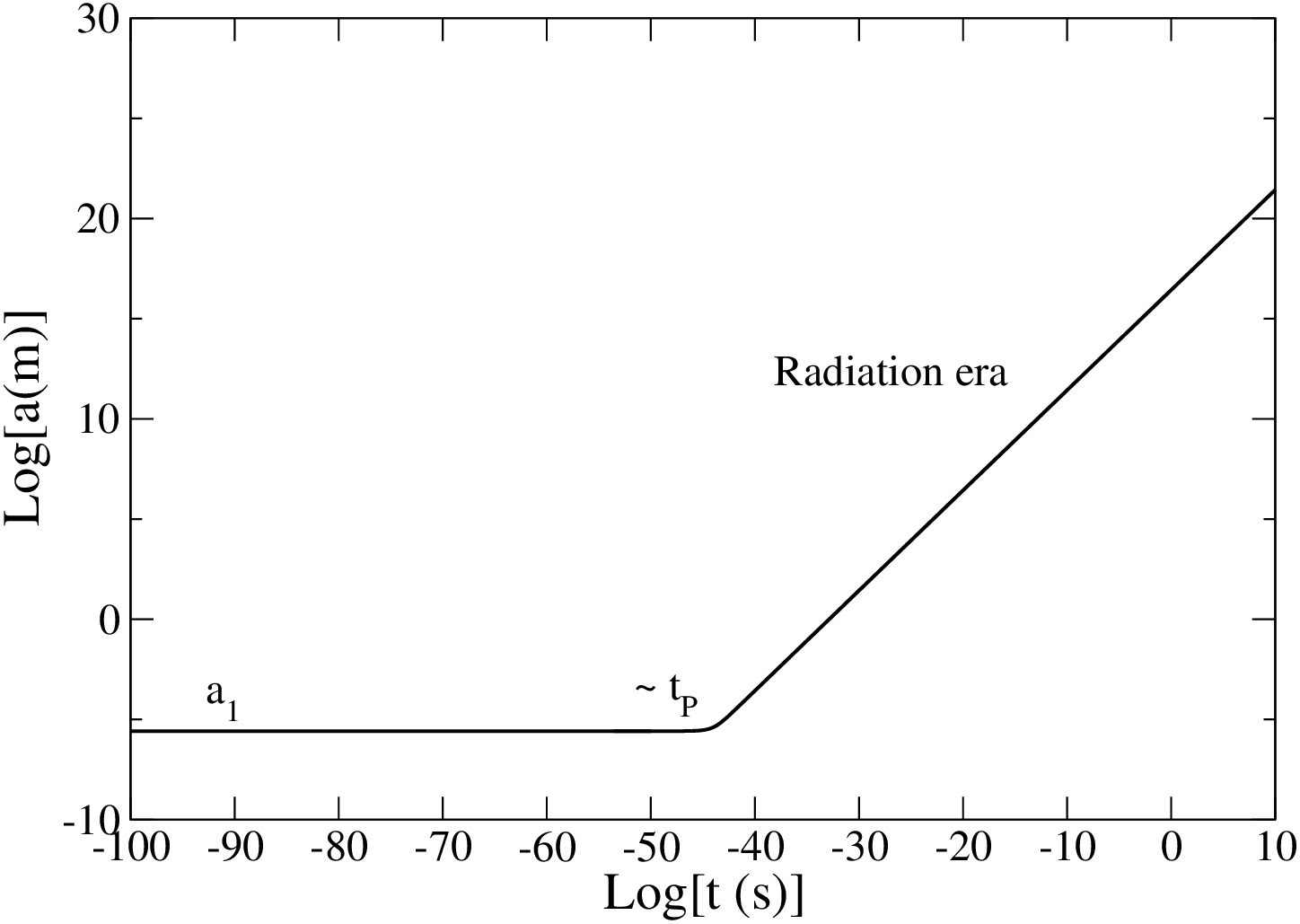}
\caption{Evolution of the scale factor $a$ with the time $t$ in logarithmic scales. There is no inflation since the universe already starts with a ``large'' radius $a_1=2.61\, 10^{-6}\, {\rm m}$. However, the initial condition is singular since the density and the temperature are infinite.}
\label{taLOGLOGPOS}
\end{center}
\end{figure}

\begin{figure}[!h]
\begin{center}
\includegraphics[clip,scale=0.3]{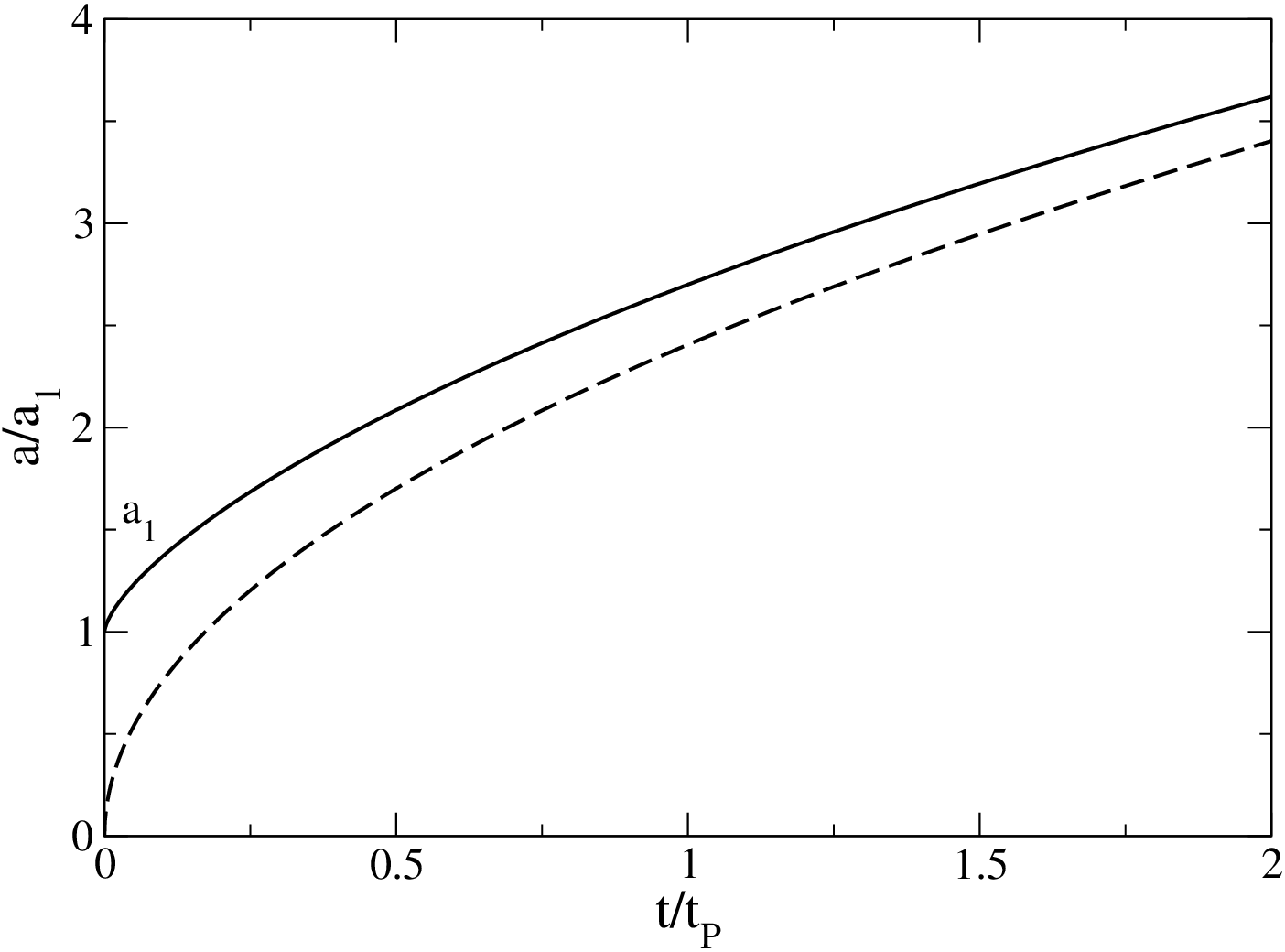}
\caption{Evolution of the scale factor $a$ with the time $t$ in linear scales. The dashed line corresponds to the standard radiation model of Sec. \ref{sec_radiation} (Big Bang).}
\label{taLINLINPOS}
\end{center}
\end{figure}

\begin{figure}[!h]
\begin{center}
\includegraphics[clip,scale=0.3]{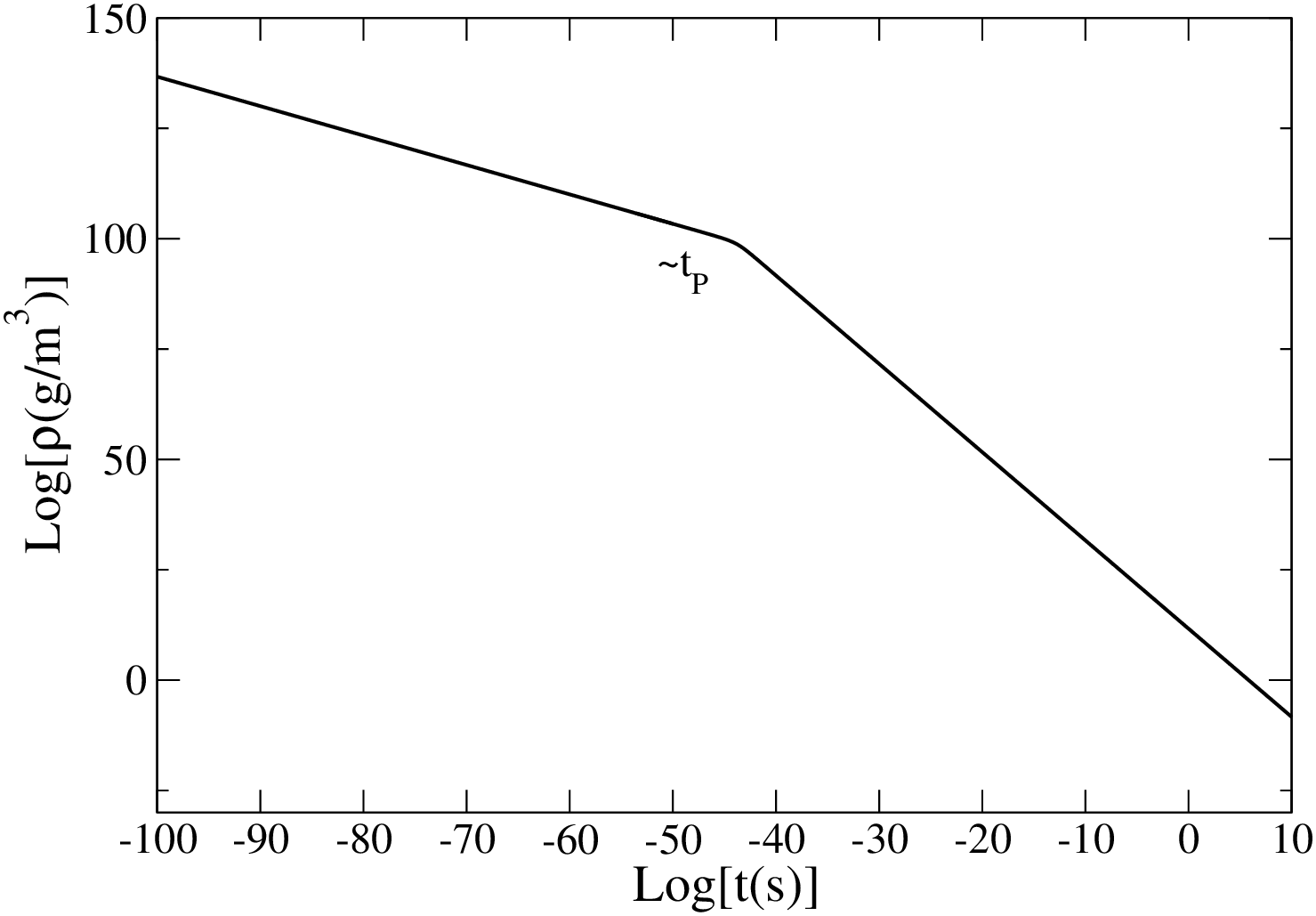}
\caption{Evolution of the density $\rho$ with the time $t$ in logarithmic scales. The slope goes from $-2/3$ in the pre-radiation era to $-2$ in the radiation era.}
\label{trhoLOGLOGPOS}
\end{center}
\end{figure}

\begin{figure}[!h]
\begin{center}
\includegraphics[clip,scale=0.3]{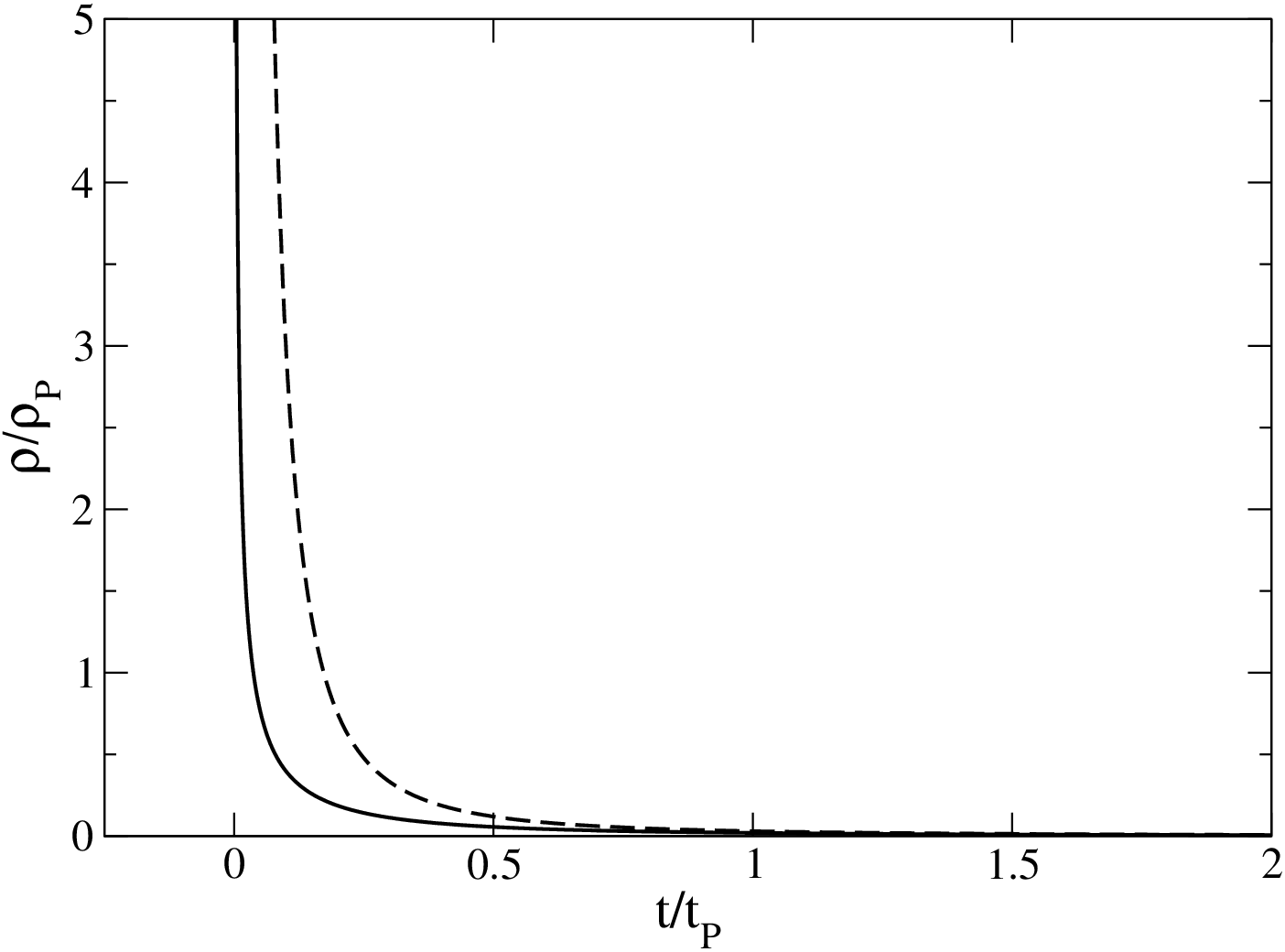}
\caption{Evolution of the density $\rho$ with the time $t$ in linear scales. The dashed line corresponds to the standard radiation model of Sec. \ref{sec_radiation}.}
\label{trhoLINLINPOS}
\end{center}
\end{figure}

\begin{figure}[!h]
\begin{center}
\includegraphics[clip,scale=0.3]{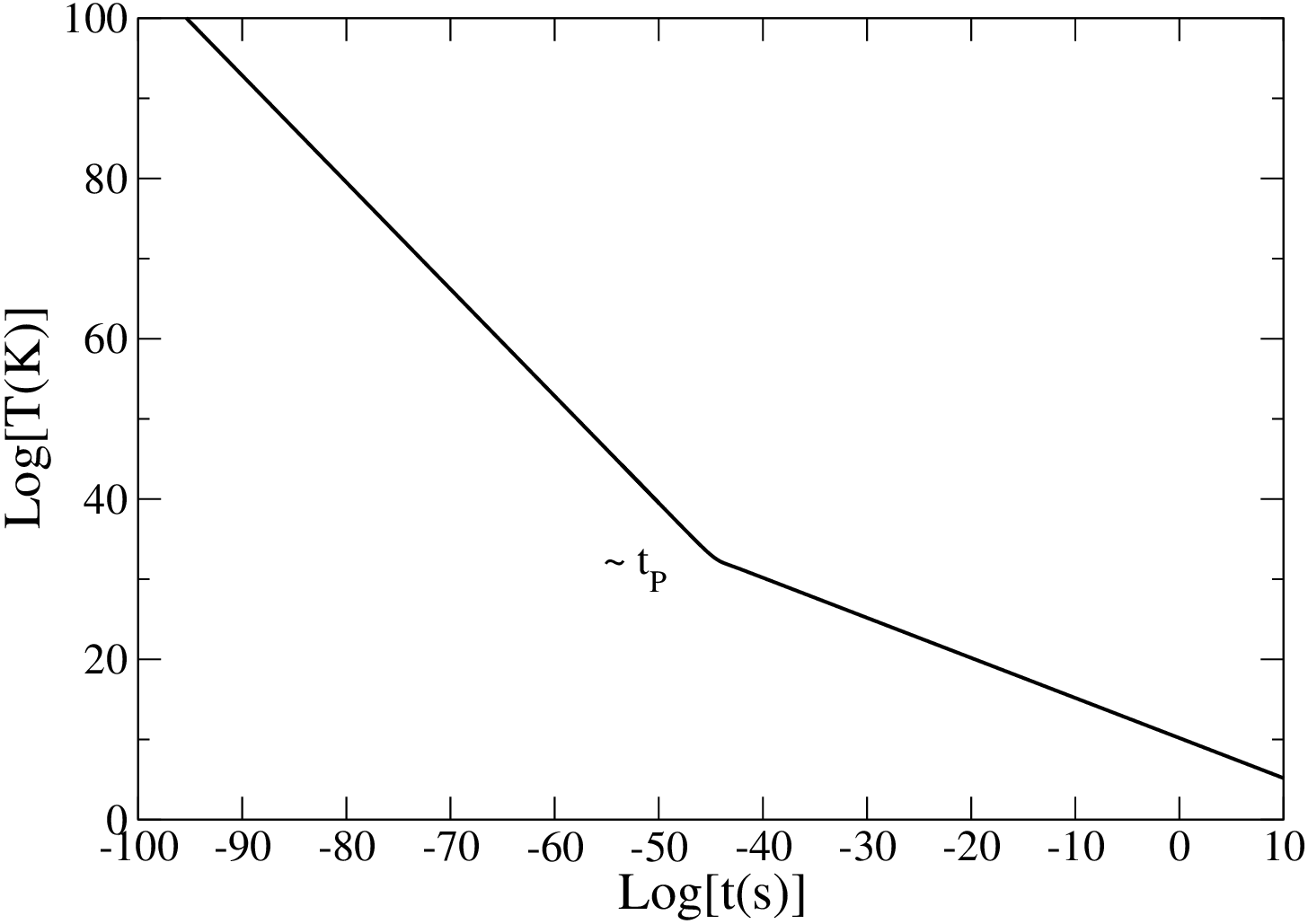}
\caption{Evolution of the temperature $T$ with the time $t$ in logarithmic scales. The slope goes from $-4/3$ in the pre-radiation era to $-1/2$ in the radiation era.}
\label{ttempLOGLOGPOS}
\end{center}
\end{figure}

\begin{figure}[!h]
\begin{center}
\includegraphics[clip,scale=0.3]{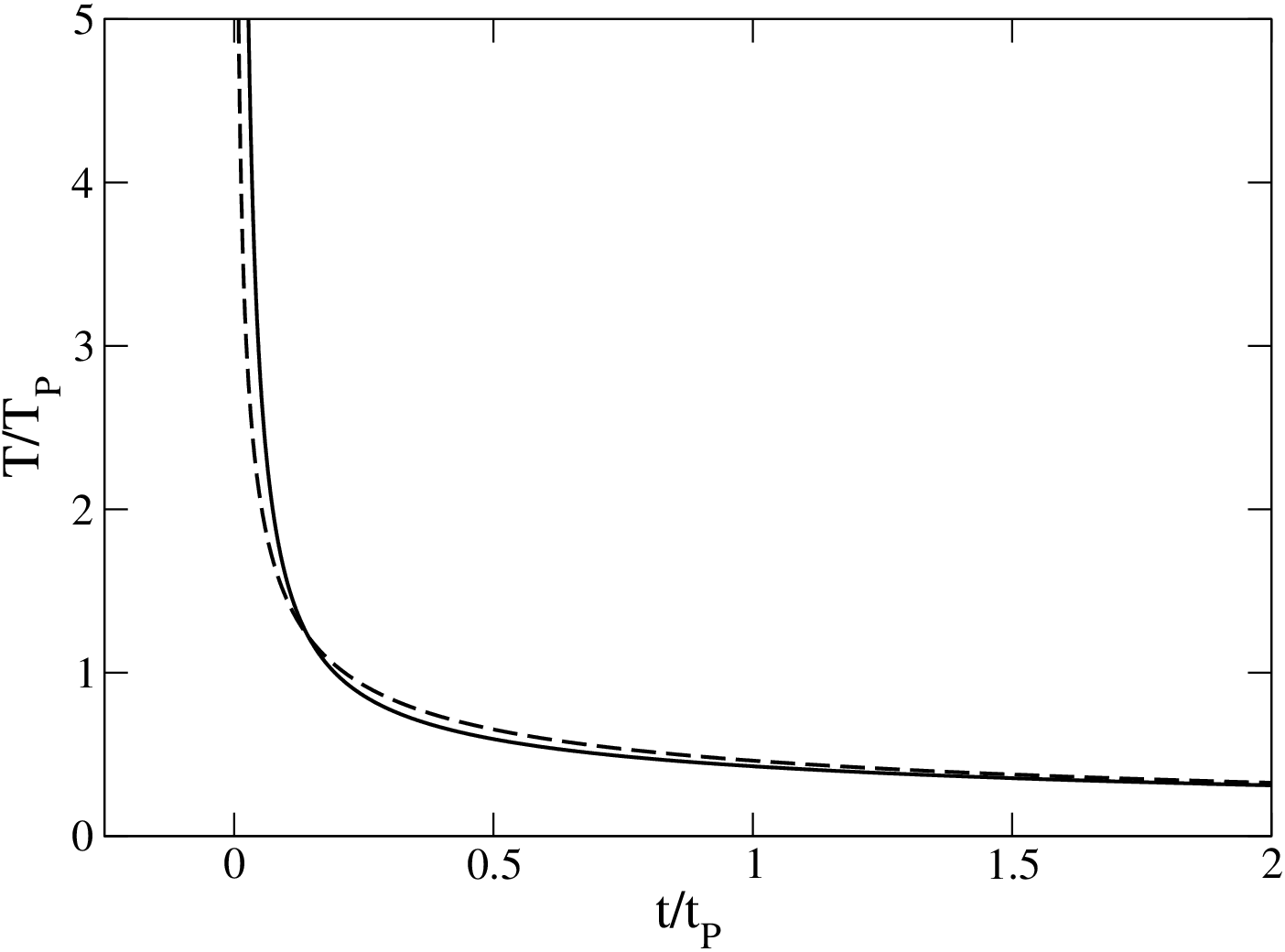}
\caption{Evolution of the temperature $T$ with the time $t$ in linear scales. The dashed line corresponds to the standard radiation model of Sec. \ref{sec_radiation}.}
\label{ttempLINLINPOS}
\end{center}
\end{figure}

For the equation of state (\ref{ni1a}), using the results of Sec. \ref{sec_darkrep}, the general solution of the Friedmann equation (\ref{f9}) is \cite{harko,c4}:
\begin{equation}
\label{ni4}
\sqrt{({a}/{a_1})^{4}-1}-\arctan \sqrt{({a}/{a_1})^{4}-1}=2\left (\frac{8\pi}{3}\right )^{1/2} \frac{t}{t_P}.
\end{equation}
The universe starts from a singularity at $t=0$. The initial radius of the universe is finite (with the value $a(0)=a_1=2.61\, 10^{-6}\, {\rm m}$) but the initial density and the initial temperature are infinite. The universe is always decelerating. The density and the temperature decrease indefinitely as the universe expands. There is no inflation since the universe starts with a ``large'' radius $a_1\gg l_P$.  When $t\rightarrow 0$,
\begin{eqnarray}
\label{ght1}
\frac{a}{a_1}\simeq 1+\left (\frac{3\pi}{2}\right )^{1/3}\left (\frac{t}{t_P}\right )^{2/3},
\end{eqnarray}
\begin{eqnarray}
\label{ght2}
\frac{\rho}{\rho_P}\sim \frac{1}{4}\left (\frac{2}{3\pi}\right )^{1/3}\left (\frac{t}{t_P}\right )^{-2/3},
\end{eqnarray}
\begin{eqnarray}
\label{ght3}
\frac{T}{T_P}\sim \left (\frac{15}{\pi^2}\right )^{1/4}\frac{1}{16}\left (\frac{2}{3\pi}\right )^{2/3}\left (\frac{t}{t_P}\right )^{-4/3}.
\end{eqnarray}
When $t\gg t_P$, the universe enters in the radiation era. In that case, we recover the results of Sec. \ref{sec_rp}.

The condition that the velocity of sound must be less than the speed of light imposes a maximum bound on the density. Using the results of Secs. \ref{sec_gessound} and \ref{sec_darkrep},  we find that the physical universe starts at a time
\begin{eqnarray}
\label{ni5}
t_i=\left (\frac{3}{8\pi}\right )^{1/2}\frac{1}{2}\left\lbrack 2-\arctan(2)\right\rbrack t_P.
\end{eqnarray}
At this time, its radius, density and temperature are
\begin{equation}
\label{ni6}
a_i=5^{1/4}a_1, \qquad \rho_i=\frac{1}{4}\rho_P,
\end{equation}
\begin{equation}
\label{ni8}
T_i=\left (\frac{15}{\pi^2}\right )^{1/4}\frac{5^{7/4}}{16}T_P.
\end{equation}
For $t<t_i$, the solution (\ref{ni4}) may be unphysical because the velocity of sound is larger than the speed of light. Numerically, $\rho_i=0.25\, \rho_P=1.29\, 10^{99}\, {\rm g}/{\rm m}^3$, $a_i=1.50\, a_1=3.90\, 10^{-6}\, {\rm m}$, $T_i=1.16\, T_P=1.64\, 10^{32}\, {\rm K}$ and $t_i=0.154\, t_P=8.32\, 10^{-45}\, {\rm s}$.

The evolution of the scale factor, density and temperature as a function of time are represented in Figs. \ref{taLOGLOGPOS}-\ref{ttempLINLINPOS} in logarithmic and linear scales.

\section{Analogy with phase transitions}
\label{sec_analogy}

The standard Big Bang theory is a classical theory in which quantum effects are neglected. In that case, it exhibits a finite time singularity: The radius is equal to zero at $t=0$ while the density and the temperature are infinite. For $t<0$, the solution is not defined and we may take $a=0$. For $t>0$ the radius increases as $a\propto t^{1/2}$.  This is similar to a second order phase transition if we view the time $t$ as the control parameter ({\it e.g.} the temperature $T$) and the scale factor $a$ as the order parameter ({\it e.g.} the magnetization $M$). It is amusing to note that the exponent $1/2$ is the same as in mean field theories of second order phase transitions ({\it i.e.} $M\sim (T-T_c)^{1/2}$) but this is essentially a coincidence.

When quantum mechanics effects are taken into account, as in the simple model of Sec. \ref{sec_inf}, the singularity at $t=0$ disappears and the curves $a(t)$, $\rho(t)$ and $T(t)$ are regularized. In particular, we find that $a=l_P>0$ at $t=0$, instead of $a=0$, due to the finite value of $\hbar$. This is similar to the regularization due to finite size effects ({\it e.g.} the system size $L$ or the number of particles $N$) in ordinary phase transitions. In this sense, the classical regime $\hbar\rightarrow 0$ is similar to the thermodynamic limit ($L\rightarrow +\infty$ or $N\rightarrow +\infty$) in ordinary phase transitions.

\begin{figure}[!h]
\begin{center}
\includegraphics[clip,scale=0.3]{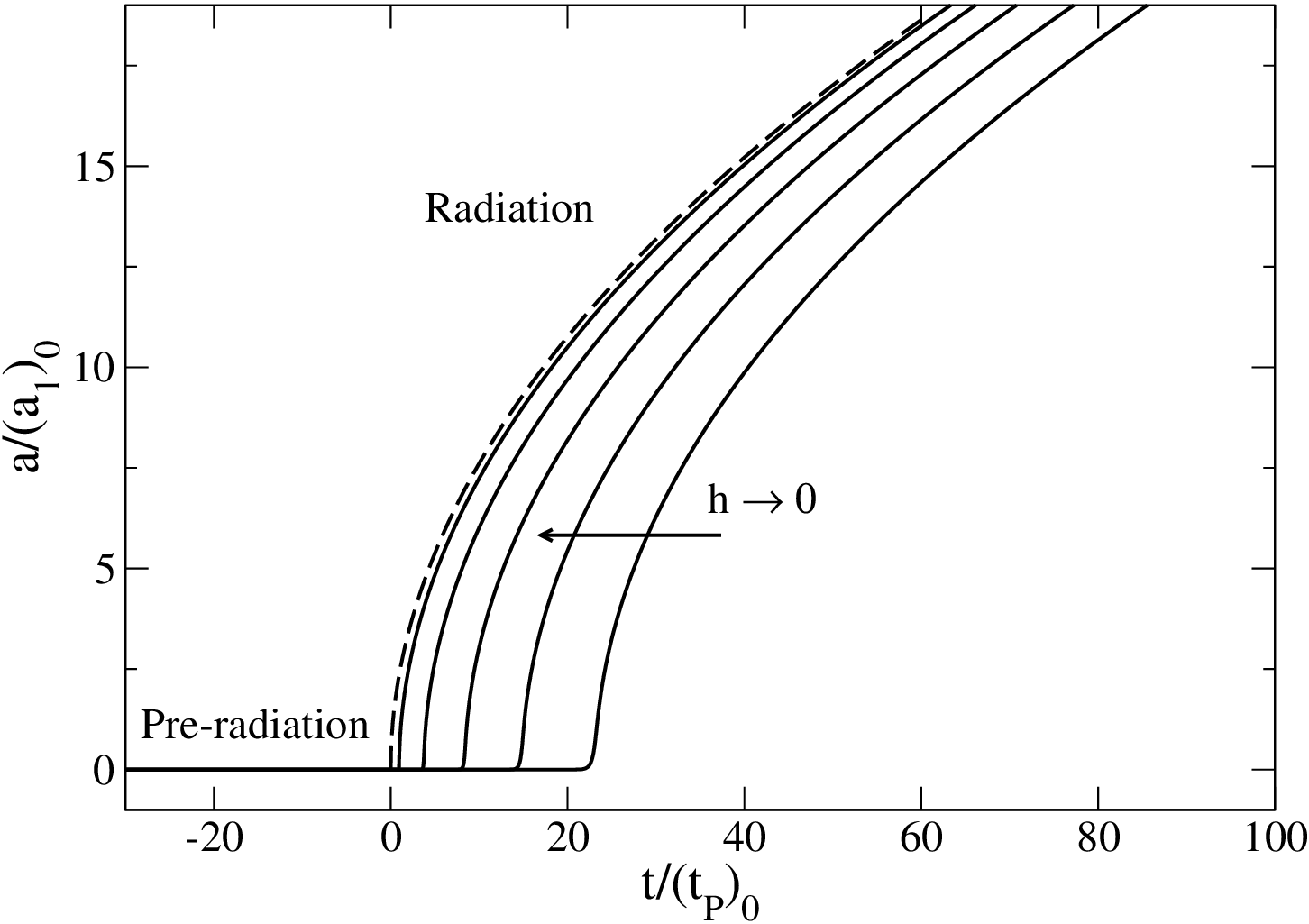}
\caption{Effect of quantum mechanics (finite value of the Planck constant) on the regularization of the singular Big Bang solution ($\hbar=0$, dashed line) in the model of Sec. \ref{sec_inf}. The singularity at $t=0$ is replaced by an inflationary expansion from the pre-radiation era to the radiation era. We can draw an analogy with second order phase transitions where the Planck constant plays the role of finite size effects (see the text for details).}
\label{phasetransition}
\end{center}
\end{figure}

\begin{figure}[!h]
\begin{center}
\includegraphics[clip,scale=0.3]{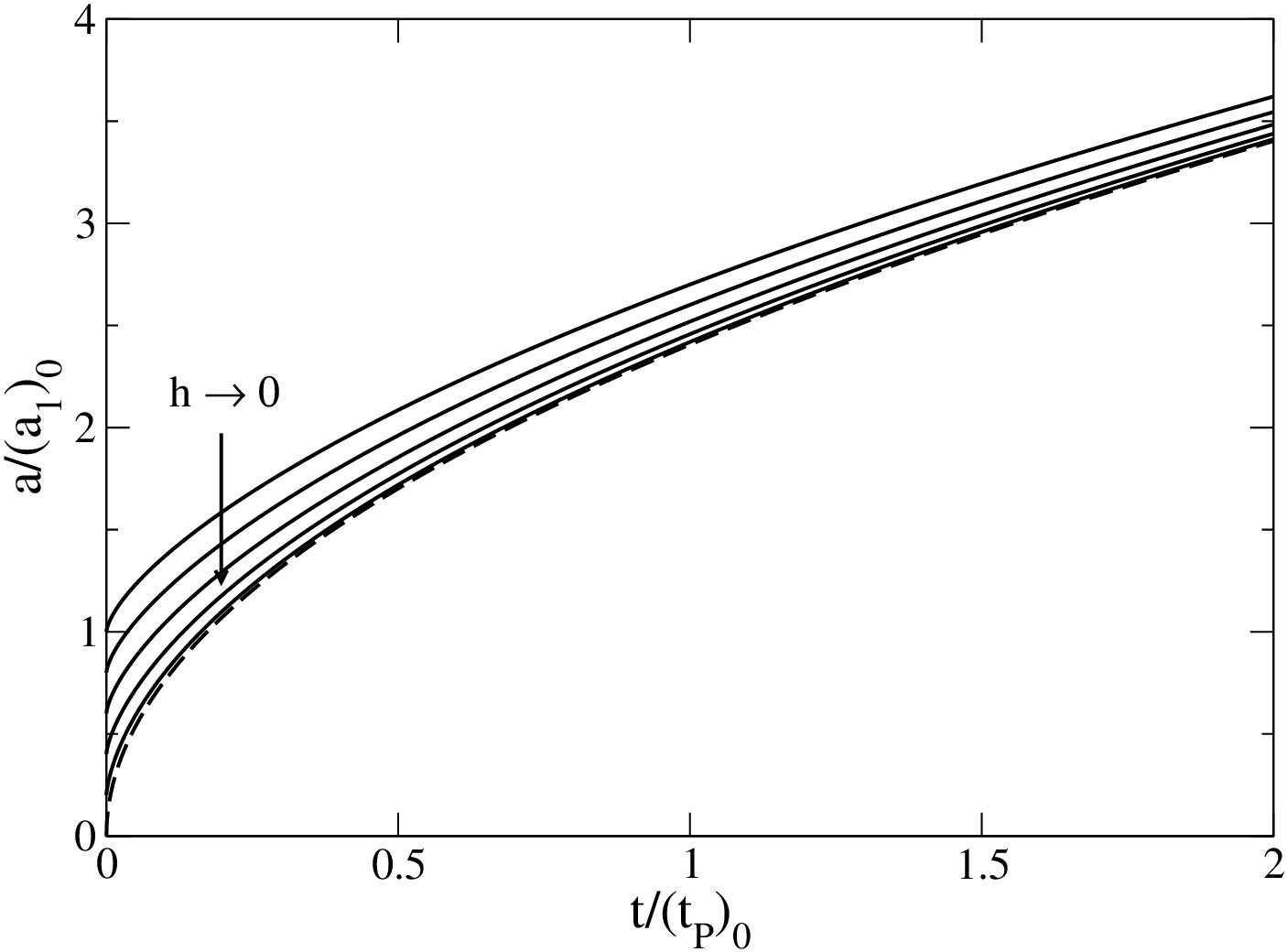}
\caption{Effect of quantum mechanics (finite value of the Planck constant) on the evolution of the scale factor $a(t)$ in the model of Sec. \ref{sec_ni}. The standard Big Bang solution ($\hbar=0$) is represented by a dashed line. For $\hbar>0$, the initial radius is strictly positive and it tends to zero as $\hbar\rightarrow 0$.}
\label{phasetransitionPOS}
\end{center}
\end{figure}

To study the convergence toward the classical Big Bang solution when $\hbar\rightarrow 0$, it is convenient to use a proper normalization of the parameters. We call $\hbar_0=1.05\, 10^{-34}\, {\rm J}\, {\rm s}$ the true value of the Planck constant (in this section, the subscript $0$ refers to true  values of the parameters). Then, we express time in terms of $t_P^0$ and lengths in terms of $a_1^0$. We introduce $\hat{a}=a/a_1^0$ and $\hat{t}=t/t_P^0$. Using $l_P=c t_P$ and Eq. (\ref{infb5}), we obtain
\begin{equation}
\label{ana1}
l_P/l_P^0=t_P/t_P^0=(a_1/a_1^0)^2=\hat{\hbar}^{1/2},
\end{equation}
where $\hat{\hbar}=\hbar/\hbar_0$.
Therefore
\begin{equation}
\label{ana2}
\epsilon/\epsilon_0=\frac{l_P/a_1}{l_P^0/a_1^0}=a_1/a_1^0=\hat{\hbar}^{1/4},
\end{equation}
where $\epsilon_0={l_P^0/a_1^0}=6.20\, 10^{-30}$. Using these relations and  Eq. (\ref{gp1}), we obtain
\begin{eqnarray}
\label{ana3}
\sqrt{\hat{a}^{4}/\hat{\hbar}+1}-\ln \left (\frac{1+\sqrt{\hat{a}^{4}/\hat{\hbar}+1}}{\hat{a}^{2}/\hat{\hbar}^{1/2}}\right )\nonumber\\
=2\left (\frac{8\pi}{3}\right )^{1/2} {\hat{t}}/{\hat{\hbar}^{1/2}}+C(\epsilon_0 \hat{\hbar}^{1/4}),
\end{eqnarray}
where $C(\epsilon)$ is defined by Eq. (\ref{gp3}). This equation describes a phase transition between the pre-radiation era and the radiation era. The normalized Planck constant $\hat{\hbar}$ plays the role of finite size effects. Finite size scalings are explicit in Eq. (\ref{ana3}). For $\hat{\hbar}=1$, we recover the non-singular model of Sec. \ref{sec_inf}. For $\hat{\hbar}=0$, we recover the singular radiation model of Sec. \ref{sec_radiation} (Big Bang). The convergence towards this singular solution as $\hat{\hbar}\rightarrow 0$ is shown in Fig. \ref{phasetransition}.

We can make the same study for the model of Sec. \ref{sec_ni} although the analogy with phase transitions with finite size effects is less clear since this model remains singular for finite values of the Planck constant. Nevertheless, we have represented the effect of quantum mechanics on the form of the curve $a(t)$ in Fig. \ref{phasetransitionPOS}. For finite values of $\hbar$, using Eq. (\ref{ni4}), the evolution of the scale factor is given by
\begin{eqnarray}
\label{ana4}
\sqrt{\hat{a}^4/\hat{\hbar}-1}-\arctan \sqrt{\hat{a}^{4}/\hat{\hbar}-1}=2\left (\frac{8\pi}{3}\right )^{1/2} \hat{t}/\hat{\hbar}^{1/2}.\nonumber\\
\end{eqnarray}
At $t=0$, the scale factor is $\hat{a}(0)=\hat{\hbar}^{1/4}$.

\section{Conclusion}

In this paper, we have carried out an exhaustive study of the generalized equation of state (\ref{ges0}) for positive indices $n>0$. For $\alpha=1/3$, $n=1$ and $k=-4/(3\rho_P)$, this equation of state describes in a unified manner the transition between the pre-radiation era ($\rho=\rho_P$) and the radiation era ($\rho\propto a^{-4}$) in the early universe. It provides a model of early inflation. The case of negative indices $n<0$ is treated in Paper II. For $\alpha=0$, $n=-1$ and $k=-\rho_{\Lambda}$, where $\rho_{\Lambda}$ is the cosmological density, this equation of state describes in a unified manner the transition between the matter era ($\rho\propto a^{-3}$) and the dark energy era  ($\rho=\rho_{\Lambda}$) in the late universe. It provides a model of late inflation. Combining these two approaches, we obtain a simple non-singular model for the whole evolution of the universe based on two symmetric polytropic equations of state (see Paper II).

When we started our study, we were mainly interested in the equation of state $p=(\alpha \rho+k\rho^2)c^2$ introduced in our previous paper \cite{c4}. This equation of state was obtained in the context of  Bose-Einstein condensates, assuming that they are the constituents of  dark matter. Although BECs should form after the radiation era when the universe has cooled sufficiently, we wondered whether this equation of state could have some sense before the radiation era. In that picture, the pre-radiation era would correspond to primordial BECs with attractive self-interaction $\lambda\sim -(m/M_P)^4$ (see Appendix \ref{sec_annbec}). Since the pre-radiation era is extremely cold (see Sec. \ref{sec_inf}), the BEC model (which assumes $T=0$) may be an interesting suggestion to develop further.

However, we progressively realized that the study of a generalized equation of state of the form $p=(\alpha \rho+k\rho^{1+1/n})c^2$ was of interest beyond the context of BECs. Therefore, our present point of view is more general. We believe that the equation of state (\ref{ges0}) may have several origins that should be discussed in future works. By studying this equation of state in full generality, we realized that the positive indices $n>0$ describe the early universe while the negative indices $n<0$ describe the late universe. Furthermore, a positive polytropic pressure ($k>0$) leads to past or future singularities (or peculiarities) while a negative polytropic pressure ($k<0$) leads to non-singular models. They exhibit phases of early and late inflation associated with a maximum density $\rho_{max}=\rho_P$ (Planck density) corresponding to the vacuum energy in the past and a minimum density $\rho_{min}=\rho_{\Lambda}$ (cosmological density) corresponding to the dark energy in the future.  Therefore, in the polytropic model, the description of the early and late universe appears to be very ``symmetric''. This result is obtained in a purely theoretical manner, without reference to observations. Strikingly, this symmetry is consistent with what we know about the real universe. This point will be further developed in Paper II.

Another motivation of our study was to discuss some aspects of the history of science and publicize (in the same vein as \cite{luminet,nb}) the important contributions of Lema\^itre in cosmology, which may not be sufficiently recognized. They include: 1. The first correct understanding of the mysterious de Sitter model \cite{lemaitre1925}; 2. The re-discovery of the Friedmann equations and their thermodynamical interpretation \cite{lemaitre1927}; 3. The instability of the Einstein static universe \cite{lemaitre1927}; 4. The interpretation of the redshifts as a results of the expansion of the universe in accord with the Einstein theory of relativity \cite{lemaitre1927};  5. The discovery of the Hubble law and the calculation of the Hubble constant two years before Hubble \cite{lemaitre1927}; 6. The basis of Newtonian cosmology \cite{lemaitreNewton,lemaitre1934}; 7. The correct place of the cosmological constant in the Poisson equation \cite{lemaitrecosmo}; 8. The importance of the cosmological constant and its interpretation in terms of a vacuum energy density \cite{lemaitre1934}; 9. The notion of a primordial singularity (primeval atom) that finally led to the Big Bang theory \cite{lemaitresingular,lemaitreNewton,lemaitre1933}; 10. The importance of quantum mechanics in the early universe \cite{lemaitrenature}.  Some other important contributions of Lema\^itre have been collected recently in \cite{luminetnew}.

\appendix

\section{Equation of state $p=(-\rho+k\rho^{\gamma})c^2$ with $n>0$ and $k>0$}
\label{sec_eosgm}

In this Appendix, we specifically study the equation of state (\ref{ges0}) with $\alpha=-1$, $n>0$, and $k>0$, namely
\begin{eqnarray}
\label{eosgm0}
p=(-\rho+k\rho^{\gamma})c^2.
\end{eqnarray}
It generalizes  the equation of state $p=-\rho c^2$ of the vacuum. For the equation of state (\ref{eosgm0}), the continuity equation (\ref{f7}) can be integrated into
\begin{eqnarray}
\label{eosgm1}
\rho=\frac{\rho_*}{\ln(a/a_*)^n},
\end{eqnarray}
where $\rho_*=(n/3k)^n$ and $a_*$ is a constant of integration. The density is defined for $a\ge a_*$. When $a\rightarrow a_*$, $\rho\rightarrow +\infty$ and $p\rightarrow +\infty$; when $a\rightarrow +\infty$, $\rho\rightarrow 0$ and $p\rightarrow 0$.

The thermodynamical equation (\ref{t2}) can be integrated into
\begin{eqnarray}
\label{eosgm2}
T=T_* \left (\frac{\rho}{\rho_*}\right )^{(n+1)/n} e^{3(\rho_*/\rho)^{1/n}},
\end{eqnarray}
where $T_*$ is a constant of integration. Combined with Eq. (\ref{eosgm1}), we obtain
\begin{eqnarray}
\label{eosgm3}
T= \frac{T_*}{\ln(a/a_*)^{n+1}}\left (\frac{a}{a_*}\right )^3.
\end{eqnarray}
When $a\rightarrow a_*$  and $a\rightarrow +\infty$, $T\rightarrow +\infty$.  The temperature has a minimum at
\begin{equation}
\label{eosgm4a}
\frac{\rho_{e}}{\rho_*}=\left (\frac{3}{n+1}\right )^n,\qquad \frac{a_{e}}{a_*}=e^{(n+1)/3},
\end{equation}
\begin{equation}
\label{eosgm4b} \frac{T_e}{T_*}=\left (\frac{3}{n+1}\right )^{n+1}e^{n+1}.
\end{equation}

\begin{figure}[!h]
\begin{center}
\includegraphics[clip,scale=0.3]{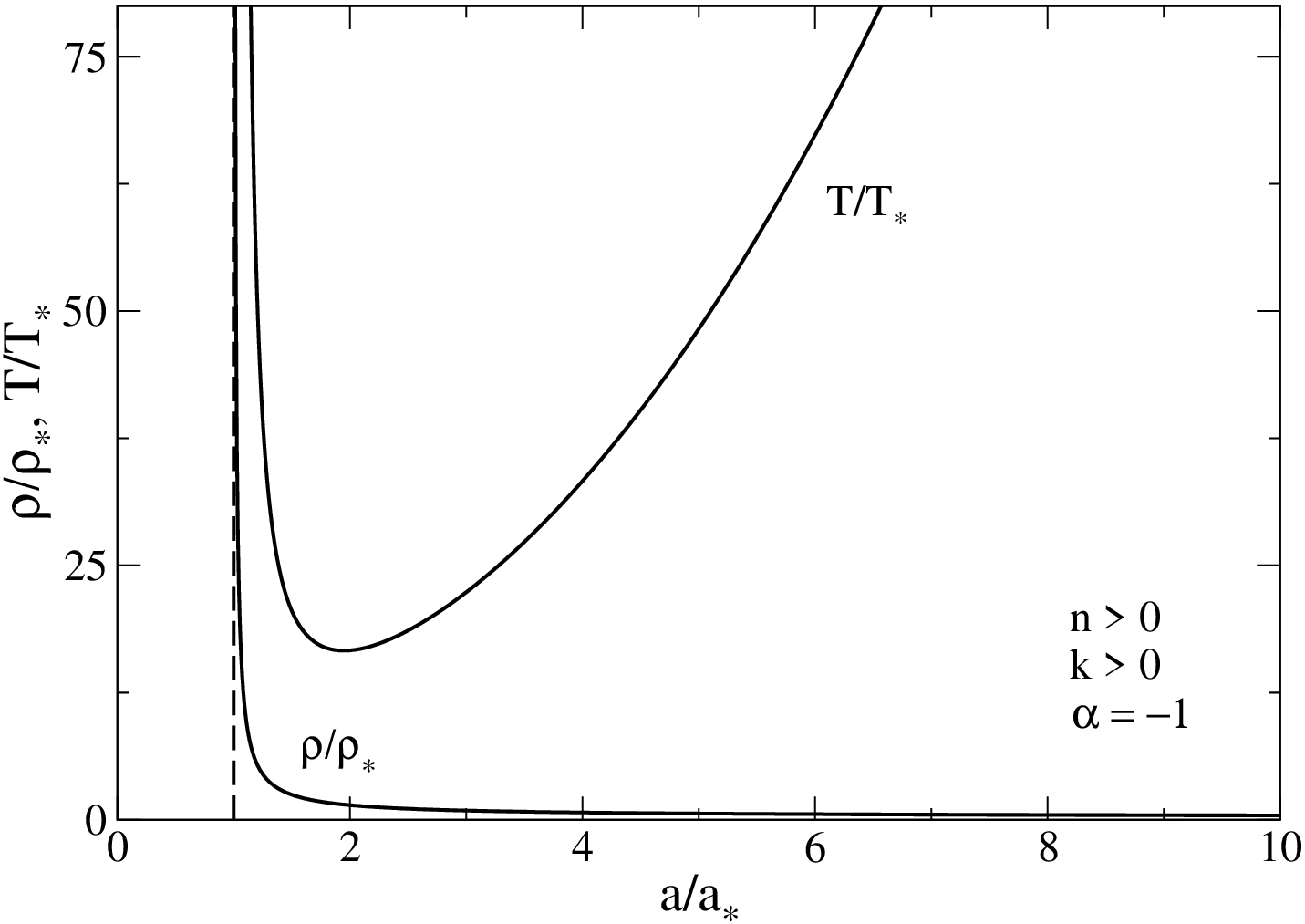}
\caption{Evolution of the density and temperature as a function of the scale factor $a$. We have taken $n=1$.}
\label{appendixarhotemp}
\end{center}
\end{figure}

The evolution of the density and temperature as a function of the scale factor is plotted  in Fig. \ref{appendixarhotemp}.

\begin{figure}[!h]
\begin{center}
\includegraphics[clip,scale=0.3]{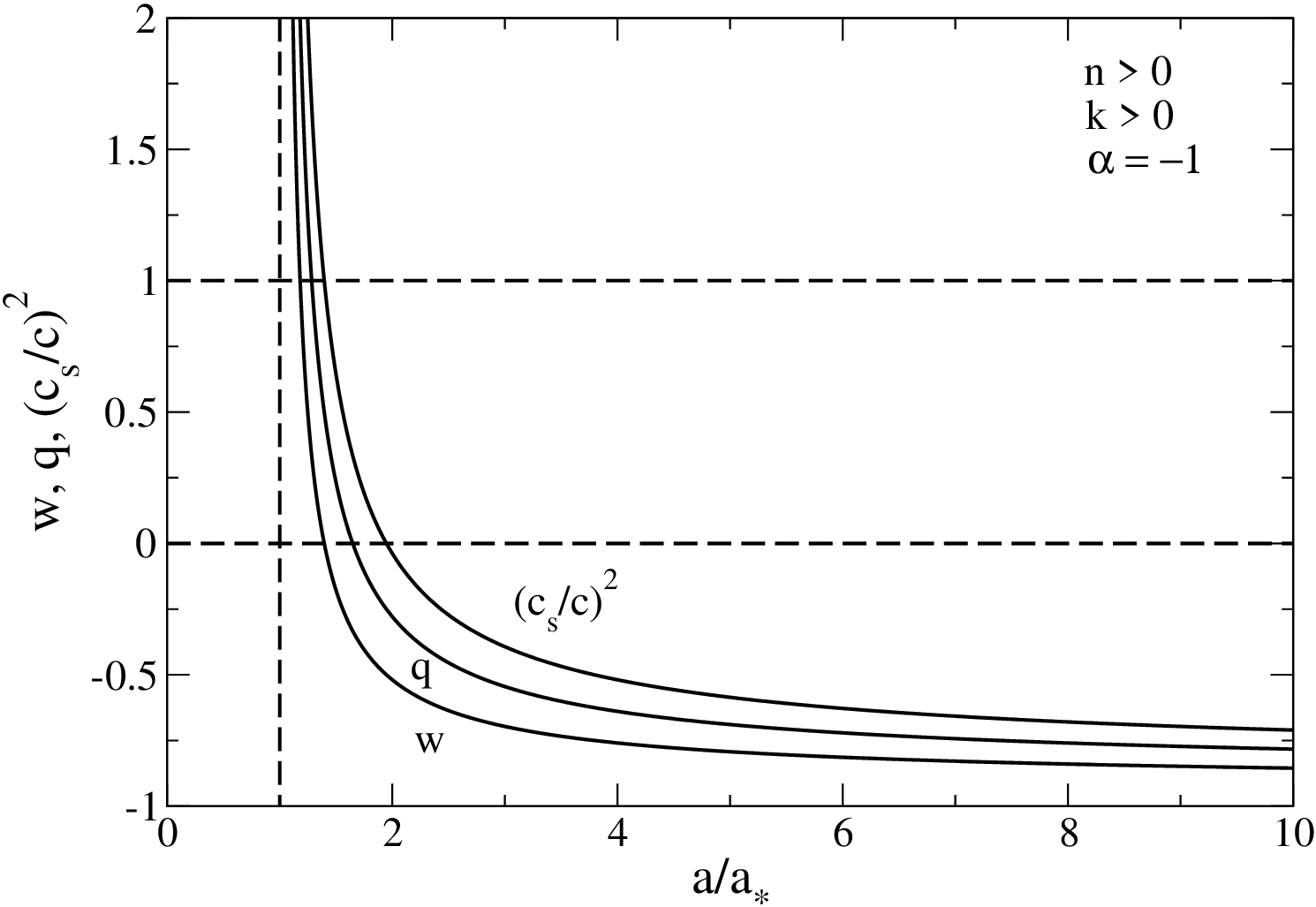}
\caption{Evolution of $w$, $q$, and $(c_s/c)^2$ as a function of the scale factor $a$. We have taken $n=1$.}
\label{appendixwqc}
\end{center}
\end{figure}

The equation of state can be written as $p=w\rho c^2$ with
\begin{equation}
\label{eosgm5}
w=-1+\frac{n}{3}\left (\frac{\rho}{\rho_*}\right )^{1/n}.
\end{equation}
The pressure vanishes ($w=0$) at
\begin{equation}
\label{eosgm6}
\frac{\rho_{w}}{\rho_*}=\left (\frac{3}{n}\right )^n,\qquad \frac{a_{w}}{a_*}=e^{n/3},\qquad \frac{T_w}{T_*}=e^n \left (\frac{3}{n}\right )^{n+1}.
\end{equation}
When $a\rightarrow a_*$, $w\rightarrow +\infty$; when $a\rightarrow +\infty$, $w\rightarrow -1$. The pressure is positive for $a_*<a<a_w$ and negative for $a>a_w$.

The deceleration parameter is given by Eqs. (\ref{dec1}) and (\ref{dec2}). Together with Eq. (\ref{eosgm5}), we obtain
\begin{equation}
\label{eosgm7}
q=-1+\frac{n}{2}\left (\frac{\rho}{\rho_*}\right )^{1/n}.
\end{equation}
The curve $a(t)$ presents an inflexion point ($\ddot a=q=0$) at
\begin{equation}
\label{eosgm8}
\frac{\rho_{c}}{\rho_*}=\left (\frac{2}{n}\right )^n,\qquad \frac{a_{c}}{a_*}=e^{n/2},\qquad \frac{T_c}{T_*}=\left (\frac{2}{n}\right )^{n+1}e^{3n/2}.
\end{equation}
When $a\rightarrow a_*$, $q\rightarrow +\infty$; when $a\rightarrow +\infty$, $q\rightarrow -1$. The universe is decelerating when $a_*<a<a_c$ and accelerating when $a>a_c$.

Finally, the velocity of sound is given by
\begin{equation}
\label{eosgm9}
\frac{c_s^2}{c^2}=-1+\frac{n+1}{3}\left (\frac{\rho}{\rho_*}\right )^{1/n}.
\end{equation}
The velocity of sound vanishes at the point defined by Eqs. (\ref{eosgm4a})-(\ref{eosgm4b}) at which the temperature reaches its minimum. At that point, the pressure is minimum with value
\begin{equation}
\label{eosgm9b}
\frac{p_e}{\rho_* c^2}=-\frac{3^n}{(n+1)^{n+1}}.
\end{equation}
When $a\rightarrow a_*$, $(c_s/c)^2\rightarrow +\infty$; when $a\rightarrow +\infty$, $(c_s/c)^2\rightarrow -1$. The velocity of sound is real when $a_*<a<a_e$ and imaginary when $a>a_e$. On the other hand, the velocity of sound is equal to the speed of light at
\begin{equation}
\label{eosgm10}
\frac{\rho_{s}}{\rho_*}=\left (\frac{6}{n+1}\right )^n,\qquad \frac{a_{s}}{a_*}=e^{(n+1)/6},
\end{equation}
\begin{equation}
\label{eosgm11}
\frac{T_s}{T_*}=\left (\frac{6}{n+1}\right )^{n+1}e^{(n+1)/2}.
\end{equation}
For $n=1$, this corresponds to the point at which the pressure vanishes ($w=0$).
The velocity of sound is larger than the speed of light when $a_*<a<a_s$ and smaller when $a>a_s$.
The evolution of $w$, $q$, and $(c_s/c)^2$ as a function of the scale factor $a$ is represented in Fig. \ref{appendixwqc}.

\begin{figure}[!h]
\begin{center}
\includegraphics[clip,scale=0.3]{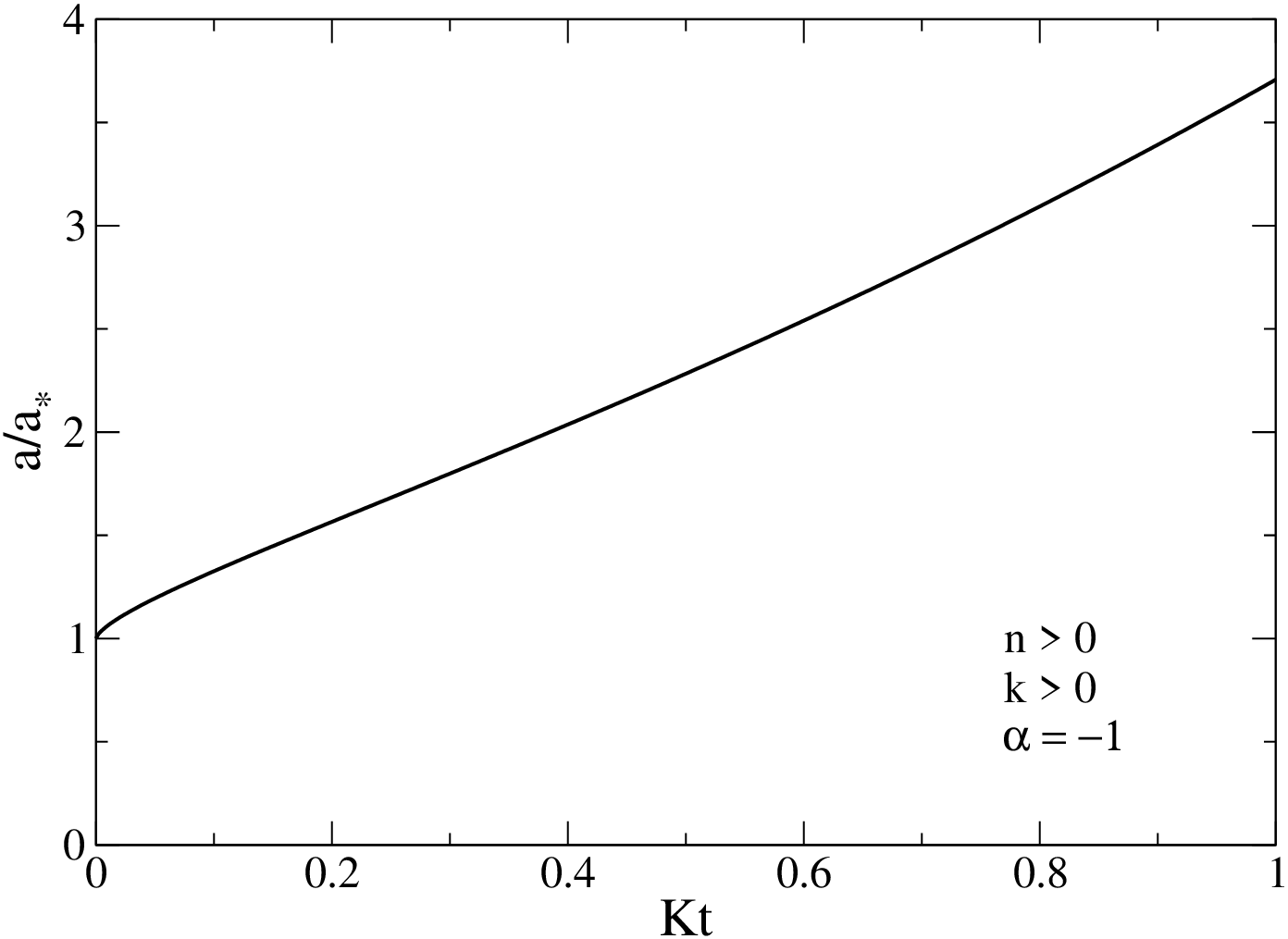}
\caption{Evolution of the scale factor with time. We have taken $n=1$. For this index, $Kt_i=0.128$, $Kt_c=0.236$, and $Kt_c=0.363$.}
\label{appendixta}
\end{center}
\end{figure}

Setting $R=a/a_*$, the Friedmann equation (\ref{f9}) can be written
\begin{eqnarray}
\label{eosgm12}
\dot R=\frac{K R}{(\ln R)^{n/2}},
\end{eqnarray}
where $K=(8\pi G\rho_*/3)^{1/2}$. Its solution is
\begin{eqnarray}
\label{eosgm13}
R(t)=e^{\left (\frac{n+2}{2} Kt\right )^{2/(n+2)}},
\end{eqnarray}
where we have determined the constant of integration such that $R=1$ at $t=0$. The density and the temperature evolve as
\begin{eqnarray}
\label{eosgm14}
\frac{\rho(t)}{\rho_*}=\left (\frac{n+2}{2} Kt\right )^{-2n/(n+2)},
\end{eqnarray}
\begin{eqnarray}
\label{eosgm15}
\frac{T(t)}{T_*}=\left (\frac{n+2}{2} Kt\right )^{-2(n+1)/(n+2)}e^{3\left (\frac{n+2}{2} Kt\right )^{2/(n+2)}}.\qquad
\end{eqnarray}
The universe starts at $t=0$ with a radius $R=1$ and an infinite density and infinite pressure. Then, the universe expands indefinitely while the density decreases to zero. The physical solution, for which the velocity of sound is smaller than the  speed of light, starts at $t=t_i$ where $Kt_i=\lbrack 2/(n+2)\rbrack \lbrack (n+1)/6\rbrack^{(n+2)/2}$. For $n=1$, this corresponds to the time at which the pressure becomes negative. The expansion of the universe is decelerating when $t<t_c$ and accelerating when $t>t_c$, where $Kt_c=\lbrack 2/(n+2)\rbrack (n/2)^{(n+2)/2}$. The velocity of sound becomes imaginary when $t>t_e$, where $Kt_e=\lbrack 2/(n+2)\rbrack \lbrack (n+1)/3\rbrack^{(n+2)/2}$. This corresponds to the time at which the temperature reaches its minimum. The evolution of the scale factor with time is plotted in Fig. \ref{appendixta}.

\section{Connection to the BEC model}
\label{sec_annbec}

At $T=0$, and in the strong coupling limit, the equation of state of  Bose-Einstein condensates (BECs) with short-range interactions is
\begin{equation}
\label{annbec1}
p=\frac{2\pi a_s\hbar^2}{m^3}\rho^{2},
\end{equation}
where $m$ is the mass of the bosons and $a_s$ is the s-scattering length \cite{dalfovo}. Therefore, a self-interacting BEC has a polytropic equation of state $p=K \rho^{\gamma}$ with index $\gamma=2$ ({\it i.e.} $n=1$) and polytropic constant $K={2\pi a_s\hbar^2}/{m^3}$. Furthermore, the scattering length $a_s$ can be positive or negative (positive scattering corresponds to repulsive self-interaction, and negative scattering to attractive self-interaction). In terrestrial BEC experiments, some atoms like $^7{\rm Li}$ have a negative scattering length \cite{fedichev}.  As a result, the pressure of a BEC can be {\it negative}. Therefore, there exist physical systems in which it is possible to obtain polytropic equations of state with negative pressure. Furthermore, the origin of these equations of state is due to quantum mechanics since the Planck constant explicitly appears in Eq. (\ref{annbec1}). This may be a justification (but not the only one) for the generalized equation of state (\ref{ges0}).

It is convenient to define a coupling constant $\lambda$  by \cite{c1}:
\begin{equation}
\label{annbec2}
\frac{\lambda}{8\pi}=\frac{a_s m c}{\hbar}.
\end{equation}
We can now use $(m,\lambda)$ as independent variables instead of $(m,a_s)$. The equation of state (\ref{annbec1}) can be rewritten as
\begin{equation}
\label{annbec3}
p=k\rho^2c^2,\qquad {\rm with} \quad k=\frac{\lambda\hbar^3}{4m^4c^3}.
\end{equation}
In the inflationary model of Sec. \ref{sec_inf}, in which the pre-radiation era is described by a ``fluid'' with  a polytropic equation of state of the form (\ref{annbec3}-a) with $k<0$, the maximum density is $\rho_*=4/(3|k|)$. If we assume that this ``fluid'' is made of BECs, then, using Eq. (\ref{annbec3}-b), we obtain
\begin{equation}
\label{annbec4}
\rho_*=\kappa\rho_P,\qquad {\rm with} \quad \kappa=\frac{16}{3}\left (\frac{m}{M_P}\right )^4 \frac{1}{|\lambda|}.
\end{equation}
Assuming that $\rho_*=\rho_{max}$ is of the order of the Planck density $\rho_P$, {\it i.e.} $\kappa\sim 1$, we find that the coupling constant of the BEC scales as  $\lambda\sim -(16/3)(m/M_P)^4$. Of course, the nature of these ``primordial'' BECs remains highly speculative. It is possible that this BEC description is just an effective one.


\begin{thebibliography}{}

\bibitem{luminet}{\small A. Friedmann, G. Lema\^itre, Essais de Cosmologie, preceded by L'invention du Big Bang by J.P. Luminet  (Source du savoir Seuil, 1997)}
\bibitem{nb}{\small H. Nussbaumer, L. Bieri, Discovering the Expanding Universe, (Cambridge, 2009)}
\bibitem{einsteincosmo}{\small A. Einstein, Sitz. K\"onig. Preu. Akad. Wiss., 142 (1917)}
\bibitem{lemaitrecosmo}{\small G.E. Lema\^itre, in {\it The Cosmological Constant,} edited by P.A. Schlipp (Open Court, La Salle, Illinois, 1997)}
\bibitem{deSitter1}{\small W. de Sitter, Proc. Akad. Wetensch. Amsterdam {\bf 19}, 1217 (1917)}
\bibitem{deSitter2}{\small W. de Sitter, Monthly Not. Roy. Astron. Soc.  {\bf 78}, 3 (1917)}
\bibitem{robertson}{\small H.P. Robertson, Rev. Mod. Phys. {\bf 5}, 62 (1933)}
\bibitem{friedmann1}{\small A. Friedmann, Zeits. f. Physik {\bf 10}, 377 (1922)}
\bibitem{friedmann2}{\small A. Friedmann, Zeits. f. Physik {\bf 21}, 326 (1924)}
\bibitem{einsteincritique1}{\small A. Einstein, Zeits. f. Physik {\bf 11}, 326 (1922)}
\bibitem{einsteincritique2}{\small A. Einstein, Zeits. f. Physik {\bf 16}, 228 (1923)}
\bibitem{lanczos1922}{\small K. Lanczos, Phys. Zeits.  {\bf 23}, 539 (1922)}
\bibitem{weyl}{\small H. Weyl, Phys. Zeits.  {\bf 24}, 230 (1923)}
\bibitem{lemaitre1925}{\small G. Lema\^itre, J. Math. and Physics (M.I.T.)  {\bf 4}, 188 (1925)}
\bibitem{lemaitre1927}{\small G. Lema\^itre, Ann. Soc. Sci. Bruxelle   {\bf 47}, 49 (1927)}
\bibitem{stromberg}{\small G. Str\"omberg, Astrophys. J.   {\bf 61}, 353 (1925)}
\bibitem{hubble}{\small E. Hubble, Proc. Nat. Acad. Sci.  {\bf 15}, 168 (1929)}
\bibitem{lemaitre1931}{\small G. Lema\^itre, Monthly Not. Roy. Astron. Soc.  {\bf 41}, 483 (1931)}
\bibitem{robertson1928}{\small H.P. Robertson, Phil. Mag.  {\bf 5}, 835 (1928)}
\bibitem{robertson1929}{\small H.P. Robertson, Proc. Nat. Acad. Sci.  {\bf 15}, 822 (1929)}
\bibitem{deSitter1930a}{\small W. de Sitter, The Observatory {\bf 53}, 37 (1930)}
\bibitem{eddington}{\small A.S. Eddington, Monthly Not. Roy. Astron. Soc.  {\bf 90}, 668 (1930)}
\bibitem{sitter1930b}{\small W. de Sitter, Bull. Astron. Neth.  {\bf 185}, 157 (1930)}
\bibitem{sitter1930c}{\small W. de Sitter, Bull. Astron. Neth.  {\bf 193}, 211 (1930)}
\bibitem{einsteinzero}{\small A. Einstein, Sitz. K\"onig. Preu. Akad. Wiss., 235 (1931)}
\bibitem{eds}{\small A. Einstein, W. de Sitter, Proc. Nat. Acad. Sci.  {\bf 18}, 213 (1932)}
\bibitem{lemaitresingular}{\small G. Lema\^itre, Monthly Not. Roy. Astron. Soc.  {\bf 91}, 490 (1931)}
\bibitem{eddington1931}{\small A.S. Eddington, Nature {\bf 127}, 447 (1931)}
\bibitem{lemaitreNewton}{\small G. Lema\^itre, L'expansion de l'espace, Revue des questions scientifiques {\bf 20}, 391 (1931)}
\bibitem{lemaitre1933}{\small G. Lema\^itre, Ann. Soc. Sci. Bruxelles   {\bf 53}, 51 (1933)}
\bibitem{hoyle}{\small F. Hoyle, Monthly Not. Roy. Astron. Soc.  {\bf 108}, 372 (1948)}
\bibitem{pw}{\small A.A. Penzias, R.W. Wilson, Astrophys. J.   {\bf 142}, 419 (1965)}
\bibitem{gamow}{\small G. Gamow, Phys. Rev.   {\bf 74}, 505 (1948)}
\bibitem{dn}{\small A.G. Doroshkevich, I.D. Novikov, Dokl. Akad. Nauk  {\bf 154}, 809 (1964)}
\bibitem{dicke}{\small R.H. Dicke, P.J.E. Peebles, P.G. Roll, D.T. Wilkinson, Astrophys. J.   {\bf 142}, 414 (1965)}
\bibitem{lemaitrenature}{\small G. Lema\^itre, Nature  {\bf 127}, 706 (1931)}
\bibitem{guth}{\small A.H. Guth, Phys. Rev. D {\bf 23}, 347 (1981); A.D. Linde, Phys. Lett. B {\bf 108}, 389 (1982); A. Albrecht, P.J. Steinhardt, M.S. Turner, F. Wilczek, Phys. Rev. Lett. {\bf 48}, 1437 (1982)}
\bibitem{linde}{\small A. Linde, Particle Physics and Inflationary Cosmology (Harwood, Chur, Switzerland, 1990)}
\bibitem{c1}{\small P.H. Chavanis, Phys. Rev. D {\bf 84}, 043531 (2011)}
\bibitem{c2}{\small P.H. Chavanis, L. Delfini, Phys. Rev. D {\bf 84}, 043532 (2011)}
\bibitem{c3}{\small P.H. Chavanis, Phys. Rev. D {\bf 84}, 063518 (2011)}
\bibitem{recent}{\small T. Harko, F. Lobo, Phys. Rev. D {\bf 83}, 124051 (2011); P.H. Chavanis, T. Harko, [arXiv:1108.3986]; T. Matos, A. Su\'arez, Europhys. Lett. {\bf 96}, 56005 (2011); D. Bettoni, S. Liberati, L. Sindoni, JCAP {\bf 11}, 007 (2011); J. Maga\~na, T. Matos, V.  Robles, A. Su\'arez, [arXiv:1201.6107]; T. Harko, E. Madarassy, JCAP {\bf  01}, 020 (2012); V. Lora, J. Maga\~na, A. Bernal, F.J. S\'anchez-Salcedo, E.K. Grebel, JCAP {\bf  02}, 011 (2012); S.L. Liebling, C. Palenzuela, Living Rev. Relat. {\bf 15} (2012);  T. Rindler-Daller, P. R. Shapiro, Monthly Not. Roy. Astron. Soc. {\bf 422}, 135 (2012); V.H. Robles, T. Matos, Monthly Not. Roy. Astron. Soc. {\bf 422}, 282 (2012); F. Briscese, M. Grether, M. de Llano, Europhys. Lett. {\bf 98}, 60001 (2012)}
\bibitem{dalfovo}{\small F. Dalfovo, S. Giorgini, L.P. Pitaevskii, S. Stringari, Rev. Mod. Phys. {\bf 71}, 463 (1999)}
\bibitem{chandra}{\small S. Chandrasekhar, An Introduction to the Study of Stellar Structure (Dover, 1958)}
\bibitem{harko}{\small T. Harko, Monthly Not. Roy. Astron. Soc. {\bf 413}, 3095 (2011)}
\bibitem{harko2}{\small T. Harko, Phys. Rev. D {\bf 83}, 123515 (2011)}
\bibitem{c4}{\small P.H. Chavanis, Astron. Astrophys. {\bf 537}, A127 (2012)}
\bibitem{kl}{\small B. Kain, H.Y.  Ling, Phys. Rev. D {\bf 85}, 023527 (2012)}
\bibitem{vw}{\small  H. Velten, E. Wamba, Phys. Lett. B {\bf 709}, 1 (2012)}
\bibitem{harko3}{\small  T. Harko, G. Mocanu, Phys. Rev. D {\bf 85}, 084012 (2012)}
\bibitem{weinberg}{\small S. Weinberg, Gravitation and Cosmology (John Wiley \& Sons, 1972)}
\bibitem{bt}{\small J. Binney, S. Tremaine, Galactic Dynamics (Princeton University Press, 2008)}
\bibitem{milne}{\small E.A. Milne, Quarterly J. Math. {\bf 5}, 64 (1934)}
\bibitem{mcCreamilne}{\small W.H. McCrea, E.A. Milne, Quarterly J. Math., {\bf 5}, 73 (1934)}
\bibitem{mcCrea}{\small W.H. McCrea,  Proc. R. Soc. London. {\bf 206}, 562 (1951)}
\bibitem{harri}{\small E.R. Harrison, Ann. Phys. (N.Y.) {\bf 35}, 437 (1965)}
\bibitem{lima}{\small J.A.S. Lima, V. Zanchin, R. Brandenberger, Monthly Not. Roy. Astron. Soc.  {\bf 291}, L1 (1997)}
\bibitem{lemaitre1934}{\small G. Lema\^itre, Proc. Nat. Acad. Sci.  {\bf 20}, 12 (1934)}
\bibitem{peeblesbook}{\small P.J.E. Peebles, The Large-Scale Structure of the Universe (Princeton University Press, 1980)}
\bibitem{sakharov}{\small A.D. Sakharov, Dokl. Akad. Nauk SSSR {\bf 177}, 70 (1967)}
\bibitem{zeldovich}{\small Ya. B. Zeldovich, Sov. Phys. Uspek. {\bf 11}, 381 (1968)}
\bibitem{weinbergcosmo}{\small S. Weinberg, Rev. Mod. Phys. {\bf 61}, 1 (1989)}
\bibitem{cst}{\small E.J. Copeland, M. Sami, S. Tsujikawa, Int. J. Mod. Phys. D  {\bf 15}, 1753 (2006)}
\bibitem{paddybook}{\small T. Padmanabhan, Theoretical Astrophysics, Volume III: Galaxies and Cosmology (Cambridge University Press, 2002)}
\bibitem{luminetnew}{\small J.P. Luminet, Gen. Rel. Grav. {\bf 43}, 2911 (2011)}
\bibitem{fedichev}{\small P.O. Fedichev, Yu. Kagan, G.V. Shlyapnikov, J.T.M. Walraven, Phys. Rev. Lett. {\bf 77}, 2913 (1996)}



\end{thebibliography}
\end{document}